\documentclass{article}

\usepackage{arxiv}

\usepackage[utf8]{inputenc} % allow utf-8 input
\usepackage[T1]{fontenc}    % use 8-bit T1 fonts

\usepackage{url}            % simple URL typesetting
\usepackage{booktabs}       % professional-quality tables
\usepackage{nicefrac}       % compact symbols for 1/2, etc.
\usepackage{microtype}      % microtypography

\usepackage{graphicx}

\usepackage[square,sort,comma,numbers]{natbib}
\usepackage{doi}

\title{Domain Decomposition of Stochastic PDEs: Development of Probabilistic Wirebasket-based Two-level Preconditioners
}

\author{
  Ajit Desai\thanks{Currently at the Bank of Canada} \\
  Department of Civil and Environmental Engineering\\
  Carleton University \\
  Ottawa, ON, Canada \\
   \And
  Mohammad Khalil\\
  Quantitative Modeling \& Analysis Department\\
  Sandia National Laboratories \\
  Livermore, CA, United States\\
   \And
  Chris L. Pettit\\
  Aerospace Engineering Department \\
  US Naval Academy\\
  Annapolis, MD, United States\\
   \And
  Dominique Poirel \\
  Department of Mechanical and Aerospace Engineering\\
  Royal Military College of Canada \\
  Kingston, ON, Canada \\
   \And
  Abhijit Sarkar \\
  Department of Civil and Environmental Engineering\\
  Carleton University \\
  Ottawa, ON, Canada\\
 \texttt{abhijit.sarkar@carleton.ca} \\
}

\usepackage{lineno}
\usepackage{lipsum}
\usepackage{amsfonts}
\usepackage{graphicx}
\usepackage{epstopdf}
\usepackage{hyperref}       % hyperlinks

\usepackage{amsmath}
\usepackage{cleveref}
\usepackage{amssymb}
\usepackage{amsopn}
\usepackage{algorithmicx}
\usepackage[ruled]{algorithm}
\usepackage{algpseudocode}
\usepackage{algpascal}
\usepackage{threeparttable}

\ifpdf
  \DeclareGraphicsExtensions{.eps,.pdf,.png,.jpg}
\else
  \DeclareGraphicsExtensions{.eps}
\fi
\modulolinenumbers[5]

\usepackage{amsmath,amsthm}
\usepackage{amsfonts,amssymb,latexsym}
\usepackage[mathscr]{euscript}
\usepackage{times}
\usepackage{enumerate}
\usepackage{enumitem}
\usepackage{cancel}
\usepackage{placeins}
\usepackage{acronym}
\usepackage{fancyhdr}
\usepackage{textcomp}
\usepackage{appendix}
\usepackage{afterpage,amsbsy}
\usepackage[center]{caption}
\usepackage[cal=boondox]{mathalfa}
\usepackage{arydshln}
\usepackage{subfig}
\usepackage{rotating}
\usepackage{tikz}
\usetikzlibrary{shapes,arrows}
\usepackage{algorithmicx}
\usepackage[ruled]{algorithm}
\usepackage{algpseudocode}
\usepackage{algpascal}
\usepackage{setspace}
\usepackage{listings}
\lstdefinestyle{customc}{
  belowcaptionskip=1\baselineskip,
  breaklines=true,
  frame=single,
  xleftmargin=\parindent,
  language=Fortran,
  showstringspaces=false,
  basicstyle=\footnotesize\ttfamily,
  keywordstyle=\bfseries\color{green!40!black},
  commentstyle=\itshape\color{purple!40!black},
  identifierstyle=\color{blue},
  stringstyle=\color{orange},
}

\lstdefinestyle{customm}{
  belowcaptionskip=1\baselineskip,
  breaklines=true,
  frame=single,
  xleftmargin=\parindent,
  language=Matlab,
  showstringspaces=false,
  basicstyle=\footnotesize\ttfamily,
  keywordstyle=\bfseries\color{green!40!black},
  commentstyle=\itshape\color{purple!40!black},
  identifierstyle=\color{blue},
  stringstyle=\color{orange},
}

\lstdefinestyle{customp}{
  belowcaptionskip=1\baselineskip,
  breaklines=true,
  frame=single,
  xleftmargin=\parindent,
  language=Python,
  showstringspaces=false,
  basicstyle=\footnotesize\ttfamily,
  keywordstyle=\bfseries\color{green!40!black},
  commentstyle=\itshape\color{purple!40!black},
  identifierstyle=\color{blue},
  stringstyle=\color{orange},
}
\Crefname{equation}{Eq.}{Eqs.}
\Crefname{figure}{Fig.}{Figs.}
\Crefname{tabular}{Tab.}{Tabs.}
\Crefname{section}{Sec.}{Secs.}

\DeclareMathAlphabet\mathbfcal{OMS}{cmsy}{b}{n}

\def\b0{\mbox{\boldmath $0$}}

\newenvironment{frcseries}{\fontfamily{frc}\selectfont}{}

\newlength{\figwidth}
\begin{document}
\maketitle

\begin{abstract}
Realistic physical phenomena exhibit random fluctuations across many scales in the input and output processes. Models of these phenomena require stochastic PDEs.
For three-dimensional coupled (vector-valued) stochastic PDEs (SPDEs), for instance, arising in linear elasticity, the existing two-level domain decomposition solvers  with the vertex-based coarse grid show poor numerical and parallel scalabilities. Therefore, new algorithms with a better resolved coarse grid are needed.
The probabilistic wirebasket-based coarse grid for a two-level solver is devised in three dimensions. This enriched coarse grid provides an efficient mechanism for global error propagation and thus improves the convergence. This development enhances the scalability of the two-level solver in handling stochastic PDEs in three dimensions.
Numerical and parallel scalabilities of this algorithm  are studied using MPI and PETSc libraries on high-performance computing (HPC) systems.
Implementational challenges of the intrusive spectral stochastic finite element methods (SSFEM)  
are addressed by coupling domain decomposition solvers with FEniCS general purpose finite element package. This work generalizes the applications of intrusive SSFEM to tackle a variety of stochastic PDEs and emphasize the usefulness of the domain decomposition-based solvers and HPC for uncertainty quantification.
\end{abstract}

%\linenumbers

%%*****%%
\section{Introduction}
The stochastic modeling  of realistic problems in engineering mechanics using SSFEM may entail solving a linear (or linearized) system with millions of unknowns exploiting HPC.  Domain decomposition method (DDM) can provide the necessary platform to develop fast and efficient solvers for large-scale stochastic systems for HPC~\cite{sarkarIJNME2009,subberJCP2014,subberCMAME2013,subber2012PhDTh,desai2017scalable,desai2019scalable}.
These challenges motivated Sarkar et al.~\cite{sarkarIJNME2009, subberJCP2014,subberCMAME2013,subber2012PhDTh} to formulate and employ one-level and two-level DDM based solvers for stochastic PDEs with a few random variables. Building upon their work, the authors extended these non-overlapping DDM algorithms for a large number of random variables involving two-dimensional stochastic PDEs~\cite{desai2017scalable,desai2019scalable}.

In this paper, however, the primary attention is given to the development and efficient parallel implementation of scalable DD solvers for three-dimensional elliptic stochastic PDEs with a large number of random variables.
The wirebasket-based coarse grid~\cite{bramble1989construction,smith1991domain} is extended to the cases of stochastic PDEs to improve the performance and scalability aspects of the DD solvers.

A large-scale system of equations arising from the application of domain decomposition algorithms to the problems in three dimensions is computationally more demanding than the problems in two dimensions. This is due to the following reasons: (a) the complex geometry of the global interface and (b) the complicated coupling among the subdomains in three dimensions.

For these reasons, the domain decomposition methods such as two-level Balancing Domain Decomposition by Constraints (BDDC) or
Neumann-Neumann with a Coarse problem (NNC) and Dual-Primal Finite Element Tearing and Interconnecting (FETI-DP) with the vertex-based coarse grid are inefficient~\cite{smith2004domain,toselli2005domain}. The effects are more pronounced for the vector-valued PDEs in three dimensions due to complex spatial coupling among the elements of the solution vector~\cite{smith2004domain,smith1991domain,smith1992optimal}.

Moreover, the probabilistic analysis using the intrusive polynomial chaos expansion (PCE) method adds an additional coupling structure in the resulting system of equations and further complicates the problem~\cite{subber2012PhDTh,ghosh2009feti}. The coupling among the PCE coefficients of the solution process and the resulting block structure of the stochastic submatrices further influence the condition number of the system resulting in SSFEM~\cite{subber2012PhDTh,ghosh2009feti,desai2017scalable}. Therefore, the direct application of the Schur complement based two-level domain decomposition algorithms developed for two-dimensional stochastic PDEs to three-dimensional problems does not give scalable performance~\cite{subber2012PhDTh,desai2019scalable}.

In the cases of domain decomposition methods for stochastic PDEs in two dimensions developed earlier in \cite{desai2017scalable,desai2019scalable},
scalable performance is achieved by providing a mechanism for global communication using vertex-based coarse grid. The coarse problem solved in the vertex-based coarse grid provides the solution only at the corner nodes (or vertices), i.e.,   the nodes at the ends of interface edge plus the nodes shared among three or more subdomains.
The solution at the remaining interface nodes, i.e., all other interface nodes excluding vertices, is computed using the available solution at the corner nodes.
In three dimensions, this procedure can introduce significant inaccuracy due to an inefficient coarse grid correction resulting from the complicated coupling among the subdomains~\cite{smith1991domain,smith2004domain}.

In the deterministic settings, the condition number bound for the vertex-based method in two dimensions is shown as ~\cite{smith1991domain,smith2004domain,toselli2005domain}
\begin{equation}\label{eq:cond_vertex2D}
\mathcal{k} \leq C(1+{\mathrm{log}}(H/h))^2,
\end{equation}
where $C$ is a positive coefficient independent of the size of subdomains ($H$) and the size of element ($h$). These bounds are quite satisfactory in two dimensions. However, for the same method in three dimensions, the condition number bound is shown to be~\cite{smith1991domain,smith2004domain,toselli2005domain}
\begin{equation}\label{eq:cond_vertex3D}
\mathcal{k} \leq C(H/h)(1+{\mathrm{log}}(H/h)).
\end{equation}
The presence of the additional multiplier $(H/h)$ in the above equation (in contrast to \Cref{eq:cond_vertex2D}) leads to poor scalability in three-dimensional case.

\begin{figure}[htbp]
\centering
 \includegraphics[width=0.6\textwidth]{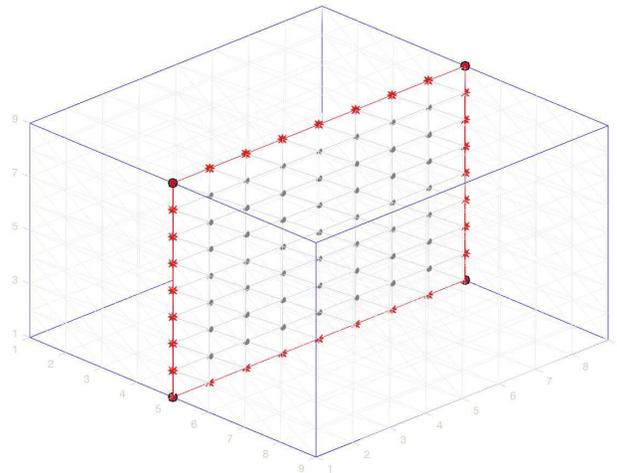}
 \caption{Schematic representation of a simple wirebasket coarse grid for a cube partitioned into two subdomains, showing ({\color{red}{-}}) for the global interface edge, ({\color{black}$\bullet$}) as vertices, ({\color{red}{$\star$}}) as interface-edges and ({\color{gray}$\bullet$}) as interface-faces.}
 \label{fig:SimpleWireBasket}
\end{figure}

\begin{figure}[htbp]
 \centering
 \subfloat[\label{subfig:grid2}]{
 \includegraphics[width=0.68\textwidth,height=0.42\textheight]{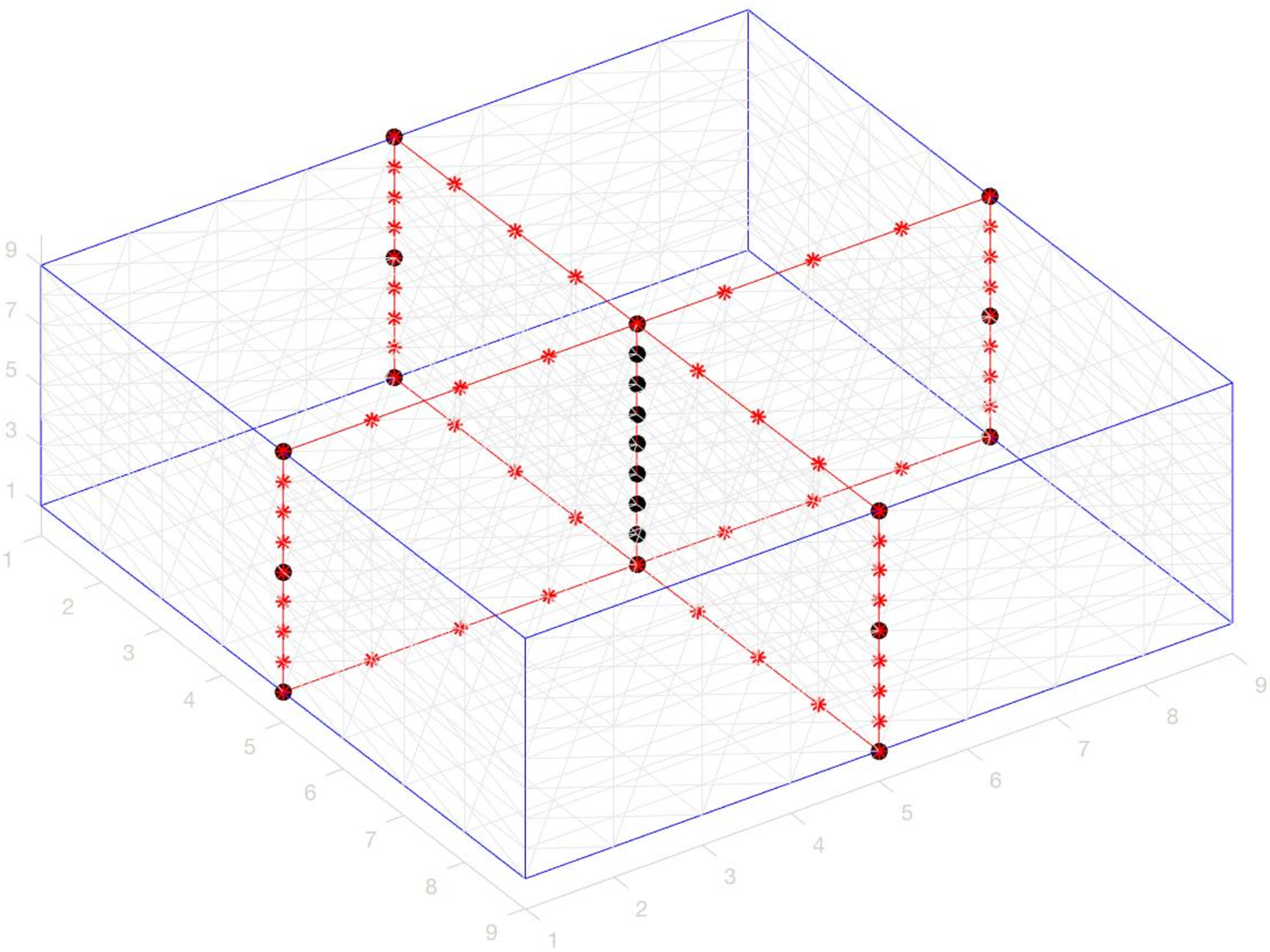}} \
 \subfloat[\label{subfig:grid3}]{
 \includegraphics[width=0.68\textwidth,height=0.42\textheight]{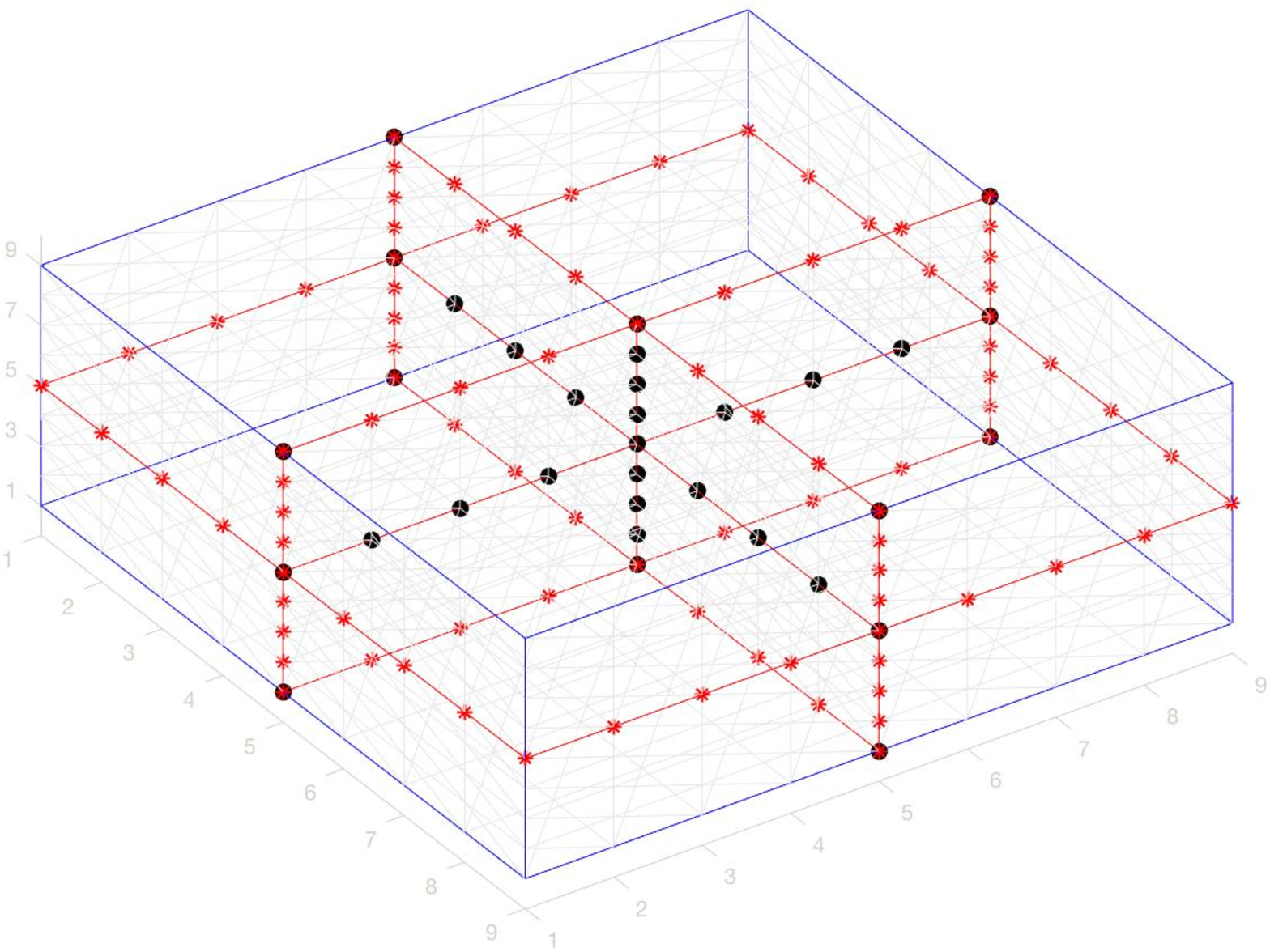}}
 \caption{\label{fig:wireBasket2} Schematic representation of wirebasket coarse grid for a cube partitioned into (a)
  four subdomains and (b) eight subdomains, showing ({\color{red}{-}}) for the global interface edge, ($\bullet$) as vertices and ({\color{red}{$\star$}}) as interface-edges.}
\end{figure}

These issues motivated Bramble~\cite{bramble1989construction}, Smith~\cite{smith1991domain}, Dryja~\cite{dryja1990some} and Mandel~\cite{mandel1990two,mandel1990iterative} to work with the wirebasket-based coarse grid as shown in~\Cref{fig:SimpleWireBasket} and~\Cref{fig:wireBasket2}.
The new coarse grid involves (a) vertices ({\color{black}$\bullet$}), i.e., the  nodes at the ends of interface edge plus the nodes shared among three or more subdomains and (b) the remaining nodes on the interface edge i.e., the remaining interface-edges (for brevity we omit the word nodes) ({\color{red}{$\star$}}). Together (a) and (b) forms a wirebasket as shown in~\Cref{fig:SimpleWireBasket} and~\ref{fig:wireBasket2}. 
The interface edge ({\color{red}{-}}) is the boundary of the interface and the nodes on the interface edge (except vertices) are referred as interface-edges for brevity. The interface-faces ({\color{gray}$\bullet$}) are the nodes shared between two subdomains excluding interface-edges.
Note that, the interface in three dimensions is a two-dimensional having a surface defined by interface face
and boundary defined by interface edge. 

The wirebasket-based coarse grid can provide an efficient mechanism for global communication of information, and therefore can help in designing more efficient coarse correction procedure~\cite{smith1991domain,smith2004domain,toselli2005domain}.
The condition number bound for the wirebasket-based approach (in the deterministic setting) is given by (refer the following articles and references therein for the theoretical proof~\cite{smith1991domain,smith2004domain,toselli2005domain}),

\begin{equation}
\mathcal{k} \leq C(1+{\mathrm{log}}(H/h))^2.
\end{equation}

In the context of stochastic PDEs, the direct application of the vertex-based coarse grid to the problems in three dimensions result in poor scaling behavior~\cite{subber2012PhDTh}.
This is due to a poorly conditioned system of equations arising in the setting of vertex-based domain decomposition methods
and weak mechanism for global error propagation~\cite{subber2012PhDTh}.
Moreover, the application of SSFEM further complicates the system due to the coupling among the PCE coefficients of the solution process. This results into block structure of the subdomain-level decomposed matrices, further affecting the condition number of resulting extended Schur complement system matrix~\cite{subber2012PhDTh}.

Taking inspiration from the wirebasket-based approaches developed for the deterministic PDEs in three dimensions~\cite{bramble1989construction,smith1991domain,toselli2005domain,mandel1990iterative}, we propose the extended (to the stochastic case) wirebasket-based coarse grid with the BDDC/NNC solver for the stochastic PDEs in three dimensions.
The BDDC/NNC solver with the extended wirebasket-based coarse grid is then employed to simulated large-scale, scalar and vector-valued stochastic PDEs with a large number of random variables.
To this end, the contributions made in this paper are broadly categorized next.

\begin{itemize}
\item The extended wirebasket-based coarse grid for BDDC/NNC solver is developed to overcome the scalability issues of two-level DD solvers with the vertex-based coarse grid in three dimensions~\cite{smith2004domain,subber2012PhDTh}.
The new coarse grid, which includes both vertices and interface-edges to form a wirebasket, strengthens the mechanism for global error propagation and correction, and thus improves the convergence.

\item The BDDC/NNC solver with the wirebasket coarse grid is shown to outperform the vertex-based coarse grid, in regards to the numerical scalability with respect to the number of subdomains and number of random variables. The superiority of the wirebasket-based coarse grid is demonstrated for the stochastic diffusion equation and the stochastic equations of linear elasticity in three dimensions.

\item  Efficient implementational strategies deployed in \cite{desai2017scalable,desai2019scalable} are used to improve the capabilities of the wirebasket-based BDDC/NNC solver to tackle three-dimensional stochastic PDEs characterized by a large number of random variables.
The three-level sparse iterative solver which employs an efficient preconditioner at each level is devised and implemented using various sparse data-structures and routines from MPI, PETSc, FEniCS and UQTk~\cite{desai2017scalable}.

\item Utilizing the subdomain-level stochastic block matrix assembly procedure presented in~\cite{desai2019scalable},  
the BDDC/NNC solver with the extended wirebasket-based coarse grid is coupled with the FEniCS deterministic finite element package~\cite{logg2012FEniCS,fenics/dolfin17}.
The coupling of BDDC/NNC solver with FEniCS reduces the implementational complexity required in the DD based intrusive SSFEM code development for the three-dimensional stochastic PDEs~\cite{le2010spectral,eldred2009comparison,reagana2003uncertainty}.

\item Finally, we also present a step-by-step approach to simplify the application of intrusive SSFEM for a coupled stochastic PDE system, such as equations of linear elasticity. In such cases, the utilization of intrusive SSFEM leads to the two-levels of couplings structure, i.e., the first is due to vector-valued solution process of the coupled PDE system, and the second is due to the interdependence of PCE coefficients of the solution process. Efforts have been made to simplify the process and provide guidance to navigate  such complexity. 

\end{itemize}

In this paper, no attempts have been made to perform the theoretical convergence analysis and condition number bound estimation of the preconditioned extended Schur complement system in the intrusive SSFEM setting.
However, extensive numerical experiments are conducted to thoroughly investigate the performance of the extended wirebasket-based BDDC/NNC solver.
Modern HPC clusters (available in compute Canada~\cite{nationalSystems}) are exploited to demonstrate the utility of BDDC/NNC solver with the wirebasket-based coarse grid to tackle large-scale stochastic PDEs with a large number of random variables.

First, the numerical scalability concerning the number of PCGM iterations to solve the extended Schur complement system is investigated with respect to the following:
(a) increasing spatial mesh resolution with a fixed number of PCE terms,
(b) increasing number of subdomains with increasing number of PCE terms for a fixed problem size per subdomains,
(c) increasing number of subdomains for a fixed number of PCE terms and a fixed mesh resolution,
(d) increasing number of random variables for a fixed order of expansion, and a fixed mesh resolution; and 
(e) increasing order of expansion for a fixed number of random variables and a fixed mesh resolution.

Secondly, the parallel scalability of the extended wirebasket-based BDDC/NNC solver is measured with respect to the strong and weak scaling~\cite{keyes1987comparison,smith2004domain}.
In the strong scalability test, 
the global problem size is kept constant and the number of cores used to solve the problem is increased to reduce the total execution time. For the weak scalability test, 
the problem size per subdomain is fixed and the global size of the problem is increased by adding more subdomains~\cite{keyes1987comparison,smith2004domain}.

Finally, the scalability study of the wirebasket-based BDDC/NNC solver concerning stochastic parameters such as the number of random variables and order of expansion is presented.
The scalability of the solver with respect to stochastic parameters is crucial for uncertainty quantification. That is because, as the number of PCE terms increases, the block-sparsity structure of the underlying intrusive SSFEM system matrix changes. 
These effects are more substantial in the cases of the three-dimensional coupled PDE system due to the additional coupling in the spatial domain. Therefore, the scalability plots concerning stochastic parameters provide useful information regarding the effectiveness 
of the solver for large-scale stochastic simulations.
The scalability study concerning stochastic parameters is somewhat unusual compared to the scalability studies commonly conducted for the deterministic FEM. That is because, in the deterministic setting, changing mesh density or the values of the system parameters, does not influence the block sparsity pattern of the underlying system matrix.

In the following sections, we first discuss the formulation of
extended Schur complement of a coupled stochastic PDE system in \Cref{sec:exSchurE}. Next the formulation of extended wirebasket-based BDDC/NNC preconditioner for the coupled stochastic PDE system is presented in \Cref{sec:NNW}.
 This is followed by the application of extended wirebasket-based BDDC/NNC solver to the three-dimensional stochastic diffusion equation to model flow through random media in \Cref{sec:3DP}.
The application of BDDC/NNC solver with the extended wirebasket-based coarse grid 
to a stochastic linear elasticity problem is presented in \Cref{sec:3DE}.
For both stochastic Poisson and elasticity problems, the random system parameters are modeled as non-Gaussian stochastic processes characterized by using a large number of RVs (up to 15 RVs). Finally, in~\Cref{sec:conclusion} we conclude our findings.  Various algorithms demonstrating implementation of the BDDC/NNC solvers are listed in Appendix~\ref{sec:ddmImplementation}.

\section{Extended Schur Complement of Coupled Stochastic PDE System}\label{sec:exSchurE}
\label{sec:schurComplement}
In this section, we formulate an extended Schur complement system for a coupled stochastic PDE system.
For demonstration we consider the equations of linear elasticity in three-dimensions to model the vector-valued stochastic displacement field.

The finite element discretization of a coupled stochastic PDE system defined over physical domain $\mathbfcal{D}(x,y,z)$ as a function of random event $\theta$, leading to the  stochastic linear system,
\begin{align}
\mathrm{\bf{A}}(\theta)\mathrm{\bf{u}}(\theta) = \mathrm{\bf{f}}
\end{align}
where $\mathrm{\bf{A}}(\theta)$ being the stochastic coefficient matrix.

The vector-valued stochastic response field $\mathrm{\bf{u}}(\theta)$ can be denoted as,
$$\mathrm{\bf{u}}(\theta)= \{ \mathrm{\bf{u}}_x(\theta), \mathrm{\bf{u}}_y(\theta), \mathrm{\bf{u}}_z(\theta) \}. $$
For simplicity, the vector $\mathrm{\bf{f}}$ is considered to be a deterministic source term. However, the methodology presented here can be easily extended to the stochastic source term~\cite{ghanem1999ingredients,ghanemSFEM1991}.

\begin{figure}[htbp]
\centering
 \includegraphics[width=0.7\textwidth]{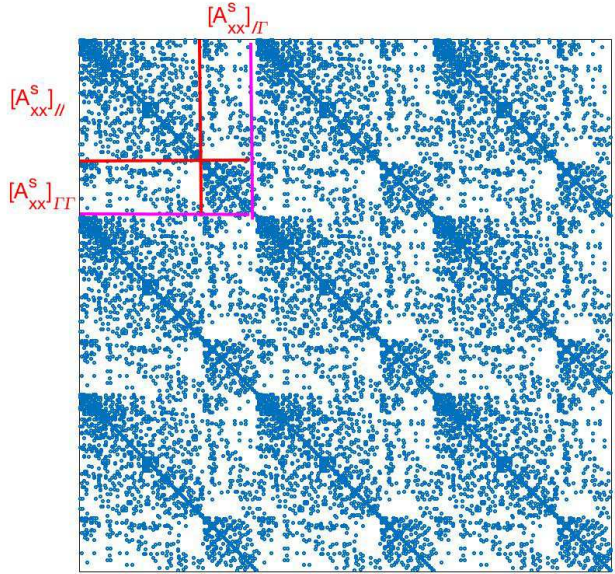}
 \caption{Subdomain-level system matrix $\mathrm{\bf{A}}^s(\theta)$
 computed using the mean system parameters, indicating coupling   among $x$, $y$ and $z$ displacements. It also displays decomposed components for $x$ displacements according to interior $I$ and interface $\it{\Gamma}$ nodes.}
 \label{fig:3DcouplingStr}
\end{figure}

To formulate the Schur complement system, we divide the domain $\mathbfcal{D}$ into $n_s$ non-overlapping subdomains with $s = 1,2, \dots, n_s$. Accordingly, the subsystem for a typical subdomain $s$  can be expressed as~\cite{subber2012PhDTh,desai2017scalable},
\begin{align}\label{eq:subEquilibrium}
\mathrm{\bf{A}}^s(\theta)\mathrm{\bf{u}}^s(\theta) = \mathrm{\bf{f}}^s.
\end{align}
The local system matrix $\mathrm{\bf{A}}^s(\theta)$ can be expressed in terms of the contribution from the $x, y$ and $z$ coordinates. Therefore, the subdomain-level stochastic equilibrium system can be expanded as

\begin{equation}\label{eq:vecCompA}
{\begin{bmatrix}
    {\mathrm{\bf{A}}^s_{xx}}(\theta)    & {\mathrm{\bf{A}}^s_{xy}}(\theta)  & {\mathrm{\bf{A}}^s_{xz}}(\theta)
\\[0.3em]
      {\mathrm{\bf{A}}^s_{yx}}(\theta)    & {\mathrm{\bf{A}}^s_{yy}}(\theta)  & {\mathrm{\bf{A}}^s_{yz}}(\theta)
\\[0.3em]
      {\mathrm{\bf{A}}^s_{zx}}(\theta)    & {\mathrm{\bf{A}}^s_{zy}}(\theta)  & {\mathrm{\bf{A}}^s_{zz}}(\theta)
\end{bmatrix}}
\begin{Bmatrix}
      {\bf{u}}_{x}^s(\theta)
\\[0.3em]
      {\bf{u}}_{y}^s(\theta)
\\[0.3em]
      {\bf{u}}_{z}^s(\theta)
 \\[0.3em]
\end{Bmatrix}  =
\begin{Bmatrix}
      {\bf{f}}_{x}^s
\\[0.3em]
      {\bf{f}}_{y}^s
\\[0.3em]
      {\bf{f}}_{z}^s
 \\[0.3em]
\end{Bmatrix}.
\end{equation}
The sparsity structure of the subdomain-level block matrix $\mathrm{\bf{A}}^s(\theta)$ in \Cref{eq:vecCompA} for a typical subdomain $s$, assembled at the mean values of the stochastic system parameters for the equations of linear elasticity is shown in \Cref{fig:3DcouplingStr}.

% %(for mean values of the stochastic system parameters).

The stochastic solution vector ${\bf{u}}^s(\theta)$ for each subdomain $s$, has three components ${\bf{u}}_x^s(\theta)$, ${\bf{u}}_y^s(\theta)$, ${\bf{u}}_z^s(\theta)$ along spatial coordinates $x,y,z$ respectively.
Let us denote the submatrix and subvector components from \Cref{eq:vecCompA} using ${\mathrm{\bf{A}}^s_{\alpha\beta}}(\theta)$ and ${\bf{u}}_{\alpha}^s(\theta)$, where the $\alpha$ and $\beta$ represent $x,y$ and $z$ components.
Next, the solution vector components ${\bf{u}}_{\alpha}^s(\theta)$ are divided into the interior vector $({\bf{u}}_{\alpha}^s)_I(\theta)$ and interface vector $({\bf{u}}_{\alpha}^s)_{\it{\Gamma}}(\theta) $.
According to this decomposition, each submatrix and subvector in \Cref{eq:vecCompA} can be written as
\begin{equation}\label{eq:vecCompAalphaBeta}
{\mathrm{\bf{A}}^s_{\alpha\beta}} =
{\begin{bmatrix}
    ({\mathrm{\bf{A}}^s_{\alpha\beta}})_{II}(\theta)    & ({\mathrm{\bf{A}}^s_{\alpha\beta}})_{I\it{\Gamma}}(\theta)
\\[0.3em]
    ({\mathrm{\bf{A}}^s_{\alpha\beta}})_{\it{\Gamma} I}(\theta) & ({\mathrm{\bf{A}}^s_{\alpha\beta}})_{\it{\Gamma}\it{\Gamma}}(\theta)
\end{bmatrix}},
\end{equation}
\begin{align}
{\bf{u}}_{\alpha}^s(\theta) = \big[({\bf{u}}_{\alpha}^s)_I(\theta), ({\bf{u}}_{\alpha}^s)_{\it{\Gamma}}(\theta)\big]^{\mathrm{T}}.
\end{align}
For example, see~\Cref{fig:3DcouplingStr} which displays the decomposed matrix components $({\mathrm{\bf{A}}^s_{\alpha\beta}})_{\gamma\delta}(\theta)$, according to interior and interface nodes for ${\mathrm{\bf{A}}^s_{xx}}(\theta)$ matrix where $\gamma$ and $\delta$ represent $I$ and ${\it{\Gamma}}$.

\begin{figure}[htbp]
\centering
 \includegraphics[width=0.7\textwidth]{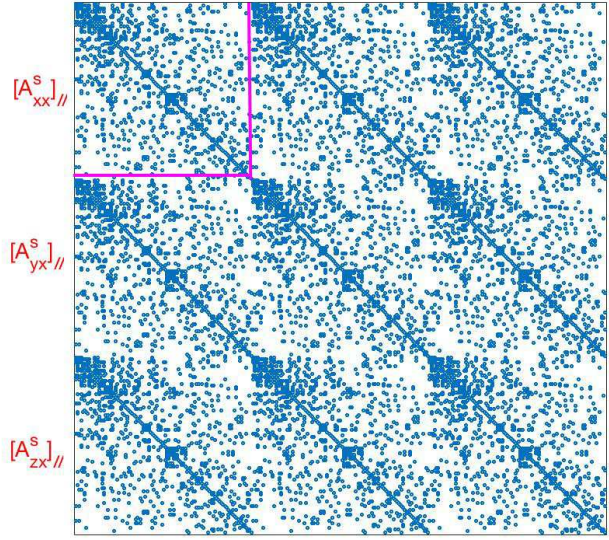}
 \caption{Subdomain-level stiffness matrix $\textbf{\textit{A}}^s_{II}(\theta)$ for interior nodes,
 computed using the mean system parameters, indicating coupling   among $x$, $y$ and $z$ displacements.}
 \label{fig:3DcouplingStrAII}
\end{figure}

The components of decomposed matrices and vectors which relate to interior nodes are given by
\begin{equation}\label{eq:vecComm_rearranged}
\textbf{\textit{A}}^s_{II}(\theta) =
{\begin{bmatrix}
 ({\mathrm{\bf{A}}^s_{xx}})_{II}(\theta)    & ({\mathrm{\bf{A}}^s_{xy}})_{II}(\theta)  & ({\mathrm{\bf{A}}^s_{xz}})_{II}(\theta)
\\[0.3em]
 ({\mathrm{\bf{A}}^s_{yx}})_{II}(\theta)    & ({\mathrm{\bf{A}}^s_{yy}})_{II}(\theta)  & ({\mathrm{\bf{A}}^s_{yz}})_{II}(\theta)
\\[0.3em]
      ({\mathrm{\bf{A}}^s_{zx}})_{II}(\theta)    & ({\mathrm{\bf{A}}^s_{zy}})_{II}(\theta)  & ({\mathrm{\bf{A}}^s_{zz}})_{II}(\theta)
\end{bmatrix}}
\end{equation}
\begin{equation}
{\textbf{\textit{u}}}_{I}^s(\theta) = \big[({\bf{u}}_x^s)_I(\theta), \big[({\bf{u}}_y^s)_I(\theta), \big[({\bf{u}}_z^s)_I(\theta) \big]^{\mathrm{T}}.
\end{equation}
\Cref{fig:3DcouplingStrAII} displays the block matrices pertaining to interior nodes for the matrix $\textbf{\textit{A}}^s_{II}(\theta)$ as in \Cref{eq:vecComm_rearranged}. % for a typical subdomain $s$.

Similar step can be taken to construct the other component matrices $({\mathrm{\bf{A}}^s_{\alpha\beta}})_{\gamma\delta}(\theta)$ and vectors $ ({\bf{u}}_{\alpha}^s)_{\gamma}(\theta) $.
%where $\gamma$ and $\delta$ represent the $I$ and ${\it{\Gamma}}$.
Consequently, the subsystem for a typical subdomain $s$ can be obtained by using rearranged component matrices and vectors as,

\begin{equation}\label{eq:subeq_rearranged}
{\begin{bmatrix}
     \textbf{\textit{A}}_{II}^s(\theta) & \textbf{\textit{A}}_{I \it{\Gamma}}^s(\theta)
\\[0.3em]
      \textbf{\textit{A}}_{\it{\Gamma}I}^s(\theta) & \textbf{\textit{A}}_{\it{\Gamma} \it{\Gamma}}^s(\theta)
\end{bmatrix}}
\begin{Bmatrix}
      \textbf{\textit{u}}_{I}^s(\theta)
\\[0.3em]
      \textbf{\textit{u}}_{\it{\Gamma}}^s(\theta)
 \\[0.3em]
\end{Bmatrix}  =
\begin{Bmatrix}
      \textbf{\textit{f}}_{I}^s
\\[0.3em]
      \textbf{\textit{f}}_{\it{\Gamma}}^s
 \\[0.3em]
\end{Bmatrix},
\end{equation}
where $\textbf{\textit{u}}_{I}^s(\theta)$ is the subdomain-level (local) interior solution vector relating to the interior nodes and $\textbf{\textit{u}}_{ \it{\Gamma}}^s(\theta)$ is the local interface solution vector pertaining to interface nodes.
The vectors $\textbf{\textit{f}}_{I}^s$ and $\textbf{\textit{f}}_{\it{\Gamma}}^s$  correspond to the source terms for interior and interface respectively.
%of the nodes shared by two or more adjacent subdomains and the physical boundary nodes as shown in fig~\ref{fig:vertexGrid}.

Next, by following these steps the extended (to the stochastic case) Schur complement system is constructed as follows:
\begin{itemize}
\item Incorporate PCE of the stochastic system matrix blocks, ${\textbf{\textit{A}}}^s_{\gamma\delta}(\theta) =  \sum^{P_{\textsc{a}}}_{i=0} {\bar{\textbf{\textit{A}}}}^s_{{\gamma\delta},i} {\Psi_i(\theta)}$, using the $P_{\textsc{a}}$ terms in the expansion and the stochastic solution process,
${\textbf{\textit{u}}}^s_{\gamma}(\theta) =  \sum^{P_{u}}_{j=0} {\bar{\textbf{\textit{u}}}}^s_{{\gamma},j}  \ {\Psi_j(\theta)}$ using $P_u$ terms in the expansion, into~\Cref{eq:subeq_rearranged}, where $\Psi_j(\theta)$ is the $j^{th}$ PC (a known  random variable) and ${\bar{\textbf{\textit{u}}}}_{{\gamma},j}$ are unknown deterministic quantities (to be computed). 
\item This is followed by the Galerkin projection along the random dimension, which involves, multiplying both sides of~\Cref{eq:subeq_rearranged} by ${\Psi_k(\theta)}$ with $k=\{ 0, 1, \dots P_u \}$ and taking expectation.
\item Enforcing the transmission conditions by performing global assembly along the interface unknowns,  the following extended Schur complement system is obtained (refer to the following articles for further details~\cite{subber2012PhDTh,desai2017scalable})
\end{itemize}
\begin{equation}\label{eq:3DE_Schur}
\mathbfcal{S} {\mathbfcal{U}}_{\it\Gamma} = \textbf{\textit{g}}_{\it\Gamma},
\end{equation}
$\mathbfcal{S}$ being the extended Schur complement system  for a coupled stochastic PDE system and $\textbf{\textit{g}}_{\it\Gamma}$ is the forcing vector. They are respectively defined as
\begin{align}\label{eq:coupledSchur}
\mathbfcal{S} &=  \sum_{s=1}^{n_s} \mathbfcal{R}_s^\mathrm{T} \Big[ \mathbfcal{A}_{{\it\Gamma} {\it\Gamma}}^s - \mathbfcal{A}_{\it{\Gamma} I}^s [{\mathbfcal{A}_{II}^s}]^{-1} \mathbfcal{A}_{I \it{\Gamma}}^s \Big] \mathbfcal{R}_s, \\ \label{eq:coupledSchur2}
\textbf{\textit{g}}_{\it\Gamma} &= \sum_{s=1}^{n_s} \mathbfcal{R}_s^\mathrm{T}  \Big[\mathbfcal{F}_{\it{\Gamma}}^s - \mathbfcal{A}_{\it{\Gamma} I}^s [\mathbfcal{A}_{II}^s]^{-1} \mathbfcal{F}_I^s \Big],
\end{align}
where the $\mathbfcal{R}_s$ denotes the restriction matrix defined later and
\begin{align}
\left[{\mathbfcal{A}_{\gamma\delta}^s}\right]_{jk} &= \sum_{i=0}^{P_{\textsc{A}}}\left< \psi_i \psi_j \psi_k \right> \bar{\textbf{\textit{A}}}_{\gamma\delta,i}^s, \label{eq:3d_Stoassembly} \\ 
\mathbfcal{U}_{\it\Gamma} &= \left[\bar{\textbf{\textit{u}}}_{\it\Gamma,0} \dots, \bar{\textbf{\textit{u}}}_{\it\Gamma,P_u}\right]^\mathrm{T}, \\
\mathbfcal{F}_{\it{\gamma},k}^s &= \left< \psi_k  \ \textbf{\textit{f}}^s_{\gamma} \right>.
\end{align}

The subdomain-level local matrices $\left[{\mathbfcal{A}_{\gamma\delta}^s}\right]_{jk}$ can be assembled by using \Cref{alg:smap} in Appendix~\ref{sec:ddmImplementation}. 

This procedure utilizes deterministic FEniCS assembly routines, which involve manipulating element stiffness matrices to handle stochasticity (see Appendix B in \cite{desai2019scalable} for further details).
The structure of the local matrices $\left[{\mathbfcal{A}_{\gamma\delta}^s}\right]$ have two levels of blocks. The first-level of blocks is due to couplings among $x,y$ and $z$ components and the second-level of blocks is due to couplings among the PCE coefficients. These couplings produce two-levels of block-sparsity structure in each of the subdomain level decomposed matrices $\left[{\mathbfcal{A}_{\gamma\delta}^s}\right]$.

The restriction operators $\mathbfcal{R}_s$ in \Cref{eq:coupledSchur} and \Cref{eq:coupledSchur2} can be obtained as
\begin{equation}
\mathbfcal{R}_s = blockdiagonal ( \textbf{\textit{R}}_{s,0},\dots, \textbf{\textit{R}}_{s,P_u}),
\end{equation}

where each of the blocks (with $i=\{ 0, 1, \dots, P_u \}$) can be obtained as
\begin{equation}\label{eq:vecRs}
\textbf{\textit{R}}_{s,i} = blockdiagonal ({\bf{R}}^{x,i}_s, {\bf{R}}^{y,i}_s, {\bf{R}}^{z,i}_s).
\end{equation}
The ${\bf{R}}^{\alpha,i}_s$ in \cref{eq:vecRs} (where $\alpha$ denotes $x, y$ or $z$) manages  gather or scatter operations  as follows,
\begin{align}
{({\bf{u}}_{\alpha,i}^s})_{\it{\Gamma}} &= \big[{\bf{R}}^{\alpha,i}_s \big] ({{\bf{u}}_{\alpha,i}})_{\it{\Gamma}}  , \\  ({{\bf{u}}_{\alpha,i}})_{\it{\Gamma}}  &= \big[{{\bf{R}}^{\alpha,i}_s} \big]^\mathrm{T} \ ({{\bf{u}}_{\alpha,i}^s})_{\it{\Gamma}}.
\end{align}

The matrix components, $\mathbfcal{A}^s_{\gamma\delta}$ in \Cref{eq:coupledSchur} and (\ref{eq:coupledSchur2}) (where $\gamma$ and $\delta$ representing $I$ or ${\Gamma}$) are considerably larger, complex and computationally intensive to assemble compared to $A^s_{\gamma\delta}$ for scalar-valued stochastic PDEs~\cite{desai2019scalable}. 
To simplify assembly procedure for stochastic matrix ${\mathbfcal{A}^s_{\gamma\delta}}$,
FEniCS-based deterministic assembly routines are employed. 
The probabilistic Schur complement system in \Cref{eq:3DE_Schur} for a stochastic PDE system is more complex and computationally expensive compared to the extended Schur complement system obtained in the case of scalar-valued stochastic PDEs~\cite{desai2019scalable}. 

For deterministic PDEs the solution of extended Schur complement system using two-level preconditioner with vertex-based coarse grid does not give scalable performance~\cite{subber2012PhDTh}. Therefore, the wirebasket-based coarse grid is required to overcome this challenge~\cite{smith1991domain,smith2004domain}. Therefore the probabilistic two-level preconditioner using extended wirebasket for stochastic PDE system is presented in the next section.

\section{Two-Level Preconditioner using Extended Wirebasket-Based Coarse Grid for SPDEs}
\label{sec:NNW}
Two-level preconditioner with a coarse problem is required  to achieve scalable performance with DD-based iterative solvers~\cite{subber2012PhDTh,desai2010analysis}.
To overcome the scalability issues of vertex-based coarse grid, we propose extended wirebasket-based coarse grid with the two-level BDDC/NNC preconditioner for SPDEs.

The wirebasket, as shown in~\Cref{fig:SimpleWireBasket},  includes the interface edges (nodes on the edges of the boundary {\color{red}{$\star$}}) along with the vertices (the nodes at the end of the interface plus the nodes shared by three or more subdomains {\color{black}$\bullet$}).
The wirebasket can provides an efficient mechanism for global error propagation during each iteration of BDDC/NNC based PCGM solver of the extended Schur complement system shown in~\Cref{eq:3DE_Schur}. 
The formulation of extended wirebasket-based BDDC/NNC preconditioner for SPDEs is briefly discussed below.

The local interface unknowns
${\bf{\mathbfcal{U}}}_{\it{\Gamma}}^s$ (for each subdomain $s$) are partitioned into faces ${\bf{\mathbfcal{U}}}_{\it{F}}^s$ and wirebasket ${\bf{\mathbfcal{U}}}_{\it{W}}^s$ such that
\begin{align}
\begin{Bmatrix}
      {\bf{\mathbfcal{U}}}_{\it{F}}^s
\\[0.3em]
      {\bf{\mathbfcal{U}}}_{\it{W}}^s
 \\[0.3em]
\end{Bmatrix}
=
\begin{Bmatrix}
      \mathbfcal{R}^F_s
\\[0.3em]
      \mathbfcal{R}^W_s
 \\[0.3em]
\end{Bmatrix}
{\bf{\mathbfcal{U}}}_{\it{\Gamma}}^s,
\end{align}
where the restriction operators are obtained as,
\begin{align}\label{eq:wb_rsmat1}
\mathbfcal{R}^F_s  &= blockdiagonal ( \textbf{\textit{R}}^F_{s,0},\dots, \textbf{\textit{R}}^F_{s,P_u}), \\ \label{eq:wb_rsmat2}
\mathbfcal{R}^W_s  &= blockdiagonal ( \textbf{\textit{R}}^W_{s,0},\dots, \textbf{\textit{R}}^W_{s,P_u}).
\end{align}
The $P_u$ is the number of PCE terms used in the characterization of solution process. The $\textbf{\textit{R}}^{\gamma}_{s,i}$ blocks in \Cref{eq:wb_rsmat1} and (\ref{eq:wb_rsmat2}) where $\gamma$ represents $F$ or $W$ and $i=\{0, 1, \dots, P_u\}$ are represented by the component matrices as follows:
\begin{align} \label{eq:wb_rsmatcomp}
\textbf{\textit{R}}^F_{s,i} &= blockdiagonal ({\bf{R}}^{Fx}_{s,i}, {\bf{R}}^{Fy}_{s,i}, {\bf{R}}^{Fz}_{s,i}), \\
\textbf{\textit{R}}^W_{s,i} &= blockdiagonal ({\bf{R}}^{Wx}_{s,i}, {\bf{R}}^{Wy}_{s,i}, {\bf{R}}^{Wz}_{s,i}).
\end{align}
The components ${\bf{R}}^{\gamma\delta}_{s,i}$ (with $\delta$ represents $x,y$ or $z$), acts as a gather or scatter operator between local faces/wirebasket and global faces/wirebasket components of the deterministic interface solution vectors.

Accordingly the global coarse problem arising in the setting of BDDC/NNC preconditioner is (refer to~\cite{subber2012PhDTh,desai2017scalable} for detailed procedure to construct coarse problem),
\begin{equation}\label{eq:coarseproblem_wb}
\mathbfcal{F}_{WW} \ \mathbfcal{U}_{W} = \mathcal{d}_{W},
\end{equation}
where $\mathbfcal{U}_{W}$ denotes global wirebasket unknowns, $\mathbfcal{F}_{WW}$ is the coarse operator and $\mathcal{d}_{W}$ is corresponding right-hand side vector, being expressed as

\begin{align}\label{eq:wirebasket_operators}
\mathbfcal{F}_{WW} &=  \sum_{s=1}^{n_s} {\mathbfcal{B}_W^s}^\mathrm{T} \Big(\mathbfcal{S}_{WW}^s - \mathbfcal{S}_{WF}^s [\mathbfcal{S}_{FF}^s]^{-1} \mathbfcal{S}_{FW}^s\Big)\mathbfcal{B}_W^s, \\
\mathcal{d}_{W} &= \sum_{s=1}^{n_s} {\mathbfcal{B}_W^s}^\mathrm{T} \Big(f_{W}^s - \mathbfcal{S}_{WF}^s [\mathbfcal{S}_{FF}^s]^{-1} f_{F}^s\Big),
\end{align}
where $\mathbfcal{S}^s_{\alpha\beta} = \mathbfcal{A}_{\alpha\beta}^s - \mathbfcal{A}_{\alpha I}^s [{\mathbfcal{A}_{II}^s}]^{-1} \mathbfcal{A}_{I \beta}^s$, where $\alpha$ and $\beta$ represent $W$ or $F$.

The $\mathbfcal{B}_W^s$ is the restriction operator that maps the global wirebasket unknowns to local wirebasket unknowns. The $\mathbfcal{B}_W^s$ can be obtained as,
\begin{equation}
\mathbfcal{B}_W^s = blockdiagonal ( \textbf{\textit{B}}^s_{W,0},\dots, \textbf{\textit{B}}^s_{W,P_u}),
\end{equation}
where each block (with $i=\{ 0, 1, \dots, P_u \}$) can be obtained as
\begin{equation}\label{eq:vecWs}
\textbf{\textit{B}}^s_{W,i} = blockdiagonal ({\bf{B}}^s_{W,x,i}, {\bf{B}}^s_{W,y,i}, {\bf{B}}^s_{W,z,i}).
\end{equation}
and  ${\bf{B}}^s_{W,\alpha,i}$ relates the global and local corner node solutions as:
\begin{equation}
{({\bf{u}}_{\alpha,i}^s})_{W} = \big[{\bf{B}}^s_{W,\alpha,i} \big] ({{\bf{u}}_{\alpha,i}})_{W}.
\end{equation}
where $\alpha$ represents $x$, $y$ and $z$.

Performing algebraic manipulations, the two-level BDDC/NNC preconditioner with the wirebasket-based coarse grid can be written as (refer~\cite{subber2012PhDTh,desai2017scalable} for further details),
\begin{equation}\label{eq:NNWP}
\mathbfcal{M}_{NNW}^{-1} = \sum_{s=1}^{n_s} \mathbfcal{R}_s^\mathrm{T} \mathbfcal{D}_s({\mathbfcal{R}_s^F}^\mathrm{T} [\mathbfcal{S}^s_{FF}]^{-1} \mathbfcal{R}_s^F) \mathbfcal{D}_s \mathbfcal{R}_s  + \mathbfcal{R}_0^\mathrm{T} [\mathbfcal{F}_{WW}]^{-1} \mathbfcal{R}_0.
\end{equation}

The $\mathbfcal{R}_0$ in \Cref{eq:NNWP} relates global fine grid and coarse grid components~\cite{subber2012PhDTh,desai2017scalable}, as given by
\begin{equation}\label{eq:NNWP_RO}
\mathbfcal{R}_0 = \sum_{s=1}^{n_s} {\mathbfcal{B}_W^s}^\mathrm{T} ({\mathbfcal{R}_s^W} - \mathbfcal{S}_{WF}^s [\mathbfcal{S}^s_{FF}]^{-1} \mathbfcal{R}_s^F) \mathbfcal{D}_s \mathbfcal{R}_s,
\end{equation}
where $\mathbfcal{D}_s$ is a diagonal scaling operator obtained as follows,
\begin{equation}
\mathbfcal{D}_s = blockdiagonal ( {\textbf{\textit{D}}}_{s,0},\dots,{\textbf{\textit{D}}}_{s,P_u})
\end{equation}
where ${\textbf{\textit{D}}}_{s,i}$ are obtained by using deterministic diagonal scaling matrices ${\bf{D}}^{\alpha}_{s,i}$ as,
\begin{equation} \label{eq:wb_rsmatcomp2}
\textbf{\textit{D}}_{s,i} = blockdiagonal ({\bf{D}}^{x}_{s,i}, {\bf{D}}^{y}_{s,i}, {\bf{D}}^{z}_{s,i}).
\end{equation}
where $i = \{ 0, 1, \dots, P_u \}$
and $\alpha$ represent $x$, $y$ and $z$.

From~\Cref{eq:NNWP}, it can be noted that the extended wirebasket-based BDDC/NNC preconditioner consists of two operators. First, the local-fine operator $({\mathbfcal{R}_s^F}^\mathrm{T} [\mathbfcal{S}^s_{FF}]^{-1} \mathbfcal{R}_s^F)$, requires a local solve $[\mathbfcal{S}^s_{FF}]^{-1}$ on each of the subdomains for each PCGM iteration for interface-face nodes.
Second, the global coarse operator $\mathbfcal{R}_0^\mathrm{T} [\mathbfcal{F}_{WW}]^{-1} \mathbfcal{R}_0$, involves a global coarse problem solve $ [\mathbfcal{F}_{WW}]^{-1}$ for each PCGM iteration for the wirebasket nodes.
This coarse operator provides an efficient mechanism for global error propagation during each application of preconditioner, therefore, accelerates the convergence of the PCGM solver~\cite{smith1991domain,smith2004domain,mathew2008domain}.

%%======
In this following sections, we utilize the extended wirebasket-based coarse grid presented in this section to solve both scalar and vector-valued stochastic PDEs in three-dimensions.
We first consider steady-state stochastic diffusion equation in three-dimensions to model the scalar-valued stochastic solution process. This is followed by the solutions of the equations of linear elasticity in three-dimensions having stochastic parameters.

%%%%%%%%%%%%%%%%%%%%%%%%%%%%%%%%%%%%%%%%%
\section{Three-Dimensional Poisson Problem}\label{sec:3DP}
We consider  a three-dimensional stochastic diffusion problem with a spatially varying random diffusion coefficient. The flow is modeled by a three-dimensional Poisson problem having a spatially varying stochastic diffusion coefficient $c_d$ on a domain $\mathcal{D}$ defined as:
\begin{align}\label{eq:3DP_spde}
-\nabla \ \cdotp \big( \ c_d(\textbf{\textit{x}},\theta) \ \ \nabla \mathcal{U}(\textbf{\textit{x}}, \theta) \ \big) &= {F}(\textbf{\textit{x}}),     \ \ \  \ \ \ \   \mathcal{D}\times \Omega, \\
\mathcal{U}(\textbf{\textit{x}}, \theta) &= 0,    \ \ \  \ \ \ \  \ \ \ \ \  \delta \mathcal{D}\times \Omega,
\end{align}
where $\nabla$ being  the gradient operator, $F(\textbf{\textit{x}})$ and $\mathcal{U}$ are the forcing and solution vectors respectively, and $\theta$ denotes the random dimension~\cite{billingsley2008probability}.

The lognormal stochastic process, $l(\textbf{\textit{x}}, \theta) = \mathrm{exp}  \big( g(\textbf{\textit{x}}, \theta) \big)$ 
defines the diffusion coefficient $c_d$ with the underlying  Gaussian process $g(\textbf{\textit{x}}, \theta)$  having a  mean  $\mu$, variance $\sigma^2$ and the exponential covariance function $C$ defined as~\cite{ghanemSFEM1991,subber2012PhDTh},
\begin{equation} \label{CoVFn3D}
C(x_1, y_1, z_1 ; x_2, y_2, z_2) = \sigma^2 \  e^{-|x_2 - x_1|/b_x -|y_2 - y_1|/b_y -|z_2 - z_1|/b_z},
\end{equation}
where $b_x$, $b_y$ and $b_z$ are the correlation lengths along $x$, $y$ and $z$ directions respectively.

\begin{figure}[htbp]
 \centering
 \includegraphics[width=0.64\textwidth,height=0.33\textheight]{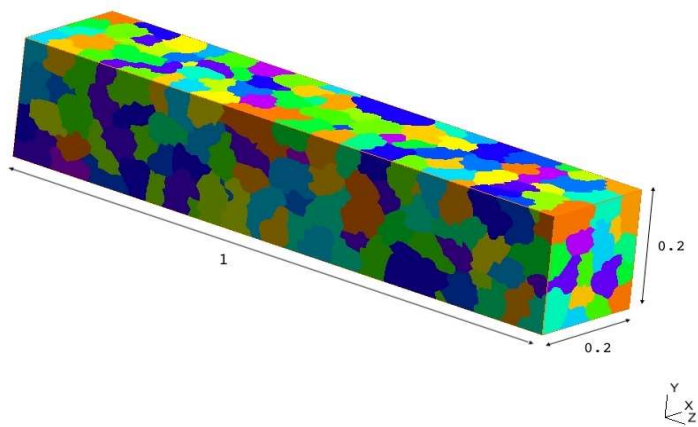}
 \caption{A typical three-dimensional finite element mesh partitioned into 320 subdomains.}
 \label{fig:fixed_beam}
\end{figure}

For simplicity, we consider deterministic source term ${F}(\textbf{\textit{x}})=1$. The correlation lengths $b_{x} = b_{y} = b_{z} = 1$ and standard deviation $\sigma$ = 0.3 are used to characterize the underlying Gaussian process.
Numerical experiments are performed for the three-dimensional computational domain shown in \Cref{fig:fixed_beam} which depicts a typical unstructured finite element mesh with $31598$ nodes and $182681$ linear (four node) tetrahedral elements partitioned into $320$ subdomains.

Similar to the two-dimensional cases, we primarily focus on  the scalabilities of the solver to tackle the high-dimensional stochastic systems. Various simulations are conducted with increasing number of random variables. We begin with  the numerical experiments in \Cref{sec:3DP_exper}, followed by the discussion on the characteristics of the solution process in \Cref{sec:3DP_solChar}. Next, in \Cref{sec:3DP_comp}, we compare the scalability plots for the extended wirebasket-based coarse grid with the vertex-based coarse grid. Finally, in \Cref{sec:3DP_scalabilites}, we present the numerical and parallel scalability plots for the wirebasket-based BDDC/NNC solvers.

\subsection{Numerical Experimental Framework}\label{sec:3DP_exper}
The PCGM \Cref{alg:pcgm} presented in Appendix~\ref{sec:ddmImplementation} is utilized to implement NCC/BDDC solver with the extended wirebasket-based preconditioner presented in \Cref{sec:NNW}. The solver is implemented by using Fortran programming language. Parallel computation is managed by  Message Passing Interface (MPI) communication routines~\cite{gropp1999MPI} in conjunction with
PETSc~\cite{petsc2016} for local (subdomain-level) sparse matrix vector operations,  GMSH~\cite{geuzaine2009gmsh,gmshWeb2017} for mesh generation along with METIS graph partitioner~\cite{karypis1995metis,metisWeb2017}.
Stochastic system matrix and vector assembly is performed by employing element-level (deterministic) assembly routines from the FEniCS/dolfin finite element software ~\cite{logg2012FEniCS}.
UQ Toolkit~\cite{debusschere2013uqtk} routines are used for computations relevant to KLE and PCE entities.
ParaView~\cite{ahrens2005ParaView,paraviewWeb2017} and Matlab~\cite{matlabGuide} are employed for post processing and visualization.

The simulations are performed on Canada's national HPC clusters managed by Compute Canada~\cite{nationalSystems}.
The nodes employed have either Intel Skylake cores running at 2.4 GHz from Niagara supercomputer~\cite{nationalSystemsNia}
or Intel E5-2683 processors, running at 2.1 GHz from Cedar and Graham HPC systems~\cite{nationalSystems}.

\subsection{Characteristics of the Solution Process}\label{sec:3DP_solChar}
\begin{figure}[htbp]
\centering
\includegraphics[width=0.63\textwidth]{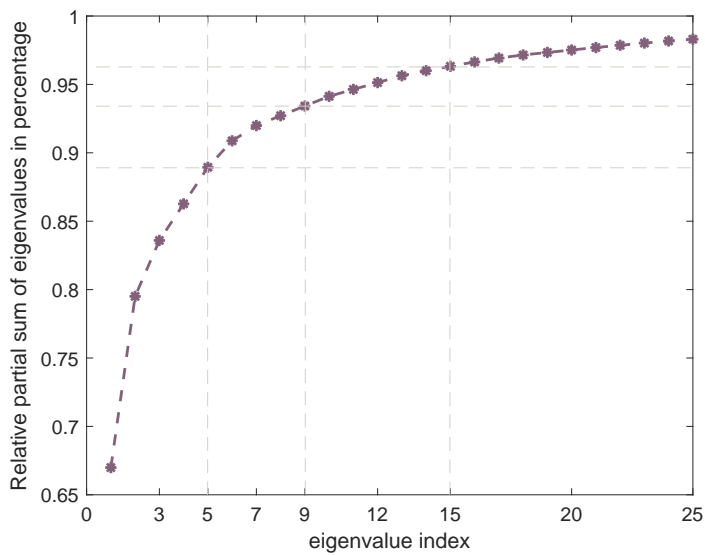}
\caption{Relative partial sum of eigenvalues with respect to eigenvalue-index.}
\label{fig:relativeEnergy3D}
\end{figure}
As a test case, consider five random variables ($L=5$) and the second order expansion ($p_{\textsc{a}} = 2$) for the PCE of the input stochastic process.
Using $5$ RVs we can capture about $89\%$ percent of the relative energy contribution for the given stochastic process.
The relative energy contribution or the decay rate of the eigenvalues  determines the number of random variables required in the KLE to approximate the underlying Gaussian process. For instance, using $9$ RVs we can capture about $94\%$ percent of the relative energy contribution and using $15$ RVs we can capture about $97\%$ of the relative energy contribution. See \Cref{fig:relativeEnergy3D} for the relative partial sum of eigenvalues for the respective eigenvalue index~\cite{ghanemSFEM1991,subber2012PhDTh}.

The PCE of the solution process with the five random variables ($L=5$) and the third order expansion $P_u=3$ is used.

For these parameters, we used 21 PCE terms to approximate the input stochastic process and 56 PCE terms to approximate the output stochastic process~\cite{ghanemSFEM1991}.

The resulting linear system is solved for the solution PCE coefficients  using the in-house parallel BDDC/NNC solvers with wirebasket-based coarse grid discussed earlier in \Cref{sec:NNW}.

\begin{figure}[htbp]
\centering
\subfloat[mean  $\mathcal{U}$ \label{subfig:Emean}]{%
\includegraphics[width=0.4\textwidth,height=0.25\textheight]{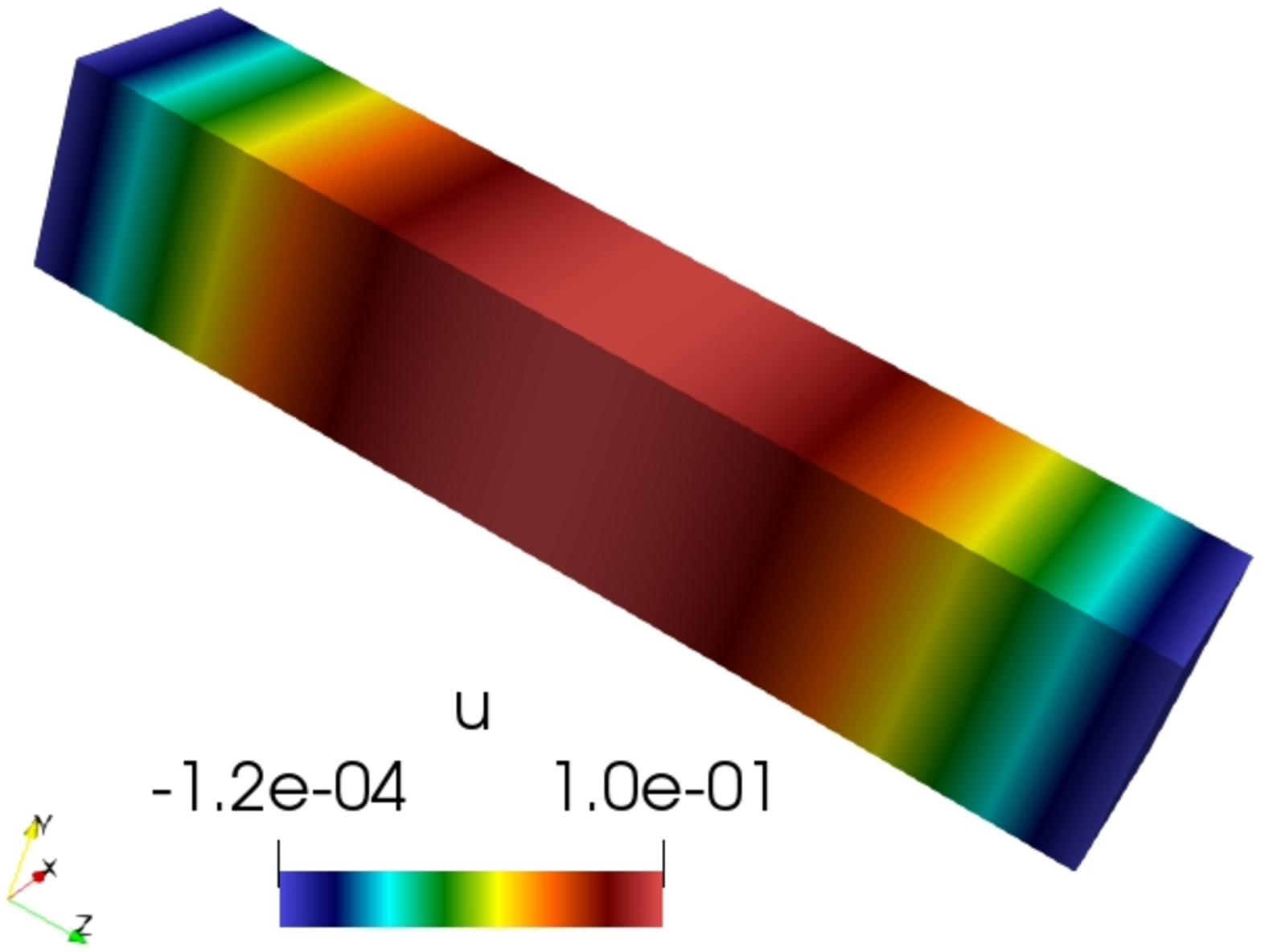}}
\subfloat[standard deviation  $\mathcal{U}$ \label{subfig:Esd}]{%
\includegraphics[width=0.4\textwidth,height=0.25\textheight]{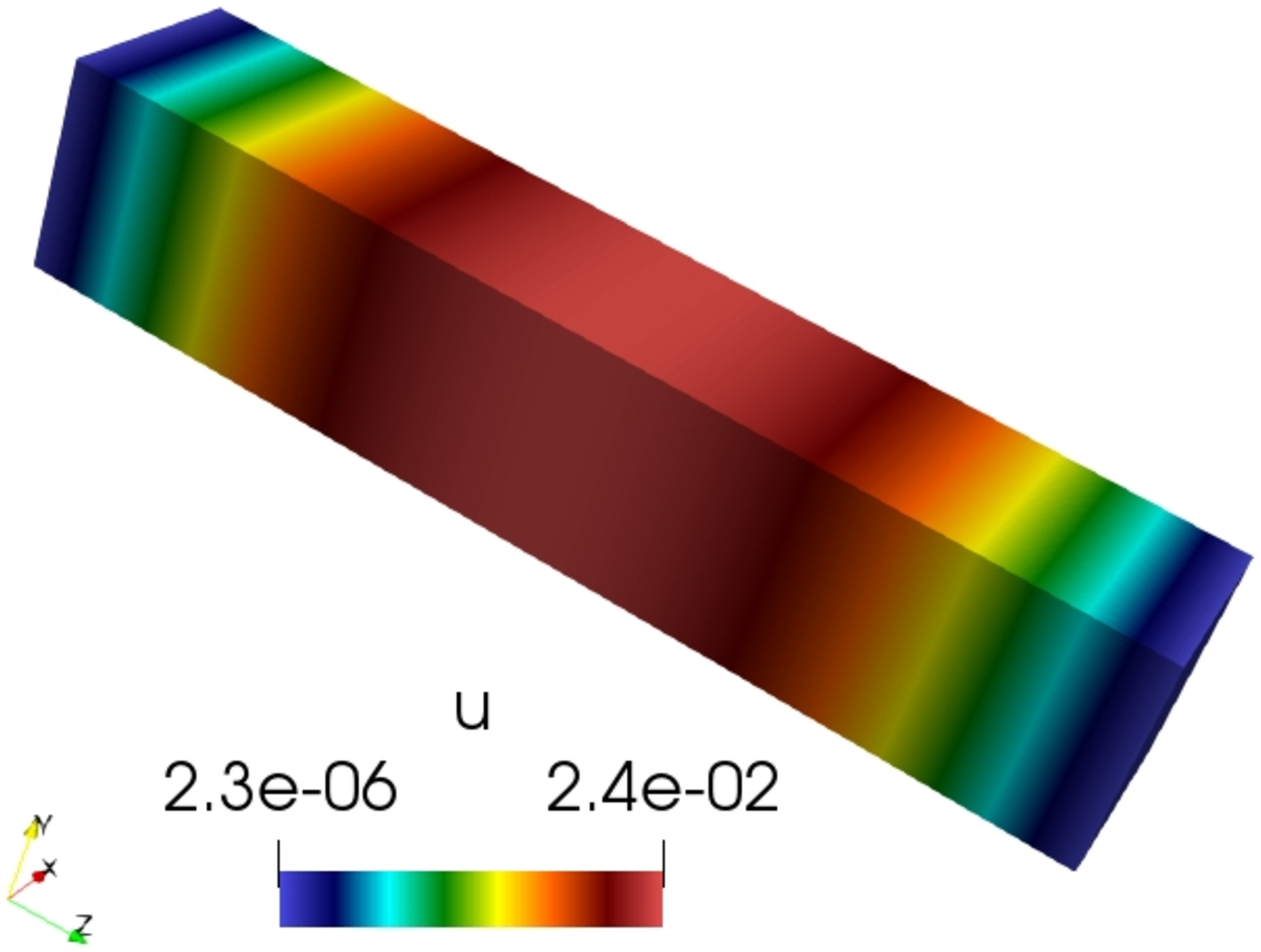}}
\caption{Mean and standard deviation of the solution process $\mathcal{U}$.}
\label{fig:EmeanSD}
\end{figure}

\begin{figure}[htbp]
 \centering
  \subfloat[PCE coefficient $u_{1}$\label{subfig:pe1}]{%
 \includegraphics[width=0.4\textwidth,height=0.25\textheight]{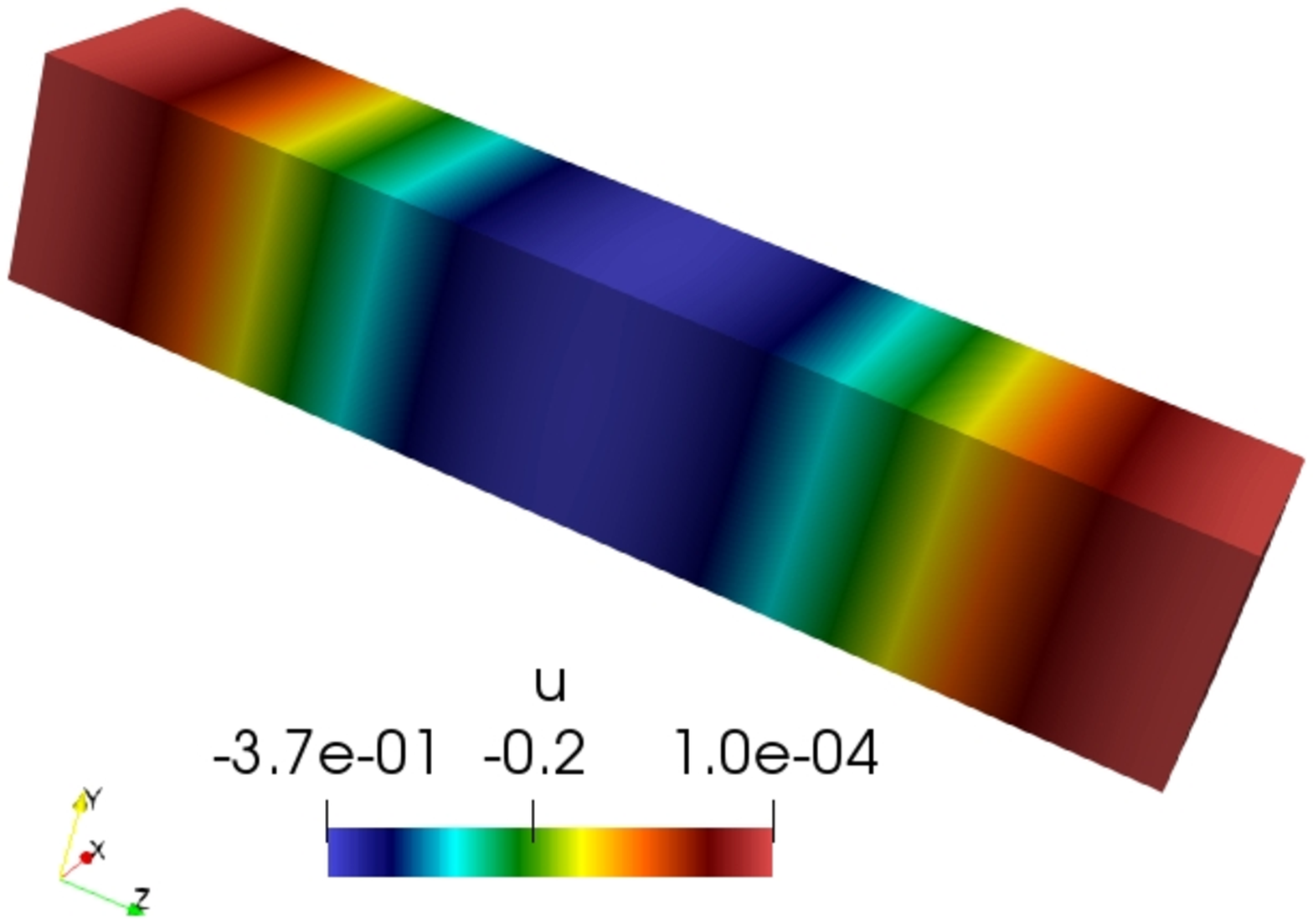}}
  \subfloat[PCE coefficient $u_{2}$\label{subfig:pe2}]{%
 \includegraphics[width=0.4\textwidth,height=0.25\textheight]{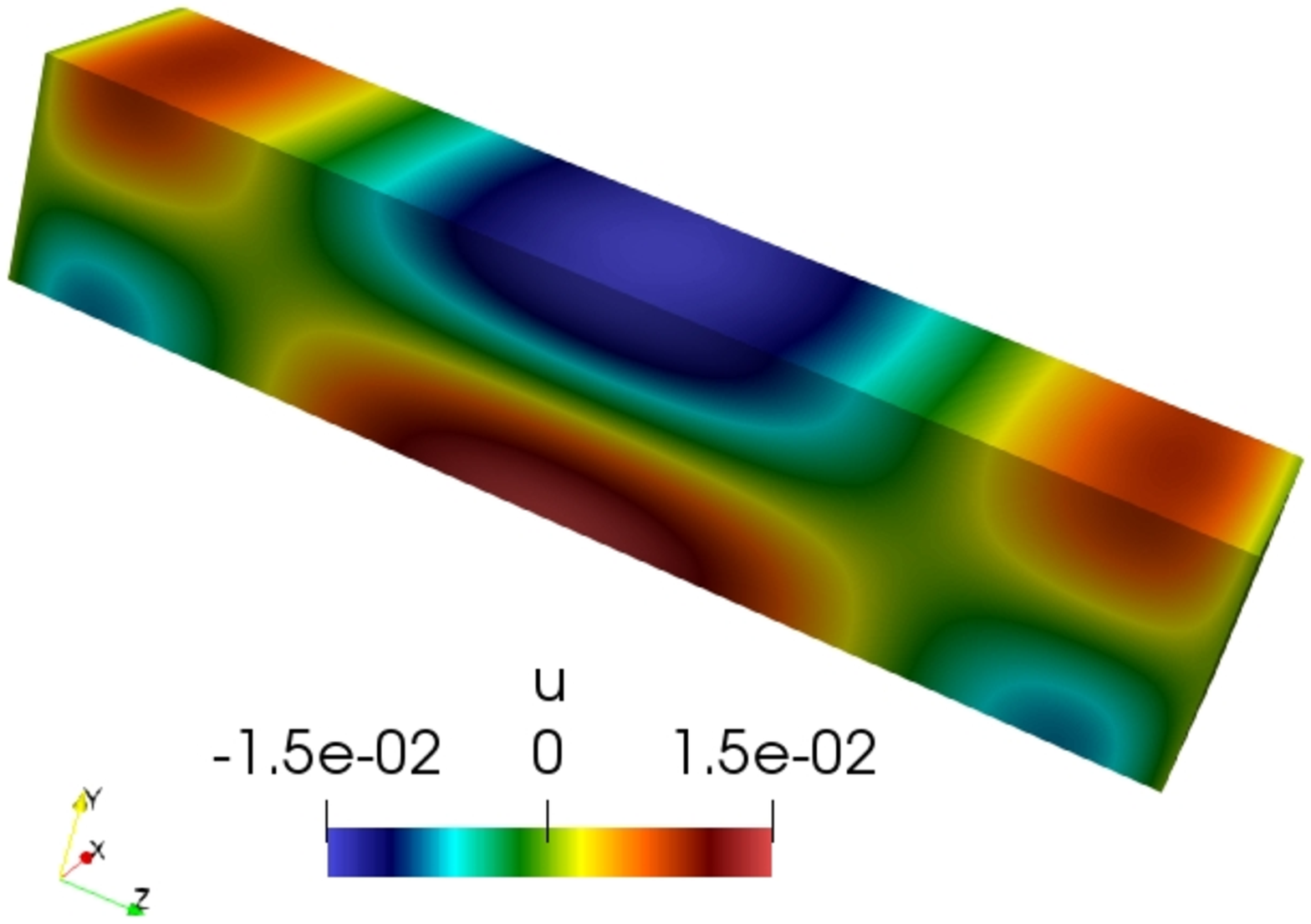}}

  \subfloat[PCE coefficient $u_{3}$\label{subfig:pe3}]{%
 \includegraphics[width=0.4\textwidth,height=0.25\textheight]{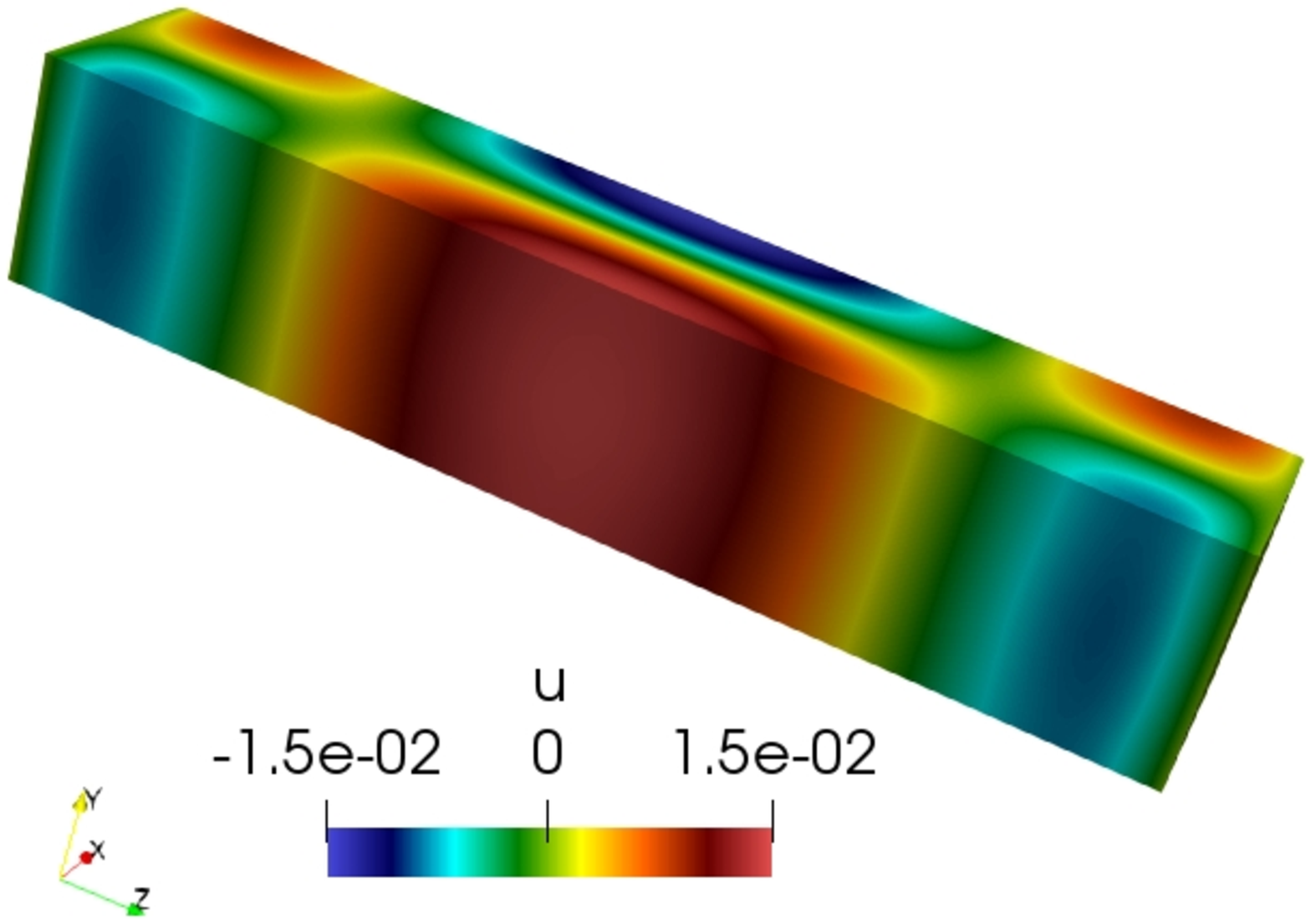}}
  \subfloat[PCE coefficient $u_{4}$\label{subfig:pe4}]{%
 \includegraphics[width=0.4\textwidth,height=0.25\textheight]{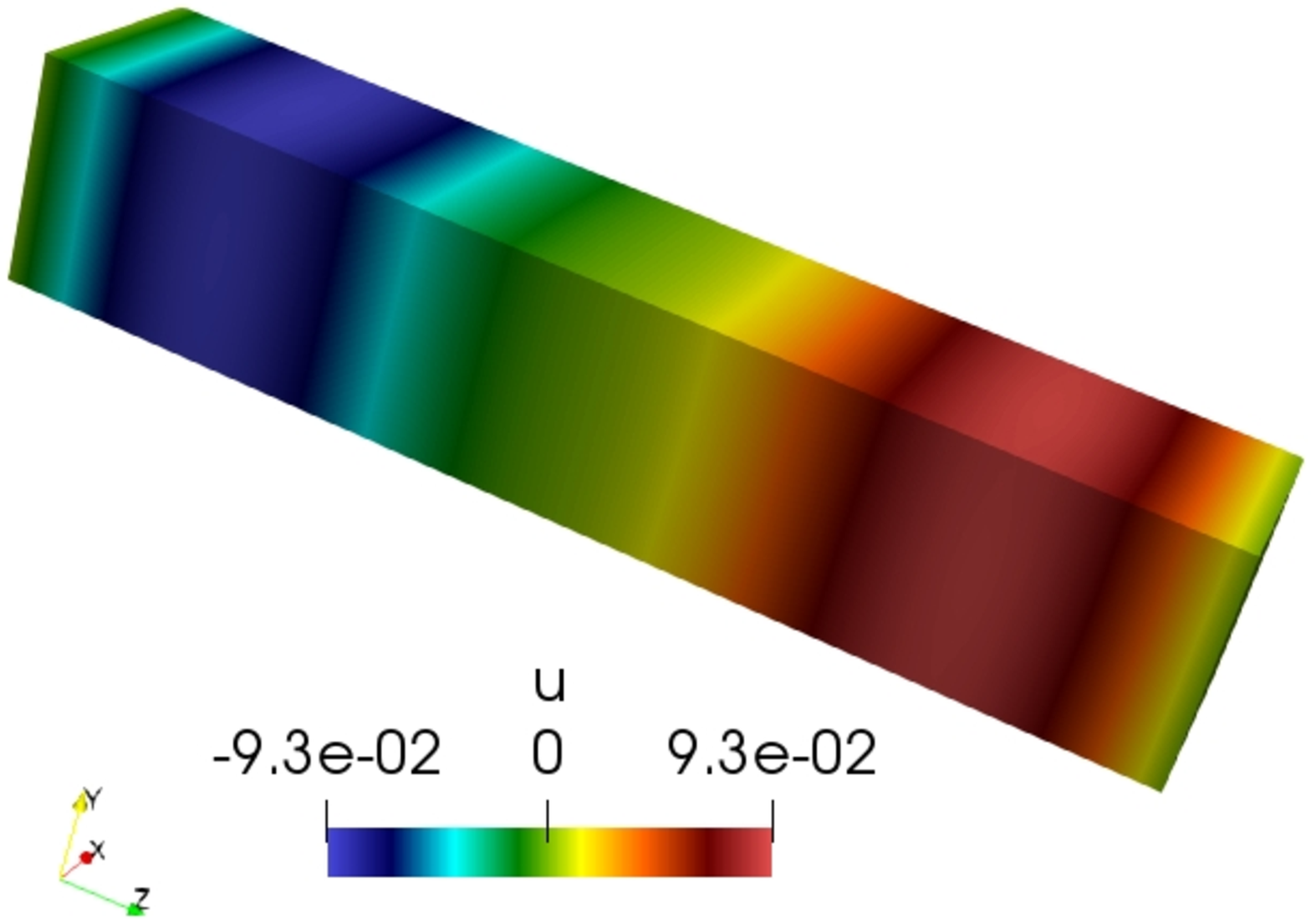}}

  \subfloat[PCE coefficient $u_{5}$\label{subfig:pe5}]{%
 \includegraphics[width=0.4\textwidth,height=0.25\textheight]{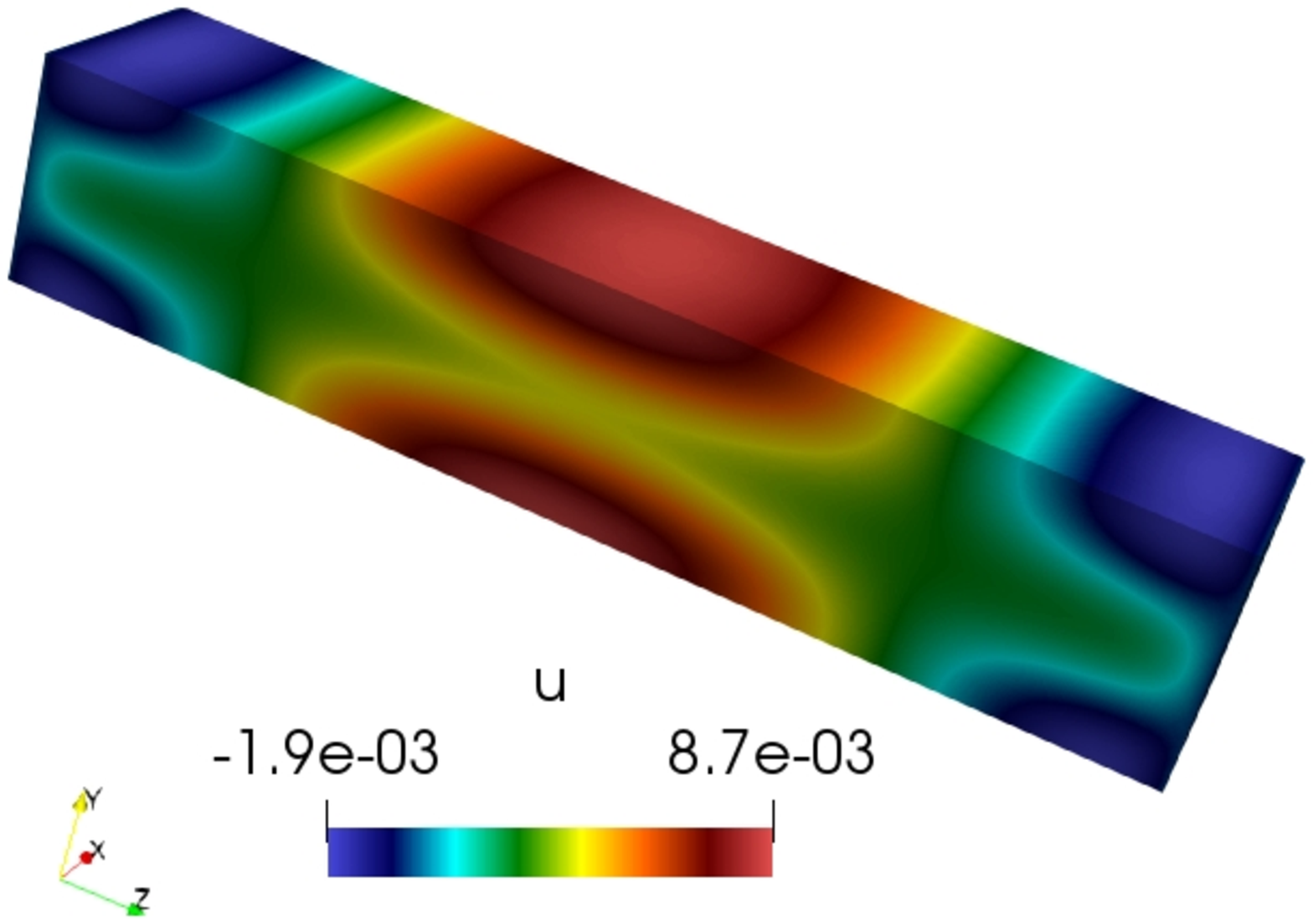}}
  \subfloat[PCE coefficient $u_{6}$\label{subfig:pe6}]{%
 \includegraphics[width=0.4\textwidth,height=0.25\textheight]{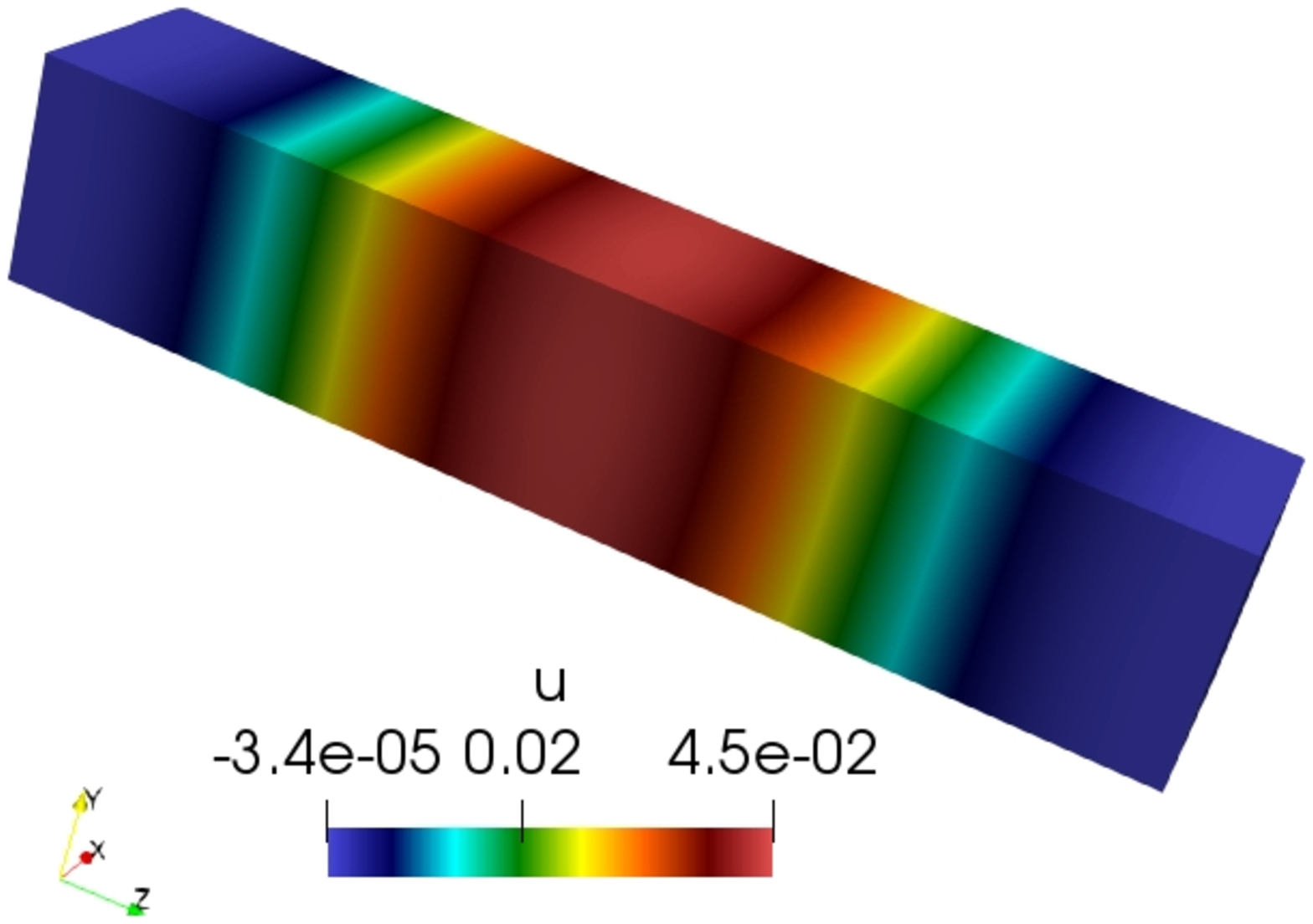}}
    \caption{Selected PCE coefficients of the solution process $\mathcal{U}$.}
 \label{fig:pce3DP1}
\end{figure}

\begin{figure}[htbp]
 \centering
 \subfloat[PCE coefficient $u_{10}$\label{subfig:pe11}]{%
 \includegraphics[width=0.4\textwidth,height=0.25\textheight]{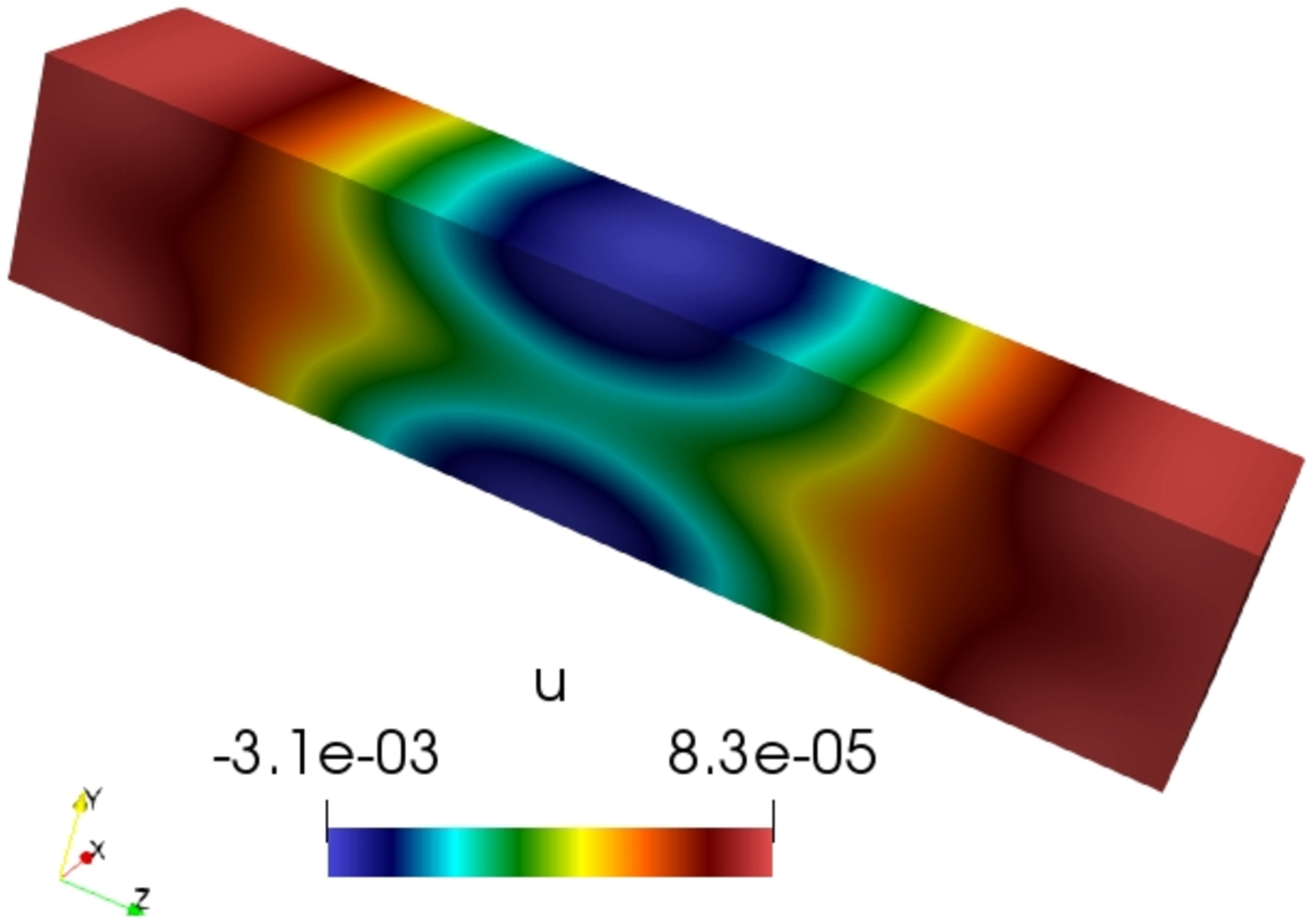}}
  \subfloat[PCE coefficient $u_{12}$\label{subfig:pe12}]{%
 \includegraphics[width=0.4\textwidth,height=0.25\textheight]{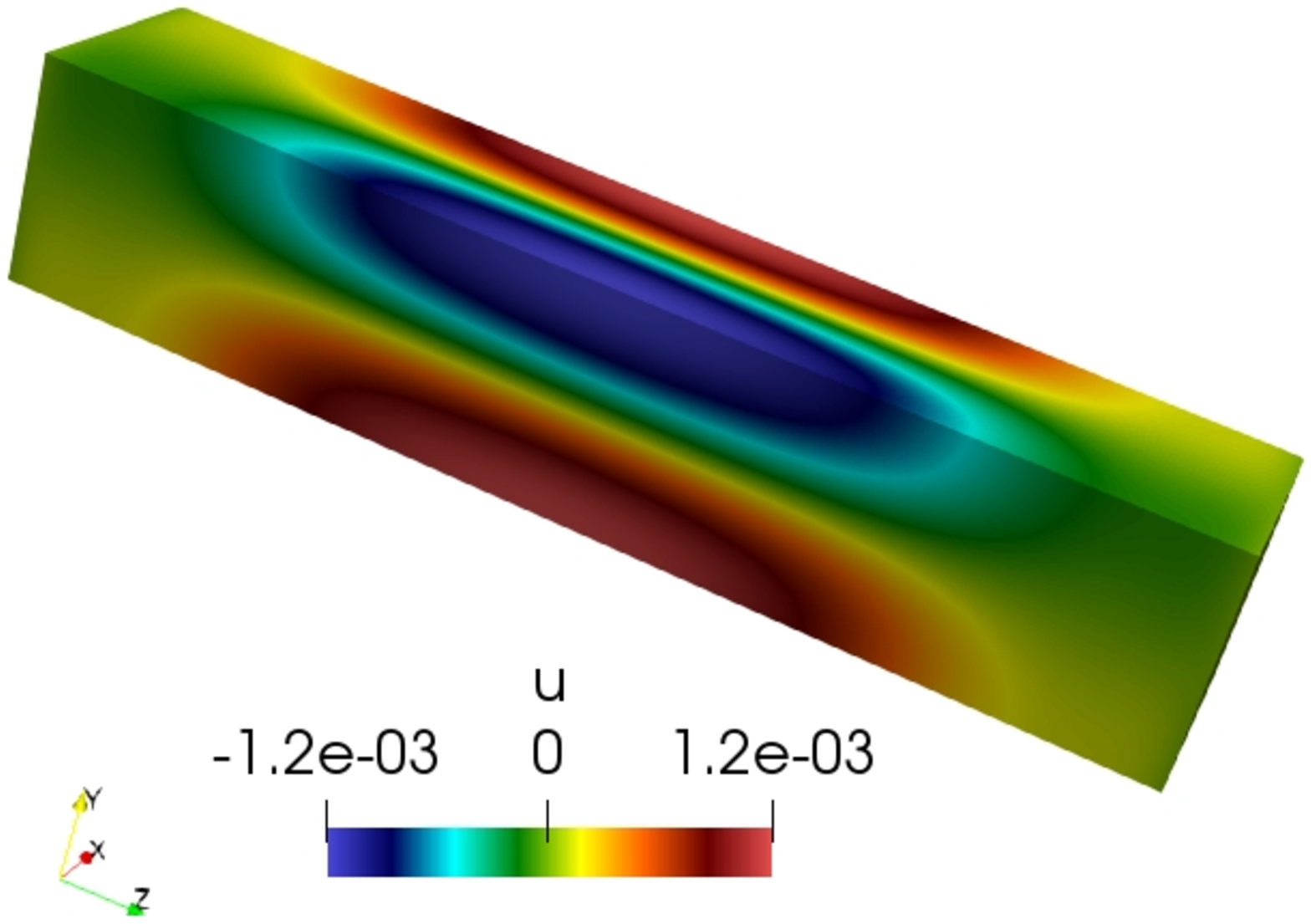}}

\subfloat[PCE coefficient $u_{13}$\label{subfig:pe13}]{%
 \includegraphics[width=0.4\textwidth,height=0.25\textheight]{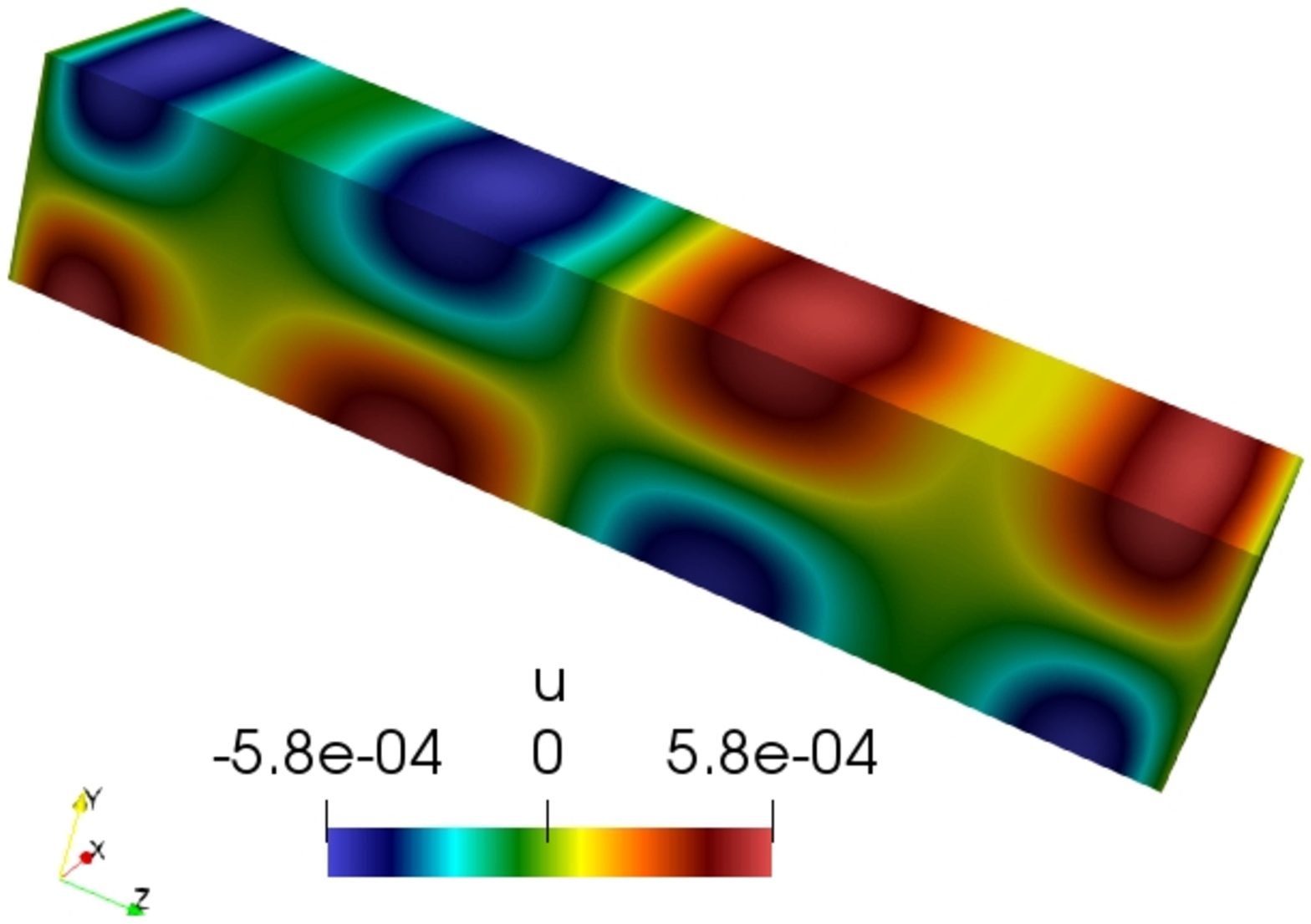}}
    \subfloat[PCE coefficient $u_{16}$\label{subfig:pe14}]{%
 \includegraphics[width=0.4\textwidth,height=0.25\textheight]{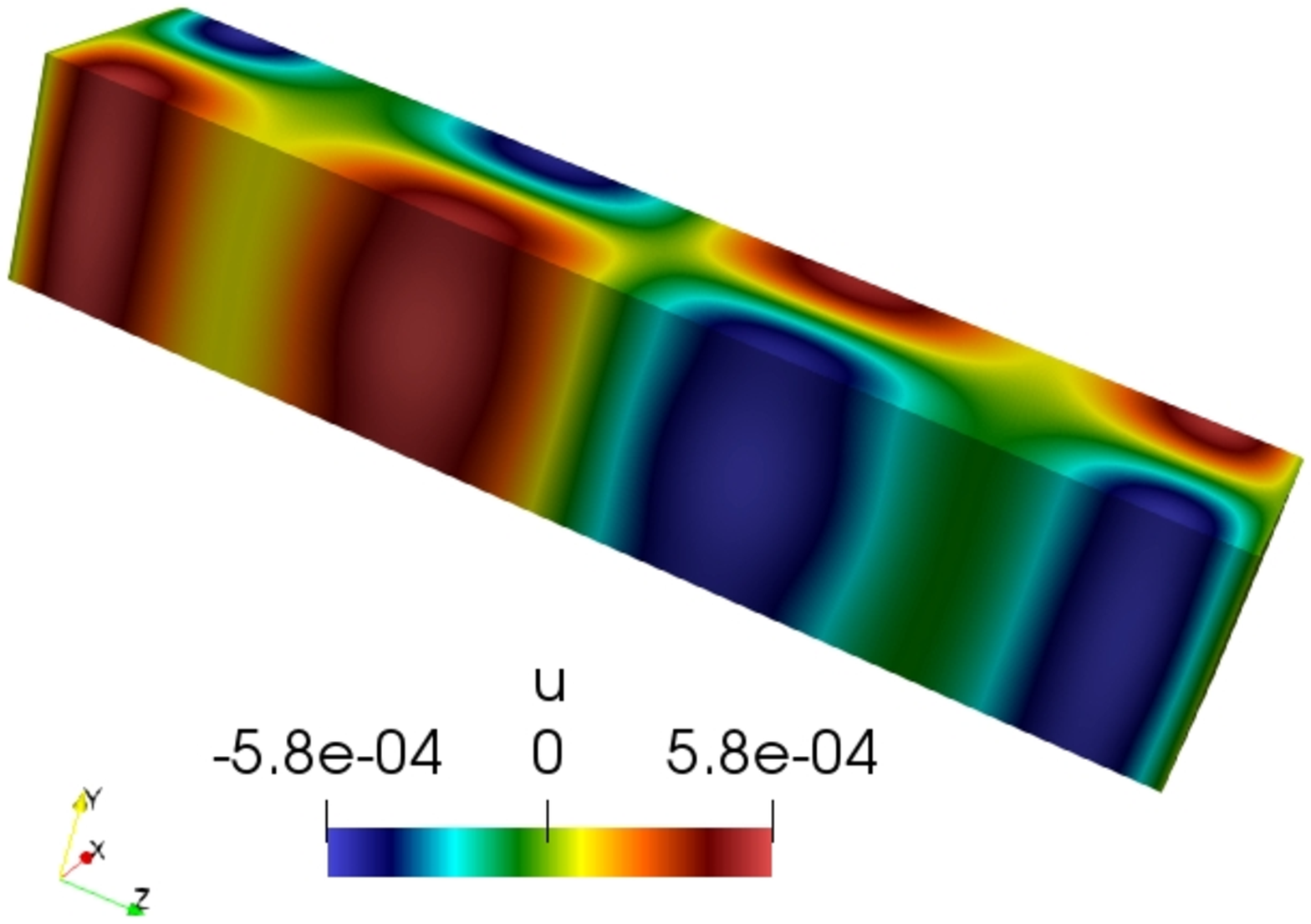}}

    \subfloat[PCE coefficient $u_{19}$\label{subfig:pe15}]{%
 \includegraphics[width=0.4\textwidth,height=0.25\textheight]{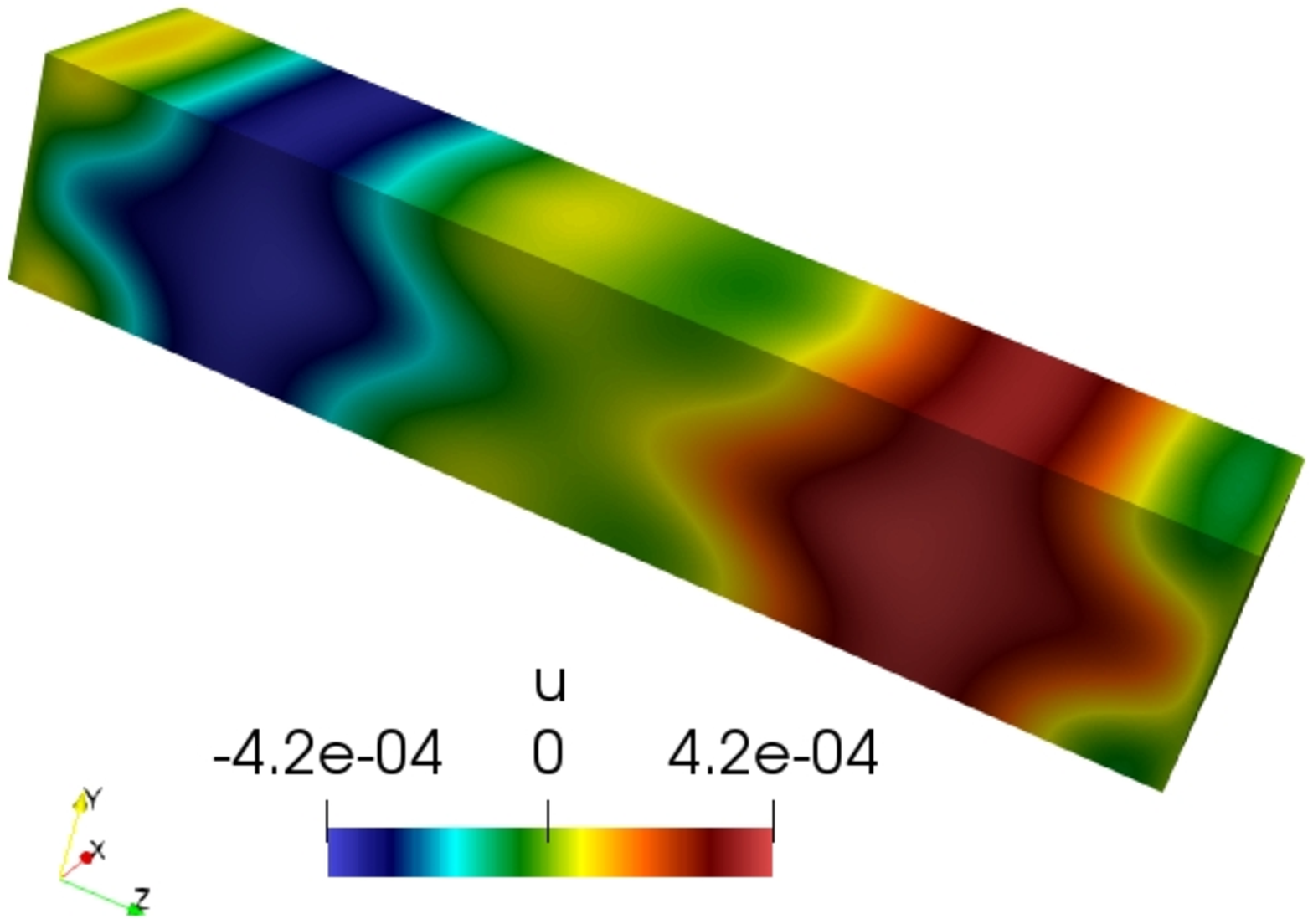}}
    \subfloat[PCE coefficient $ u_{27}$ \label{subfi:pe16}]{%
 \includegraphics[width=0.4\textwidth,height=0.25\textheight]{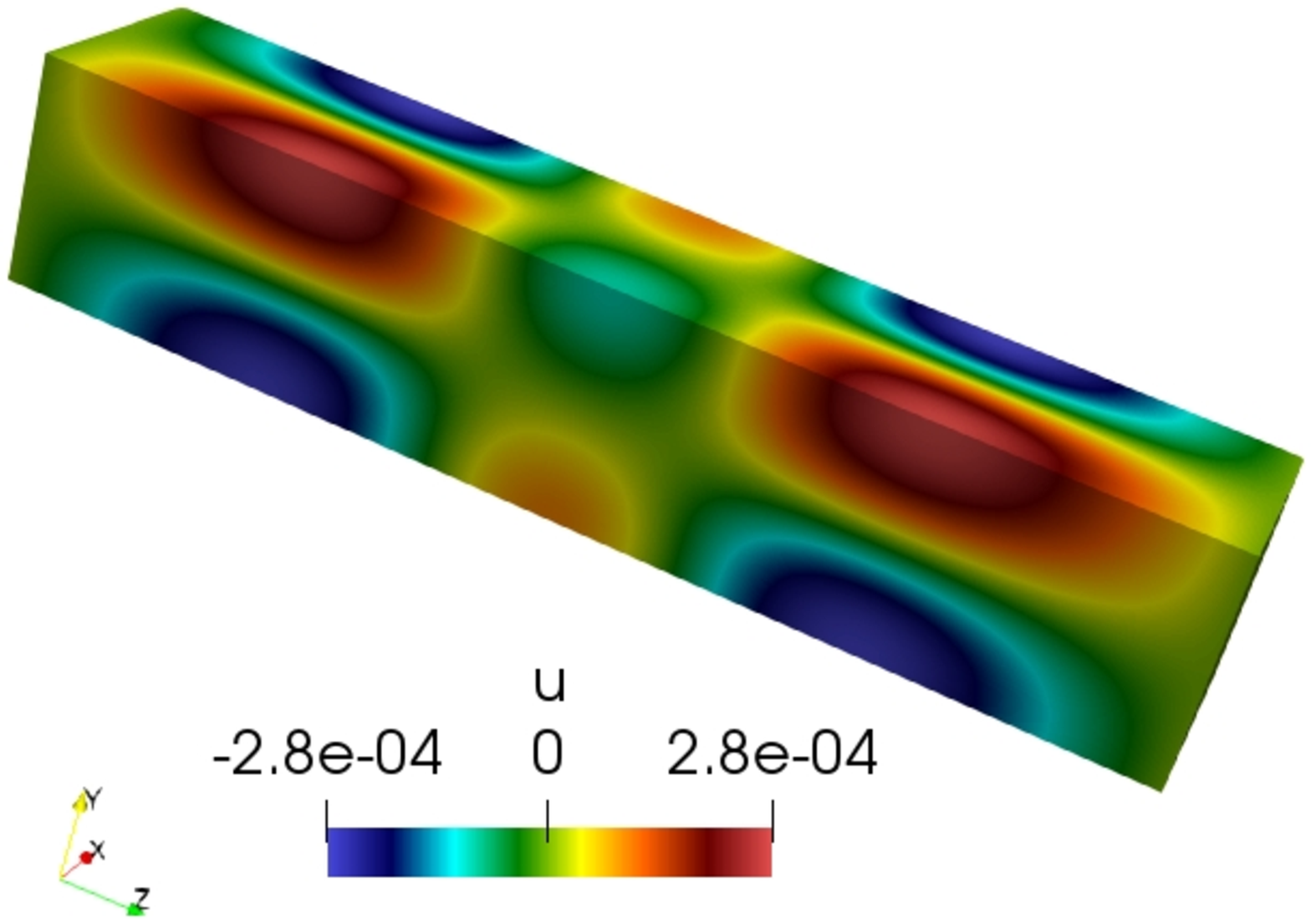}}
    \caption{Selected PCE coefficients of the solution process $\mathcal{U}$.}
 \label{fig:pce3DP2}
\end{figure}

\begin{figure}[htbp]
 \centering
    \subfloat[PCE coefficient $u_{32}$ \label{subfig:pe17}]{%
 \includegraphics[width=0.4\textwidth,height=0.25\textheight]{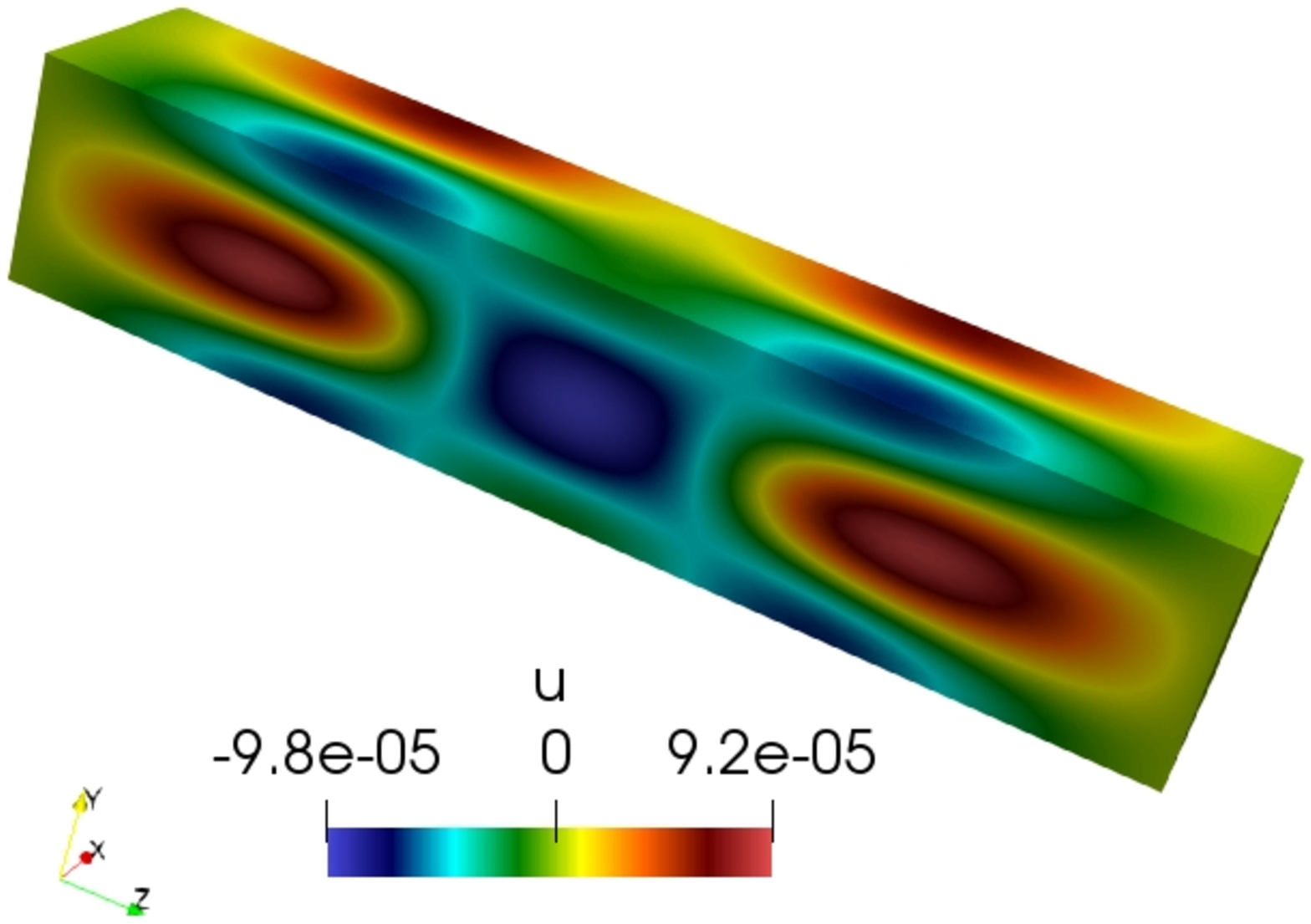}}
    \subfloat[PCE coefficient $ u_{41}$ \label{subfig:pe18}]{%
 \includegraphics[width=0.4\textwidth,height=0.25\textheight]{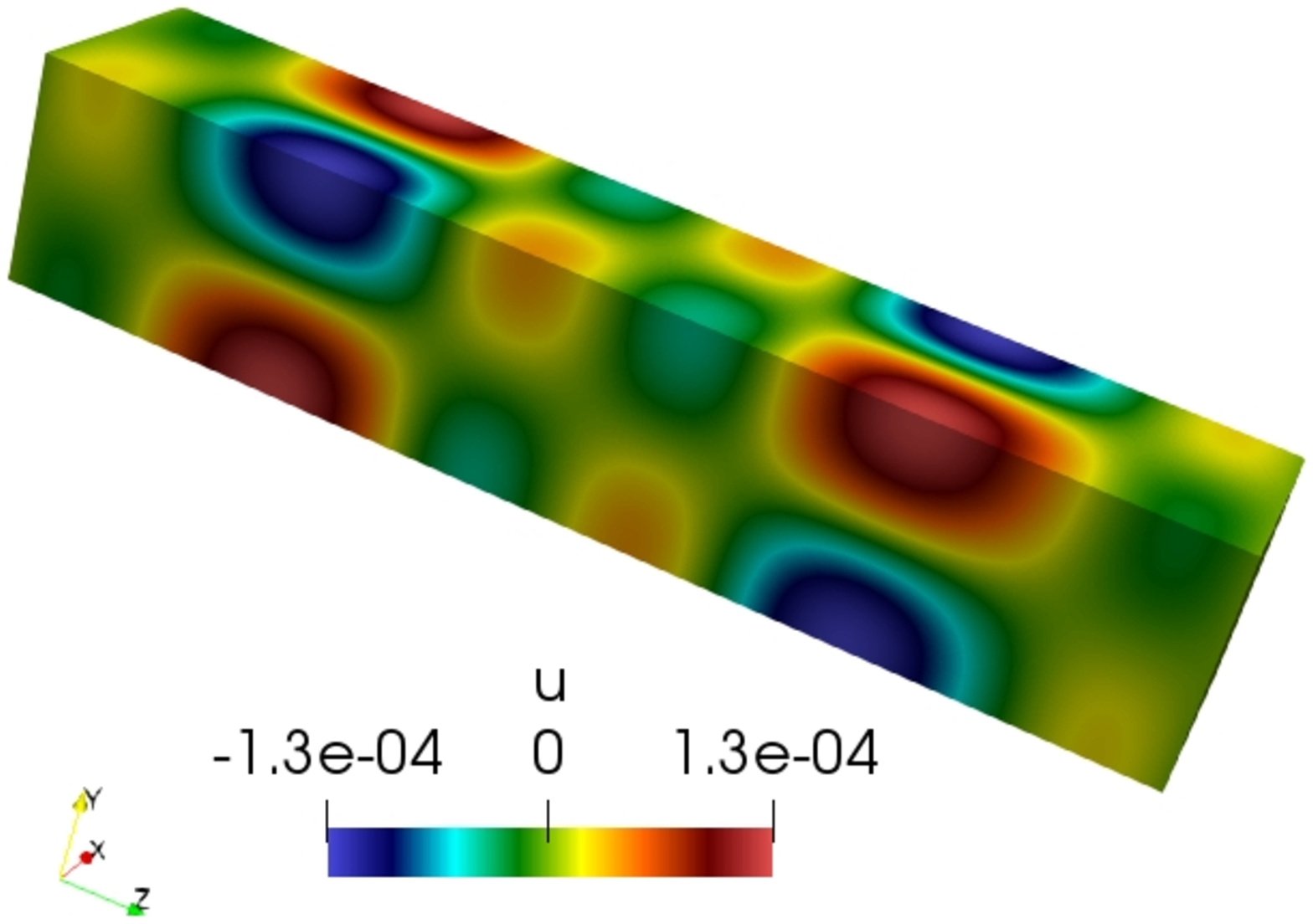}}

    \subfloat[PCE coefficient $ u_{45}$ \label{subfig:pe19}]{%
 \includegraphics[width=0.4\textwidth,height=0.25\textheight]{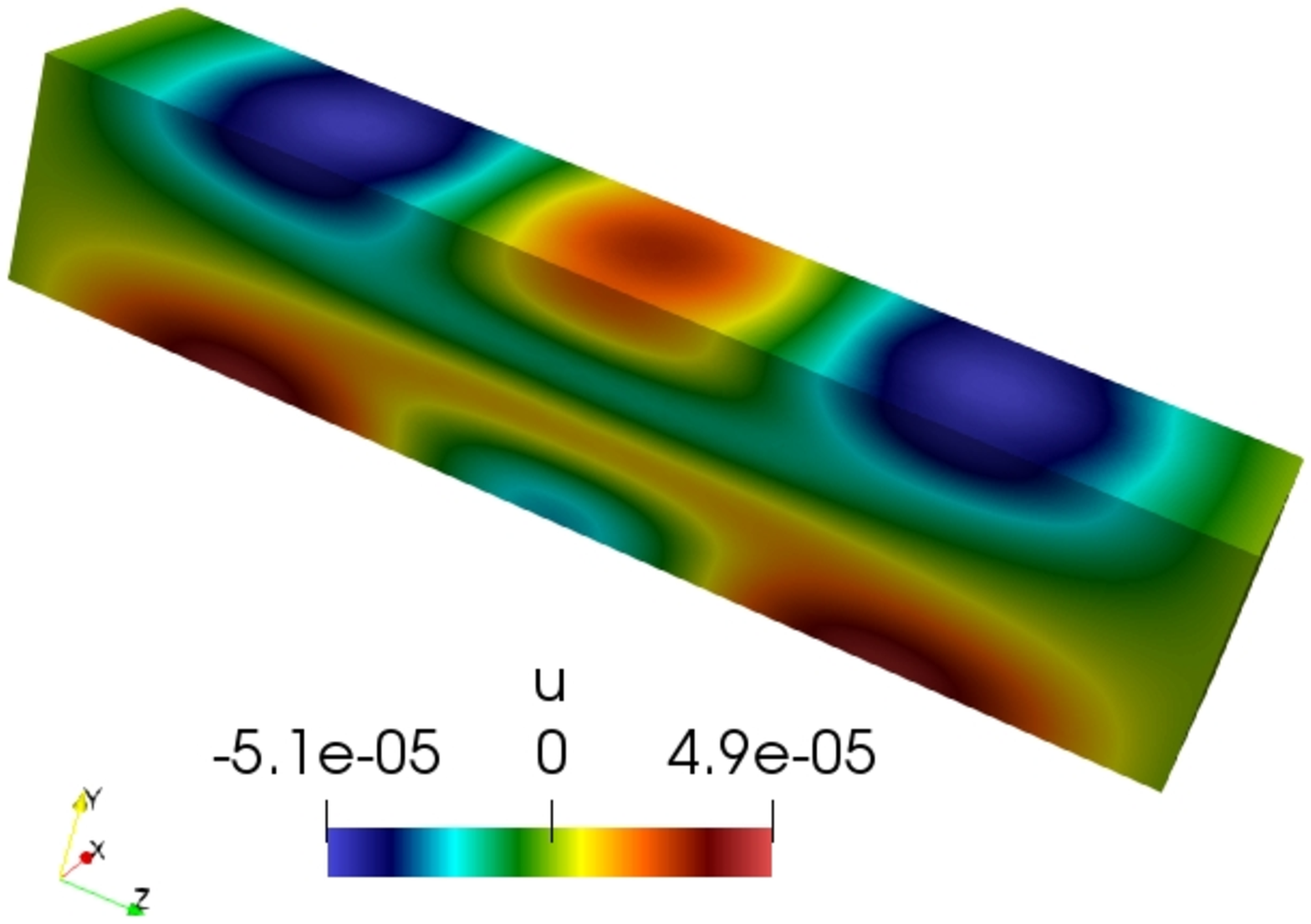}}
    \subfloat[PCE coefficient $ u_{48}$ \label{subfig:pe20}]{%
 \includegraphics[width=0.4\textwidth,height=0.25\textheight]{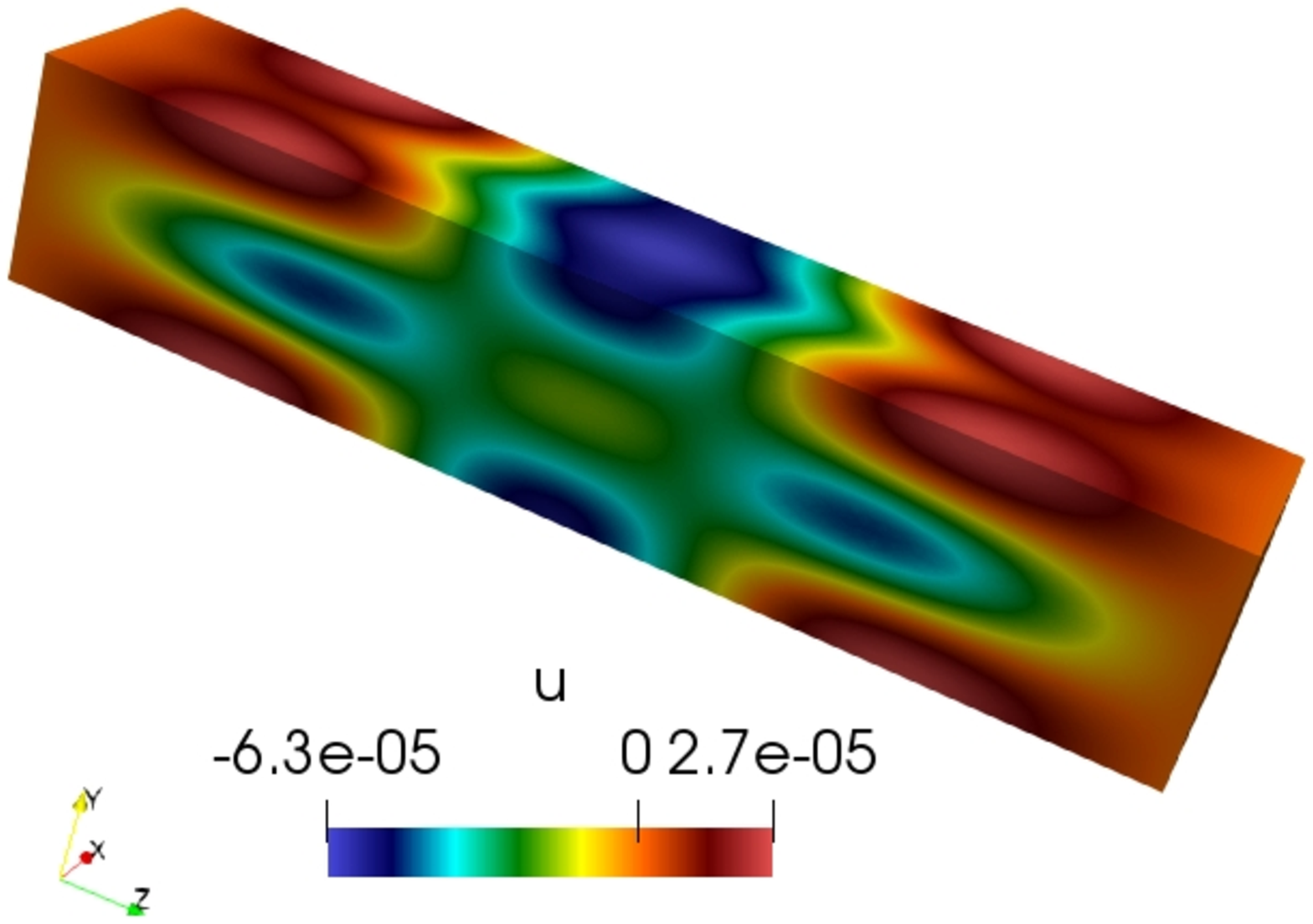}}

 \subfloat[PCE coefficient $ u_{50}$ \label{subfig:pe211}]{%
 \includegraphics[width=0.4\textwidth,height=0.25\textheight]{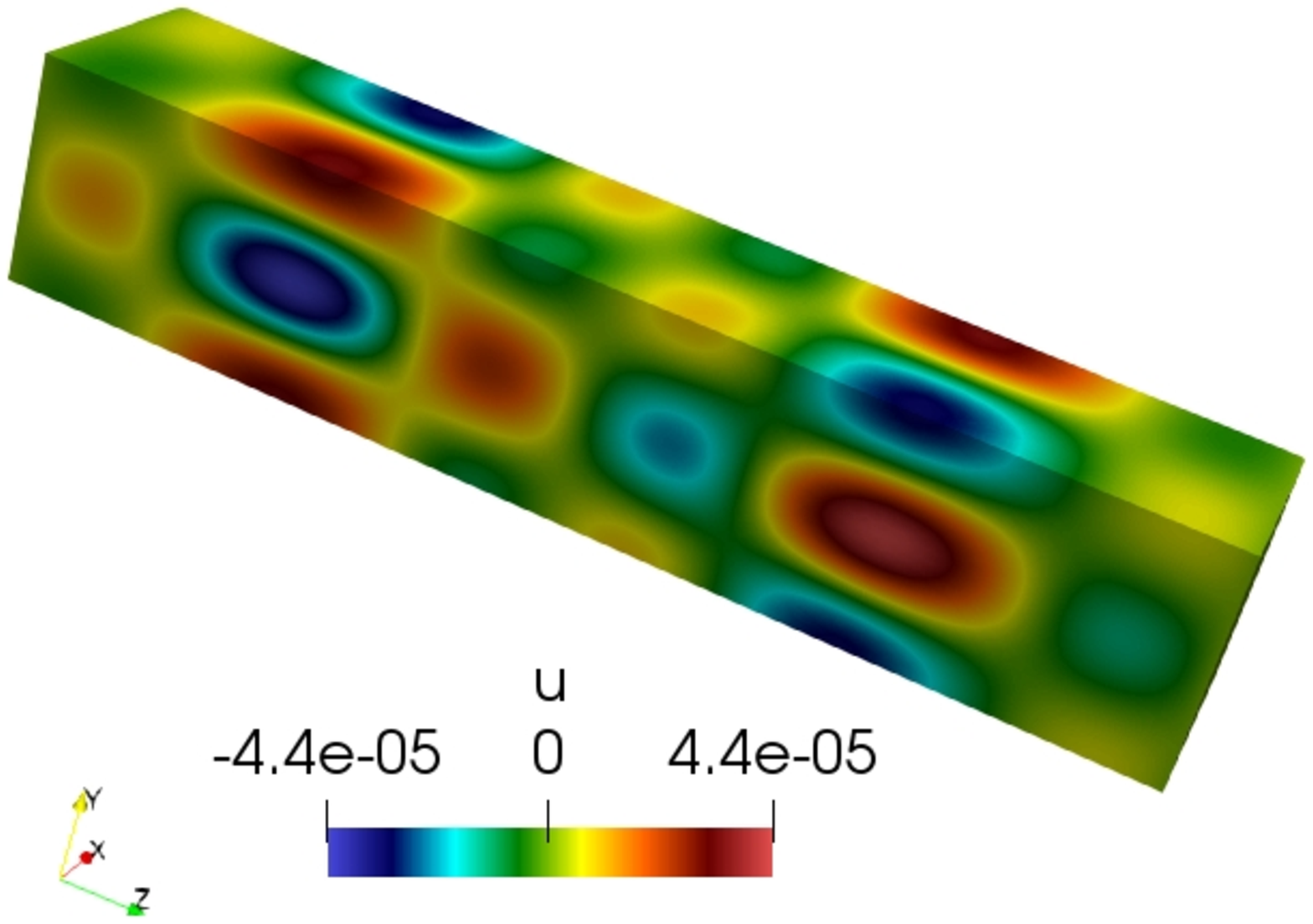}}
 \subfloat[PCE coefficient $ u_{53}$ \label{subfig:pe212}]{%
 \includegraphics[width=0.4\textwidth,height=0.25\textheight]{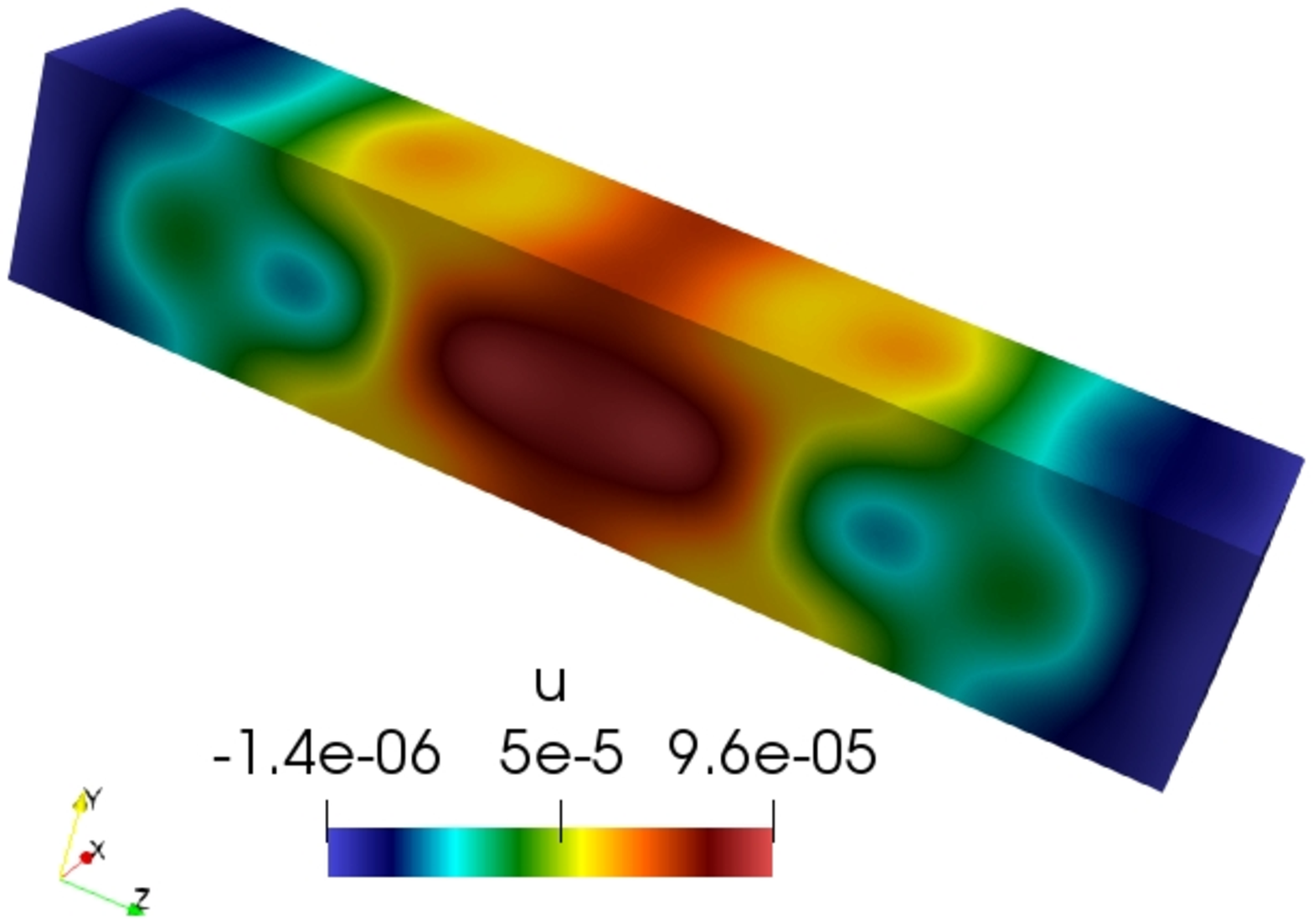}}
 \caption{Selected PCE coefficients of the solution process $\mathcal{U}$.}
 \label{fig:pce3DP3}
\end{figure}
The mean and standard deviation of the solution field $\mathcal{U}$ has the same trend, but, of course, their magnitudes are different. See \Cref{fig:EmeanSD} (a) and (b) respectively showing mean and standard deviation (SD) of the solution process.
The maximum coefficient of variation in the centre of the domain is about $25\%$.
To get the further insights into the stochastic aspects of the solution process, the first few PCE coefficients are plotted in \Crefrange{fig:pce3DP1}{fig:pce3DP3}.
One can see that, the PC coefficients exhibit oscillatory behavior and the magnitude of the chaos coefficients $u_{j}$ decreases with increasing PCE index $j$. 
Among these PCE coefficients, the first order coefficients contain Gaussian contributions and the higher order coefficients contain the non-Gaussian effects.
As the order of expansion increases, the mean and standard deviation of the solution process converge as shown in \Cref{fig:EmeanSD}.

\subsection{Comparison of Extended Wirebasket-based and Vertex-based Algorithms}\label{sec:3DP_comp}
The performance of extended wirebasket-based coarse grid preconditioner devised earlier is compared to that of the extended vertex-based coarse grid. In the current investigation, we primarily focus on to the performance comparison concerning numerical scalabilities of the solvers,

i.e., the number of PCGM iterations necessary for the convergence with the given tolerance: $$\frac{\| \mathcal{U}_{{\it \Gamma}_{i+1}} - \mathcal{U}_{{\it \Gamma}_{i}} \|_2}{\| \mathcal{U}_{{\it \Gamma}_{i}} \|_2} \le tol = 10^{-5},$$
where $\|. \|_2$ represents the $\mathrm{L_2}$ norm and the subscript $i$ indicating the PCGM iteration number. 

The numerical scaling is essential to understanding the convergence behavior of the solver. Therefore the numerical scalability is considered crucial for the utility of the solver for the large-scale applications.
Note that, the finite element mesh with the fixed resolution having $31598$ nodes and $182681$ (linear tetrahedral) elements is used for all simulations in this section. The maximum size of the linear system tackled in this section is about $\approx 7$ million, i.e., $31598$ nodes $\times$ 220 PCE terms.

\begin{figure}[htbp]
\centering
\includegraphics[width=0.67\textwidth]{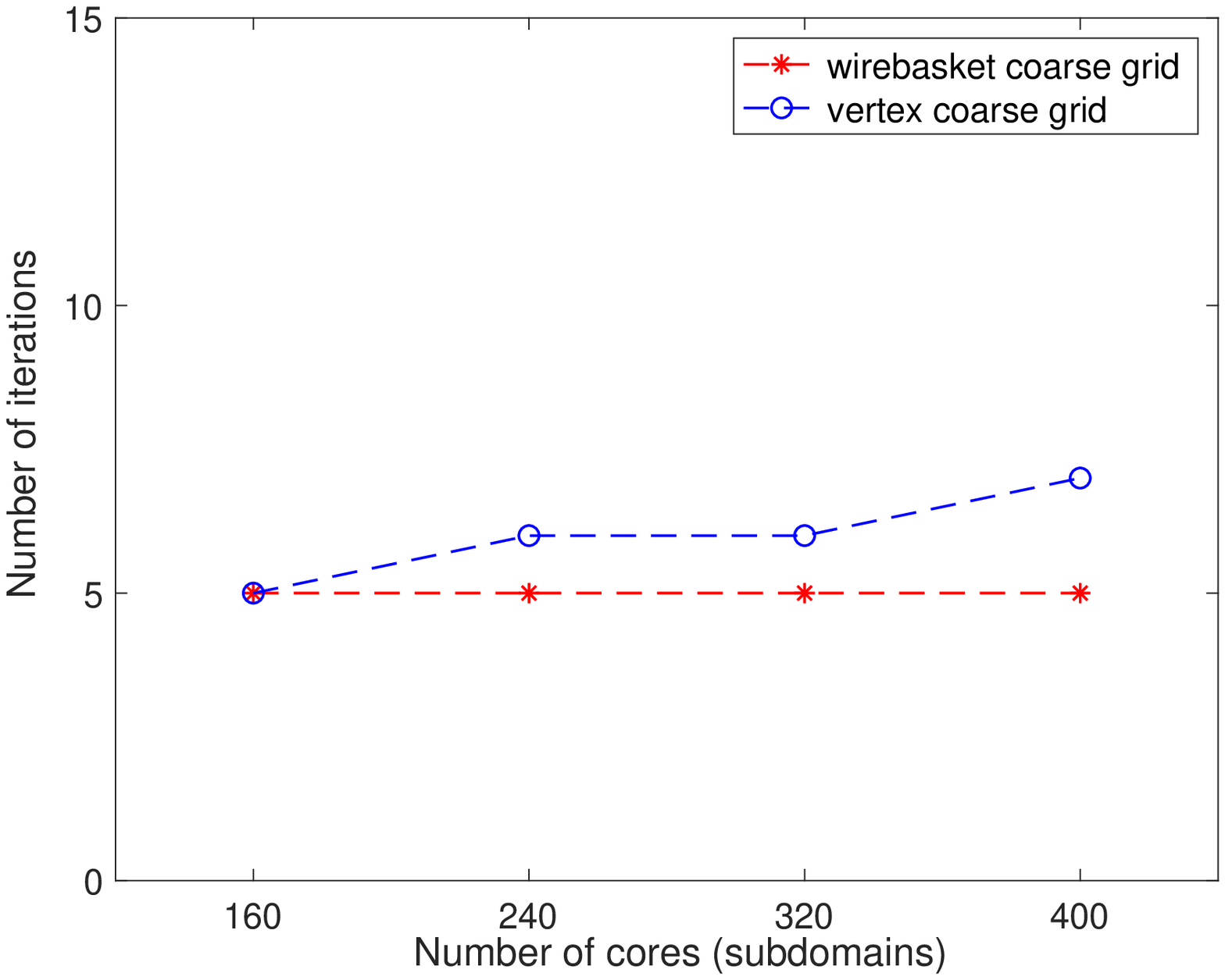}
\caption{Iteration count versus number of subdomains for the fixed mesh resolution with fixed number of PCE terms.}
\label{fig:3DP_comp_nIterVnProc_stoStrong}
\end{figure}

For a fixed mesh resolution and the fixed  PCE terms ($P_u=56$), and increasing number of subdomains (accordingly number of cores), the number of PCGM iteration increases faster with the vertex-based coarse grid compared to that of the wirebasket-based coarse grid. The BDDC/NNC solver with wirebasket-based coarse grid
showed better numerical scaling with respect to the number of subdomains compared to the vertex-based coarse grid (\Cref{fig:3DP_comp_nIterVnProc_stoStrong}).

\begin{figure}[htbp]
\centering
\includegraphics[width=0.67\textwidth]{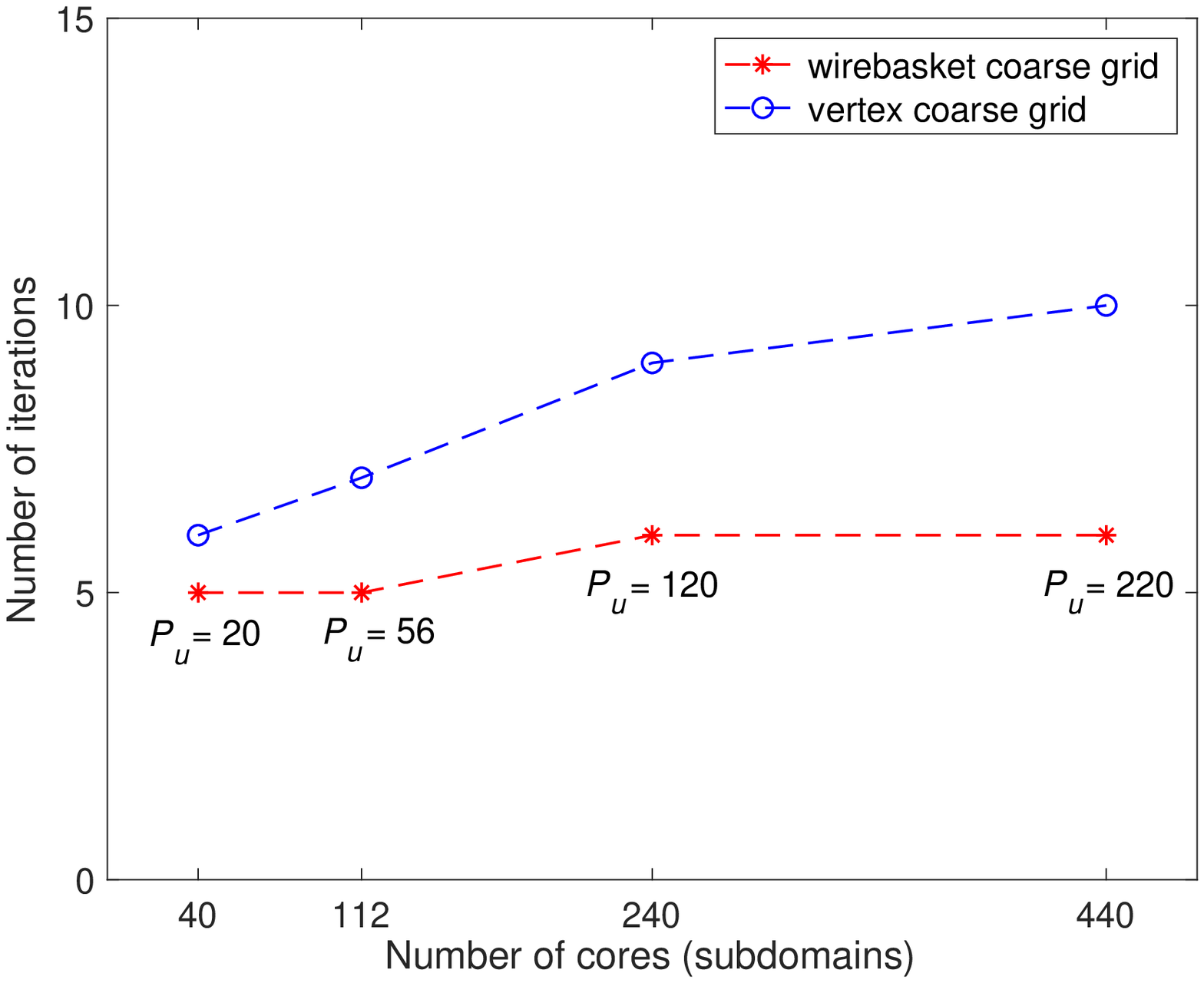}
\caption{Number of iterations versus number of subdomains for fixed problem size per subdomain with increasing number of PCEs (fixed mesh resolution).}
\label{fig:3DP_comp_nIterVnProc_stoWeak}
\end{figure}
% Time comparison between WG vs VG
\begin{figure}[htbp]
\centering
\includegraphics[width=0.67\textwidth]{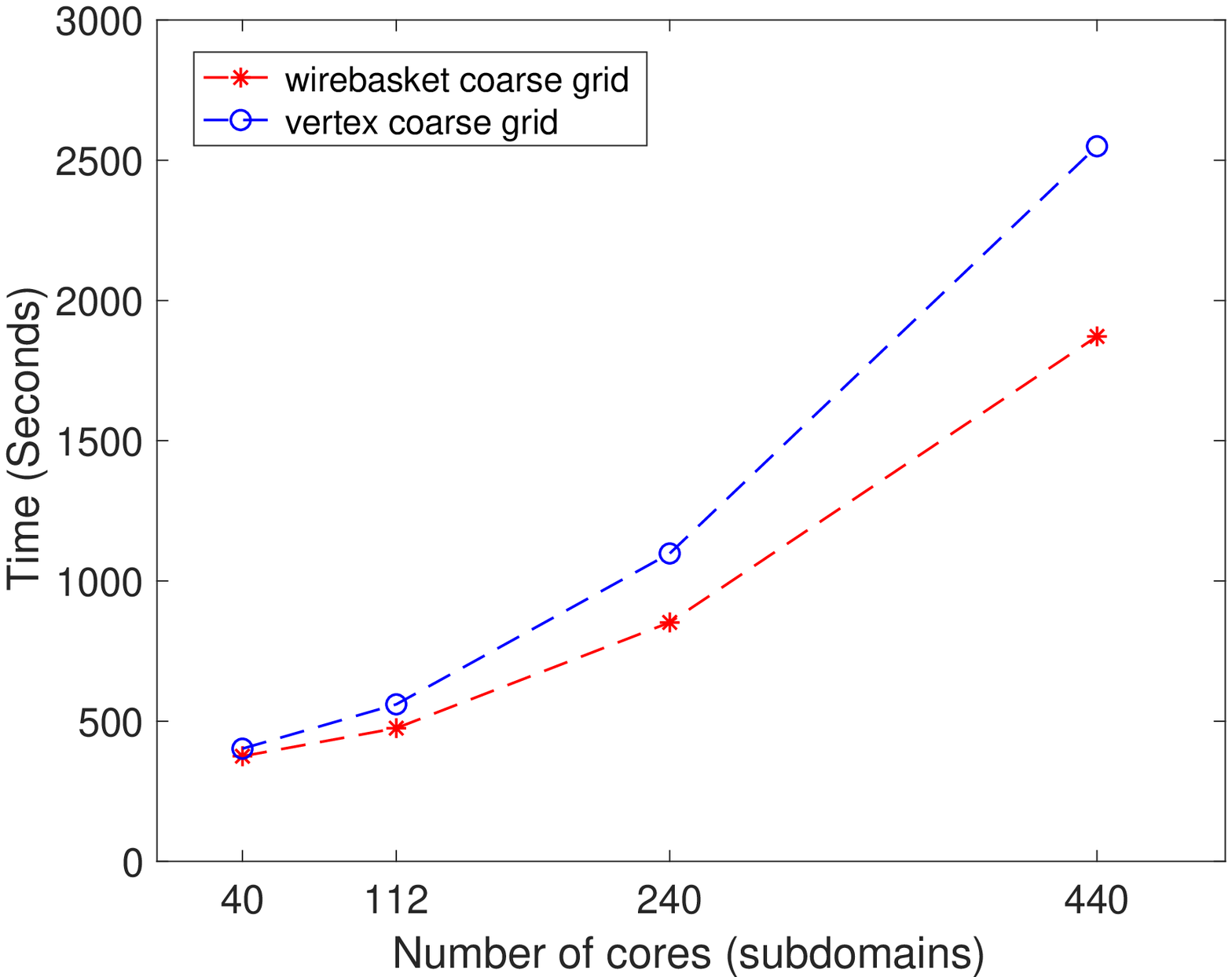}
\caption{Execution time versus number of subdomains for increasing number of PCEs with fixed problem size per subdomain (fixed mesh resolution).}
\label{fig:3DP_comp_weak_timeVnProcs}
\end{figure}

For a fixed mesh resolution, the global problem size relates to the number of PCE terms (i.e., $P_u$ = 20, 56, 120 and 220, corresponding to $L$ = 3, 5, 6 and 9, respectively).
For roughtly the fixed problem size per core ($\approx 15000$),
the resulting linear system is tackled using more cores.
For the fixed problem size per core, the PCGM iteration count for the wirebasket-based preconditioner against subdomain counts  remains almost the same (\Cref{fig:3DP_comp_nIterVnProc_stoWeak}).
On the other hand, for the same case, the number of iterations increases with the vertex-based coarse grid.
Therefore, the wirebasket-based coarse grid BDDC/NNC solver scales better with respect to the number of PCE terms compared to the vertex-based coarse grid.
For the same case as in \Cref{fig:3DP_comp_nIterVnProc_stoWeak}, the total execution time for the wirebasket-based coarse grid and vertex-based coarse grid BDDC/NNC solvers against the number of subdomains (with the fixed problem size per subdomain) are compared in~\Cref{fig:3DP_comp_weak_timeVnProcs}.
Total execution time for the BDDC/NNC solver with the vertex-based coarse grid increases faster compared to that of the wirebasket-based coarse grid. This is primarily due to more PCGM iteration required with the vertex-based coarse grid.

\begin{figure}[htbp]
\centering
\includegraphics[width=0.65\textwidth]{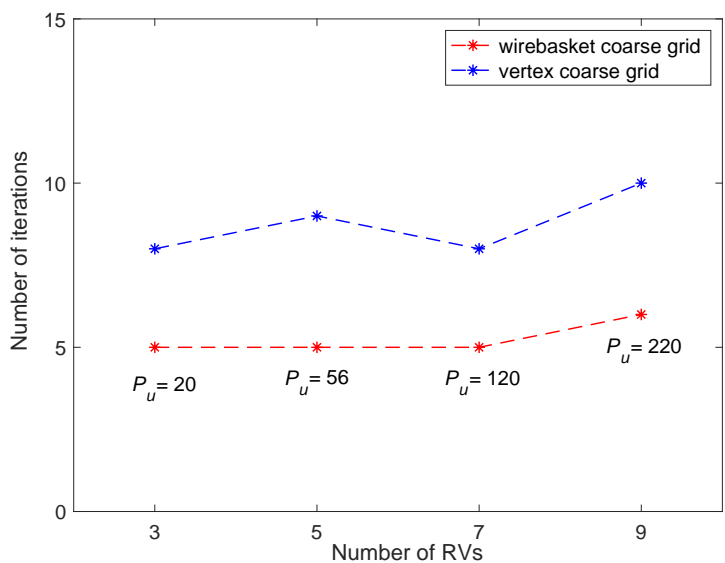}
\caption{Iteration count versus number PCE terms for the fixed mesh resolution with fixed number of subdomains.}
\label{fig:3DP_comp_nIterVnRVs}
\end{figure}

For a fixed mesh and number of subdomains ($n_s=384$), 
the PCGM iteration counts for the wirebasket-based preconditioner remain nearly constant against
increasing PCE terms (by adding more random variables, but
with the third-order expansion)
 as shown in \Cref{fig:3DP_comp_nIterVnRVs}. Conversely, for the same case, the number of PCGM iterations increases with the vertex-based coarse grid. Therefore, the BDDC/NNC solver with the extended wirebasket-based coarse grid is superior to vertex-based coarse grid concerning numerical scaling with respect to the number of subdomains and the number of random variables.

\subsection{Scalabilities Studies for Stochastic Simulations}\label{sec:3DP_scalabilites}

Next we investigate the performance of the proposed wirebasket-based 
preconditioner for high dimensional stochastic space. Therefore, for most of the simulations, the fixed mesh with $59741$ nodes and  $350113$ elements is used. For some cases, the mesh size is increased up to $159193$ nodes and $947046$ elements.

The scalability plots for various cases by selecting the number of random variables ranging between 3 to 12 are presented. The relative contribution of the KLE eigenvalues is used to select these cases  (see \Cref{fig:relativeEnergy3D}).

The orders of expansion $p_{\textsc{a}} = 2$ and $p_u = 3$ are used for the input and output PCE representations, respectively.
Note that the orders of expansion, $p_{\textsc{a}}$ and $p_u$
are kept constant for most of the experiments.

For some cases $p_u$ is varied between $2$ to $5$.
The number of solution PCE terms  varies from 20 to 455~\cite{ghanemSFEM1991}.
The maximum size of the system of linear equations simulated in this case is about $75$ million, i.e., using finite element mesh with approximately $160$ thousand node points and $455$ PCE terms (12-RVs). To test the numerical scalability of the solver,  we study the PCGM iterations counts for the convergence with the tolerance $tol = 10^{-5}$.

Next, we analyze the (strong and weak) parallel scalabilities in terms of execution time against the number of subdomains. Finally, we present the numerical and parallel scalability plots with respect to the stochastic parameters such as the number of random variables and the order of expansion.

\subsubsection{Numerical Scalability}

\begin{figure}[htbp]
\centering
\includegraphics[width=0.6\textwidth]{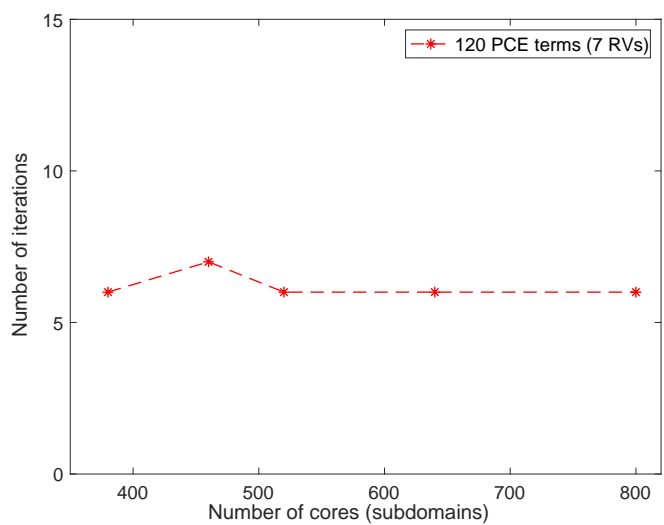}
\caption{Iteration count versus number of subdomains for the fixed number of PCE terms and the fixed mesh resolution.}
\label{fig:3DP_nIterVnProc}
\end{figure}

For a fixed problem size (59741 nodes and 350113 elements and PCE terms $P_u=120$ for $7$ RVs), the PCGM iteration count almost remains constant with increasing the number of subdomains indicating the numerical scalability of the wirebasket-based BDDC/NNC-PCGM solver (\Cref{fig:3DP_nIterVnProc}).

\begin{figure}[htbp]
\centering
\includegraphics[width=0.6\textwidth]{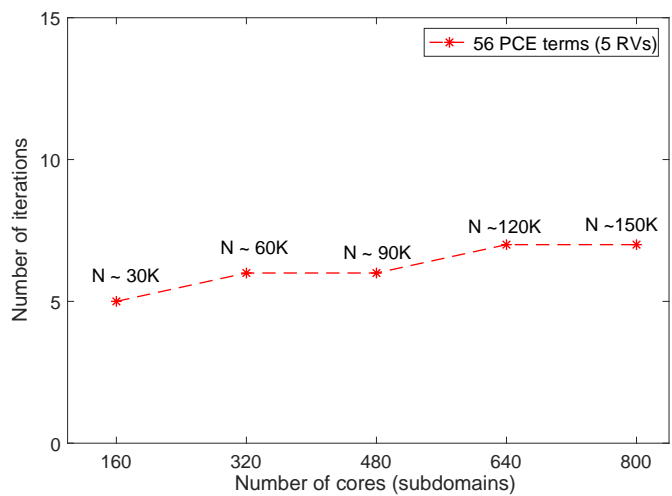}
\caption{Iteration count versus number of subdomains for the fixed problem size per core with increasing mesh resolution (fixed number of PCE terms).}
\label{fig:3DP_weak_IterVnProc}
\end{figure}
Next, for the fixed number of PCE terms ($P_u=56$ for $5$ RVs), the global problem size is increased by increasing mesh resolution for a fixed problem size per subdomain $(\approx 10500)$.
As the sizes of the total problem and coarse problem  increase while adding new subdomains,
the PCGM iteration count increases moderately with subdomain counts. \Cref{fig:3DP_weak_IterVnProc}  shows reasonable numerical scalability of BDDC/NNC-PCGM solver  with respect to mesh resolution and fixed problem size per subdomain.

\subsubsection{Parallel Scalabilities}
\begin{figure}[htbp]
\centering
\includegraphics[width=0.65\textwidth]{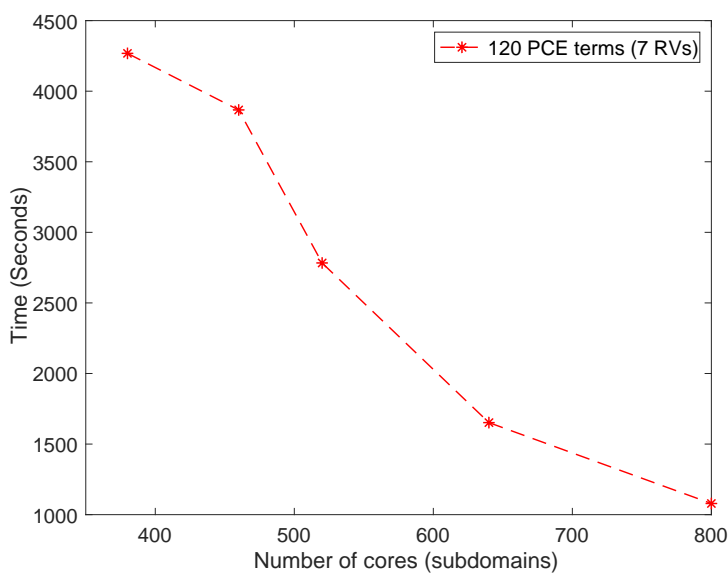}
\caption{Execution time versus number of subdomains with the number of PCE terms and the fixed mesh resolution.}
\label{fig:3DP_timeVnProcs}
\end{figure}

\Cref{fig:3DP_timeVnProcs} shows the strong scaling plots of the BDDC/NNC-PCGM solver.
For the fixed number of PCE terms ($P_u = 120$ for $7$ RVs) and a fixed spatial mesh resolution, increasing core counts cut the execution time as expected.
\Cref{fig:3DP_timeVnProcs} indicates excellent strong scaling of the BDDC/NNC solver with the wirebasket-based coarse grid.
Similar behavior is observed in the cases with three and five random variables (although not shown here).

\begin{figure}[htbp]
\centering
\includegraphics[width=0.65\textwidth]{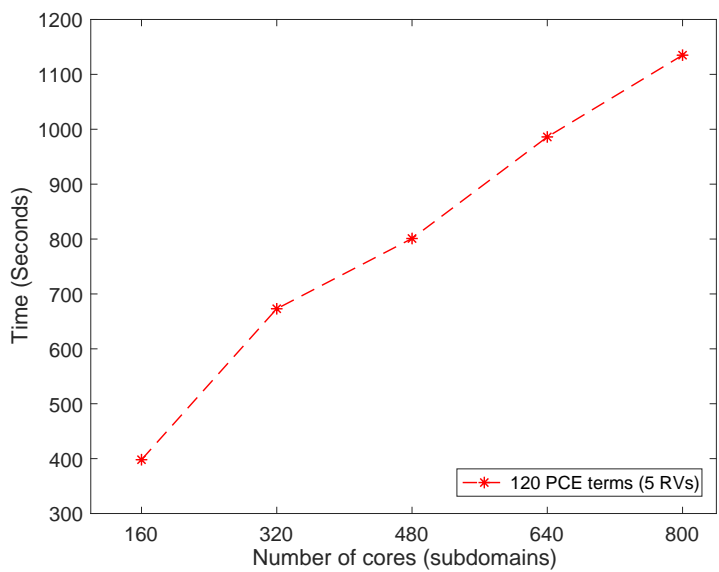}
\caption{Execution time versus number of subdomains for the fixed problem size per core with increasing mesh resolution (fixed number of PCE terms).}
\label{fig:3DP_weak_timeVnProcs}
\end{figure}
The weak scaling of the wirebasket-based  solver is shown in \Cref{fig:3DP_weak_timeVnProcs}. For the fixed number of PCE terms ($P_u=56$ for $5$ RVs), the global problem size is increased by increasing mesh resolution while keeping a fixed problem size per subdomain $(\approx 10500)$.
The growth in the total execution time implies that the probabilistic BDDC/NNC solver shows poor weak scaling.
Similar behavior is also observed in the cases of probabilistic BDDC/NNC solver for the two-dimensional stochastic diffusion equations~\cite{desai2019scalable}. 
The suboptimal weak scaling of the solver is mainly because of the increased parallel overhead with the number of cores caused by the MPI collective communication~\cite{desai2017scalable,subber2012PhDTh}.
The wirebasket coarse problem increases substantially with 
the number of subdomains, demanding more computational efforts.

\subsubsection{Scalability with respect to Stochastic Parameters}
 
\begin{figure}[htbp]
\centering
\includegraphics[width=0.65\textwidth]{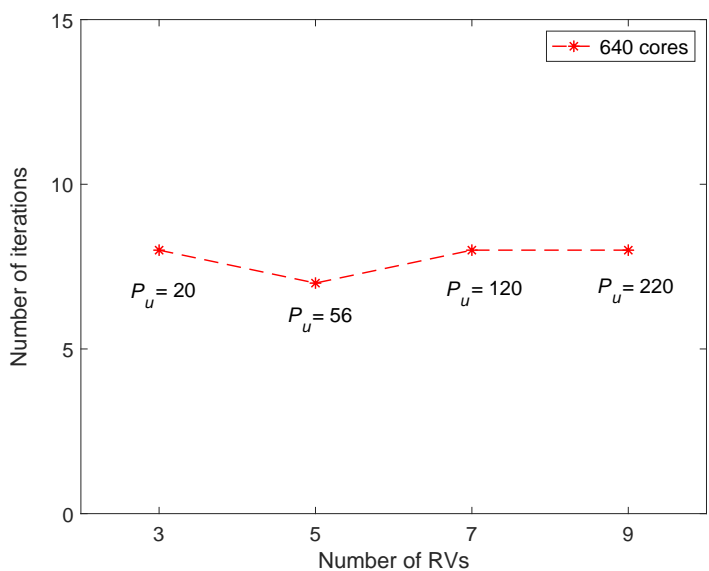}
\caption{Iteration count versus number PCE terms for the fixed mesh resolution with fixed number of subdomains.}
\label{fig:3DP_nIterVnPCE}
\end{figure}

\begin{figure}[htbp]
\centering
\includegraphics[width=0.65\textwidth]{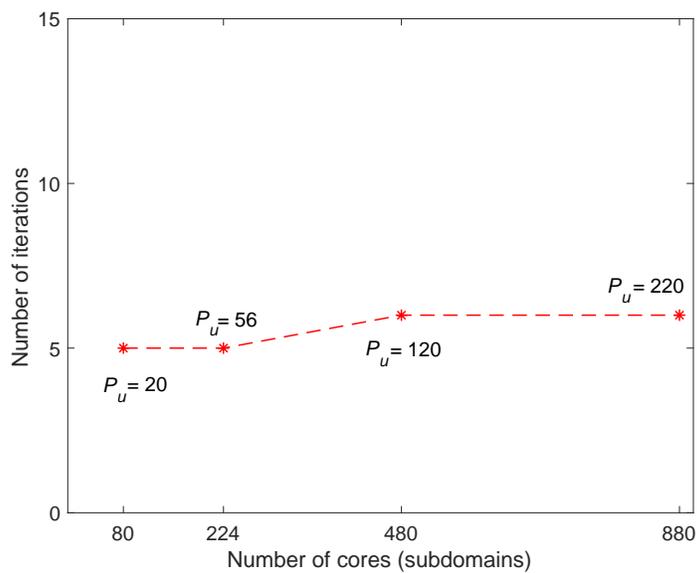}
\caption{Iteration count versus number of subdomains for the fixed problem size per core with increasing number of PCE terms (fixed mesh resolution).}
\label{fig:3DP_StoWeak_IterVnProc}
\end{figure}

\begin{figure}[htbp]
\centering
\includegraphics[width=0.65\textwidth]{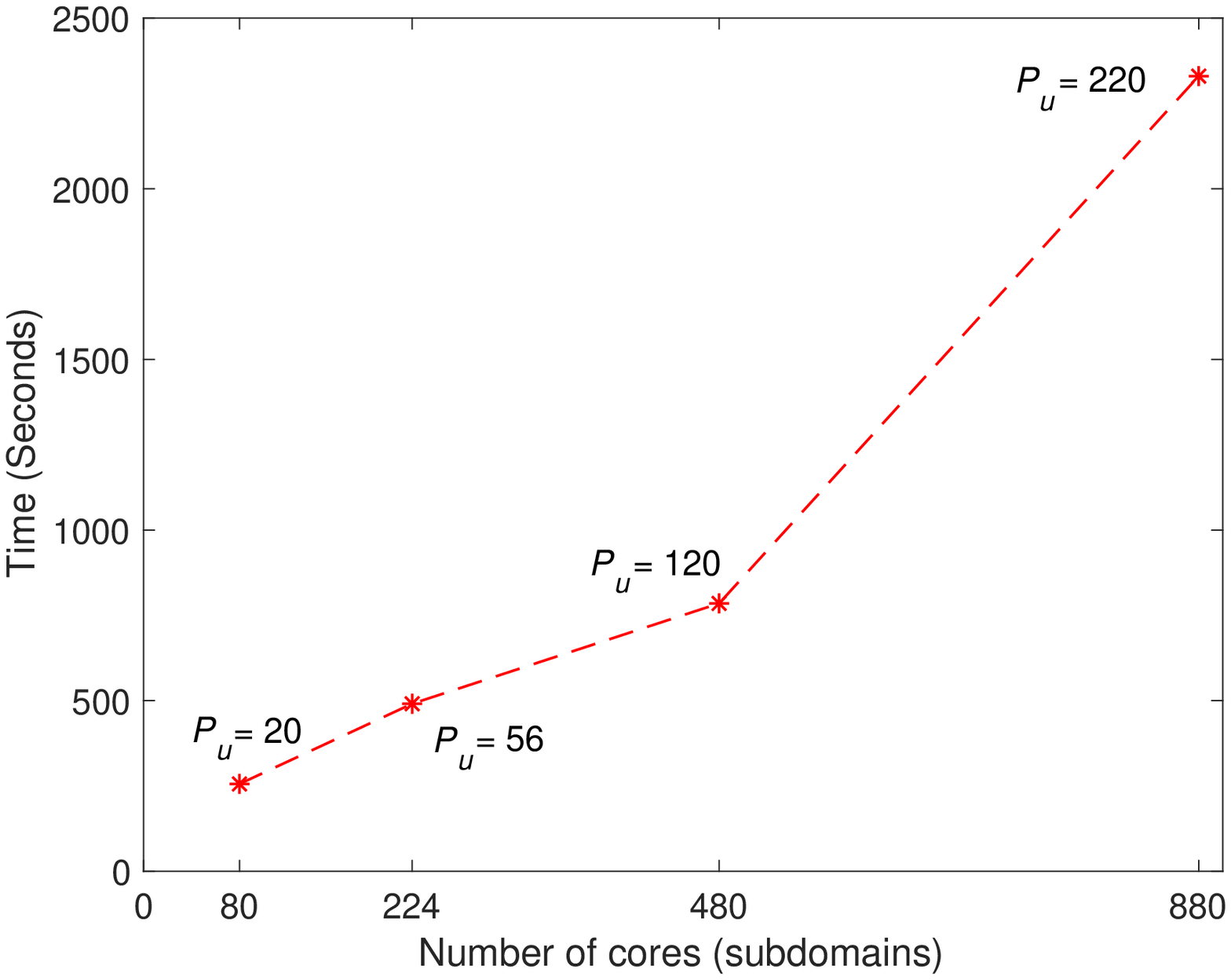}
\caption{Execution time versus number of subdomains for fixed problem size per subdomain  with increasing  number of PCEs (fixed mesh resolution).}
\label{fig:3DP_StoWeak_TimeVnProc}
\end{figure}

To study the stochastic aspect of the solver, the numerical and parallel scaling against PCE terms and the order of expansions are presented in this section. 
The four points in \Crefrange{fig:3DP_nIterVnPCE}{fig:3DP_StoWeak_TimeVnProc} for PCE terms 20, 56, 120 and 220 correspond to the input random variables 3, 5, 7 and 9 respectively, with the mesh having 59741 nodes and 350113 elements.

For a fixed mesh  and subdomains ($n_s=640$), an increase in the number of PCE terms by only increasing the number of random variables, i.e,  maintaining the fixed order of expansion, requires  almost constant PCGM iterations (\Cref{fig:3DP_nIterVnPCE}).
This demonstrate that, a larger problem due to increasing PCE terms can be tackled with nearly constant PCGM interation counts.
 This fact demonstrates the numerical scalability of the wirebasket-based BDDC/NNC-PCGM solver against the number of random variables.

\begin{figure}[htbp]
\centering
\includegraphics[width=0.65\textwidth]{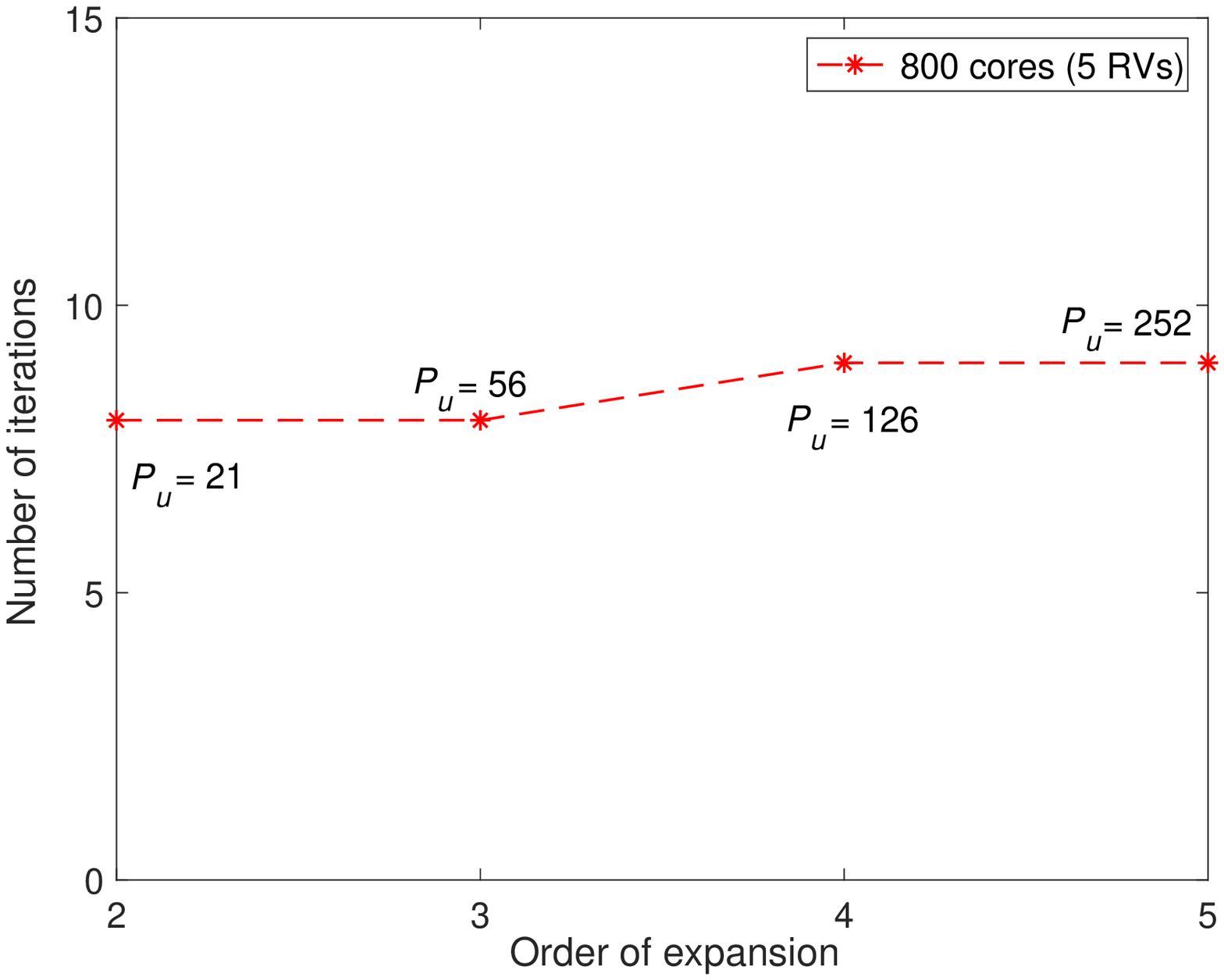}
\caption{Iteration count versus order of expansion for the fixed mesh resolution with fixed number of subdomains (fixed number of RVs).}
 \label{fig:3DP_nIterVnOrd}
\end{figure}

\Cref{fig:3DP_StoWeak_IterVnProc} demonstrates that the wirebasket-based BDDC/NNC-PCGM solver's numerically scalability with respect to the number of PCE terms. For a fixed mesh, the larger linear system with increasing PCE terms
is solved by adding more cores, maintaining approximately the same problem size per core  ($\approx 15000$). In this case, The PCGM iteration counts against the number of subdomains (with the fixed problem size per subdomain) remain nearly the same.

For the third order output PCE expansion with 3, 5, 7 and 9 RVs, \Cref{fig:3DP_StoWeak_TimeVnProc} shows the execution time of the solver for a fixed problem size per subdomain. The increase in the execution time is primarily due to increased MPI collective communication overhead. Additionally, as pointed out earlier in \cite{desai2019scalable}, 
the increasing PCE terms leads to a larger set of coupled PDEs with a more complex coupling structure.

 For a fixed mesh with fixed number of subdomains ($n_s=800$), while maintaining the fixed number of random variables $L=5$, an increase in the number of PCE terms by increasing the order of expansion $p_u$, results in almost constant PCGM iteration counts (\Cref{fig:3DP_nIterVnOrd}). It suggests the numerical scalability of the  solver against  the order of PCE expansion.
Note that, the coupling structures resulting from increasing number of PCE terms by increasing the number of random variables are more complex and computationally demanding than those resulting from increasing order of expansion~\cite{desai2019scalable}.

\section{Three-Dimensional Linear Elasticity Problem}\label{sec:3DE}
Application of deterministic finite element method to the equations of linear elasticity results in an extremely large, coupled system of linear equations for high resolution meshes. This linear system is often much more computationally expensive than those resulting from finite element discretization of scalar-valued elliptic PDEs, such as diffusion equation~\cite{smith2004domain,toselli2005domain}. Therefore solution of such system demands extensive computational efforts~\cite{smith1992optimal,smith2004domain,toselli2005domain}.
Moreover, extending the problem in stochastic space by using polynomial chaos expansion based intrusive SSFEM adds an additional coupling in the resulting system of equations and further complicates the system~\cite{subber2012PhDTh}.
This coupling among the polynomial chaos coefficients of the vector-valued solution process results in complicated block structure of the stochastic system matrix. Therefore, this can further influence the condition number of the global Schur complement system resulting from non-overlapping domain decomposition method. To tackle such system, a simple vertex-based coarse grid is inefficient~\cite{subber2012PhDTh}, as we need a much stronger mechanism such as a wirebasket coarse grid to communicate information globally~\cite{subber2012PhDTh,smith1992optimal}. In this section, we employ the two-level BDDC/NNC preconditioner using extended wirebasket-based coarse grid discussed in~\Cref{sec:NNW}, to simulate the stochastic system of coupled equations in linear elasticity.

To simplify the application of DD-based intrusive SSFEM to a coupled stochastic PDE system, we have presented a detailed formulation for the extended Schur complement system for coupled stochastic PDE system in \Cref{sec:exSchurE}. The efforts are made to explain the two-levels of couplings arising in the setting of DD-based intrusive SSFEM due to PC expansion of vector-valued solution process.
In \Cref{sec:3DE_exper}, an experimental framework for the numerical simulations of a cantilever beam is discussed. This is followed by the discussion on the characteristics of the stochastic solution process in \Cref{sec:3DE_solchar}. The \Cref{sec:3DE_compare} is dedicated to comparing the numerical scalability of BDDC/NNC solver with the extended wirebasket-based coarse grid against the vertex-based coarse grid. Finally in \Cref{sec:3DE_scalabilites}, the numerical and parallel scalability results of wirebasket-based BDDC/NNC solver are presented for a cantilever beam deformed due to self-weight.

In this section, we exploit extended wirebasket-based BDDC/NNC preconditioners to solve the Schur complement system for a coupled stochastic PDE system exemplified through linear elasticity problems. For an exposition of the methodology, we consider the equations of linear elasticity to model the vector-valued stochastic displacement field \ $\mathbfcal{U}(\textbf{\textit{x}},\theta)$ of a three-dimension body defined over the volume $\mathbfcal{D}({x,y,z})$. The resulting equations can be written as~\cite{smith1991domain,smith1992optimal,smith2004domain},
\begin{align}
-\nabla \cdot \sigma\big(\mathbfcal{U}(\textbf{\textit{x}},\theta) \big) &= F({\textbf{\textit{x}}}) \ \ \ \   in \ \ \ \ \ \mathbfcal{D}, \label{3Delasticity1} \\
\sigma\big(\mathbfcal{U}(\textbf{\textit{x}},\theta) \big) \cdot \bf{\hat{n}} &= b_T  \ \ \ \ \  on \ \ \ \ \ \Gamma_{1}  = \delta{\mathbfcal{D}}\backslash \Gamma_{0}, \\
\mathbfcal{U}(\textbf{\textit{x}},\theta) &= 0 \ \ \ \ on \ \ \ \ \  \Gamma_{0},
\end{align}
where $\sigma$ is the stress tensor, $F$ is the body force vector per unit volume, $\bf{\hat{n}}$ is the outward unit normal on the boundary $\Gamma$ and $b_T$ is the traction on the boundary $\Gamma_{1}$.
For homogeneous and isotropic linear elasticity the stress tensor $\sigma$ can be written as~\cite{smith2004domain,logg2012FEniCS}
\begin{equation}
\sigma \big(\mathbfcal{U}(\textbf{\textit{x}},\theta) \big) = \lambda \big( \nabla \cdot \mathbfcal{U}(\textbf{\textit{x}},\theta) \big) I + 2 \mu \epsilon \label{3Delasticity3}, %\Big( {\mathbfcal{U}} (\textbf{\textit{x}},\theta) \Big)
\end{equation}
where $I$ is the identity matrix, the $\mu$ and $\lambda$ are Lam\'e constants and $\epsilon$ is the symmetric strain tensor expressed by
\begin{align}
  \epsilon = \frac{1}{2} \Big( \nabla \mathbfcal{U}(\textbf{\textit{x}},\theta) + \big(\mathbfcal{U}(\textbf{\textit{x}},\theta) \big)^{\mathrm{T}}  \Big).
\end{align}

To construct the variational form we take

inner product of \Cref{3Delasticity1} with the test function $ \mathbfcal{V}$ and integrate over the domain  $\mathbfcal{D}$,

\begin{align}
\int_{\mathbfcal{D}} \big(-\nabla \cdot \sigma(\mathbfcal{U}) \big) \cdot \mathbfcal{V} dx & = \int_{\mathbfcal{D}} F \cdot \mathbfcal{V} dx.
\label{3Delasticity_weak}
\end{align}
Integrating by parts, \Cref{3Delasticity_weak} becomes (note $\textbf{\textit{x}}$ and $\theta$ of ${\mathbfcal{U}}$ are dropped for brevity),
\begin{align}
\int_{\mathbfcal{D}} \sigma(\mathbfcal{U}) : \nabla\mathbfcal{V} dx & = \int_{\mathbfcal{D}} F \cdot \mathbfcal{V} dx + \int_{\Gamma_{1}} b_{T} \cdot \mathbfcal{V} ds,
\label{3Delasticity_weak2}
\end{align}
where the operator $:$ is the inner product between tensors.
$b_t$ is prescribed on a part $\Gamma_{1}$ of the boundary.
Therefore, the integral on the remaining part of the boundary $\Gamma_{0}$ vanishes due to a Dirichlet boundary condition.
\Cref{3Delasticity_weak2} can be re-written more concisely using {\it{bilinear}} and {\it{linear forms}} as,
\begin{align}
  a(\mathbfcal{U}, \mathbfcal{V}) = \mathbfcal{L}(\mathbfcal{V}).
\end{align}

Lam\'e constants arising in the stress-strain relationship for the material can be written as the  functions of Poisson ratio $\nu$ and Young's modulus $E$ as follows~\cite{smith2004domain,logg2012FEniCS},
\begin{equation}\label{eq:lameparams}
\lambda = \frac{E\nu}{(1+\nu)(1-2\nu)}, \ \ \ \ \  \mu = \frac{E}{2(1+\nu)}.
\end{equation}
For simplicity, we consider Poisson's ratio as a spatially invariant parameter and model Young's modulus as the lognormal stochastic process representing spatially varying uncertainty,
\begin{equation}\label{eq:stoE}
E({\textbf{\textit{x}}},\theta) = E_0({\textbf{\textit{x}}}) \mathrm{exp}\big(g(\textbf{\textit{x}},\theta)\big),
\end{equation}
where $E_0({\textbf{\textit{x}}})$ represents the mean and $g(\textbf{\textit{x}},\theta)$ denotes the underlying Gaussian process with the variance $\sigma^2$ and the exponential covariance function $C$ defined as~\cite{subber2012PhDTh,ghanemSFEM1991}.
\begin{equation} \label{CoVFn}
C(x_1, y_1, z_1 ; x_2, y_2, z_2) = \sigma^2 \  e^{-|x_2 - x_1|/b_x -|y_2 - y_1|/b_y -|z_2 - z_1|/b_z},
\end{equation}
where $b_x$, $b_y$ and $b_z$ are the correlation lengths along $x$, $y$ and $z$ directions respectively.
While the assumption of the spatial invariance on the Poisson's ratio simplifies the analysis, the spatial variability of both Poisson's ratio and Young's modulus can be handled concurrently as the proposed framework is general~\cite{doostan2007stochastic,ghanem2008probabilistic,ghanemSFEM1991}.

\subsection{Numerical Experimental Framework}\label{sec:3DE_exper}
For the extended wirebasket-based BDDC/NNC solver, the PCGM presented in \Cref{alg:pcgm}~\cite{desai2019scalable} 
is implemented using Fortran programming language
with MPI communication routines~\cite{gropp1999MPI}.

Subdomain-level sparse matrix-vector storage,  operations and local system
solves are implemented using  PETSc~\cite{petsc2016}.
GMSH~\cite{gmshWeb2017} is used for unstructured finite element mesh generation and  METIS graph partitioner~\cite{metisWeb2017} is used for mesh partitioning. The stochastic system matrix and vector assemblies are performed by employing element-level (deterministic) assembly routines from the FEniCS/dolfin~\cite{logg2012FEniCS}. The  KLE and PCE related computations are performed using  the UQ Toolkit~\cite{debusschere2013uqtk}.
ParaView~\cite{ahrens2005ParaView,paraviewWeb2017} and Matlab~\cite{matlabGuide} are employed for post processing and visualization.
The simulations are performed on Canada's national HPC clusters managed by Compute Canada~\cite{nationalSystems}.
The nodes employed have either Intel Skylake cores running at 2.4 GHz from Niagara supercomputer~\cite{nationalSystemsNia}
or Intel E5-2683 processors, running at 2.1 GHz from Cedar and Graham HPC systems~\cite{nationalSystems}.
Similar to the earlier cases, the primary attention is given to study the efficacy of the NNC/BDDC solver in tackling high-dimensional coupled stochastic systems.
In all the simulations we use (four node) linear tetrahedral elements. Such low order linear finite elements are shown to be sufficient for compressible elastic materials~\cite{klawonn2006dual,klawonn2006parallel}.
In the following sections, we will  briefly discuss the characteristics of the solution process and,  numerical and parallel scalabilities of the  solver by varying the number of random variables.

\subsection{Characteristics of the Solution Process}\label{sec:3DE_solchar}
\begin{figure}[htbp]
 \centering
 \includegraphics[width=0.65\textwidth,height=0.38\textheight]{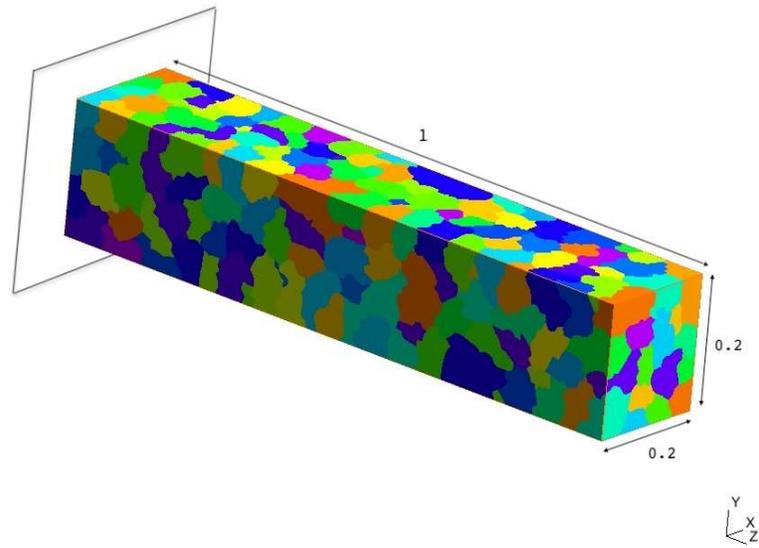}
 \caption{A typical three-dimensional finite element mesh for a clamped beam partitioned into 320 subdomains.}
 \label{fig:clampedBeam}
\end{figure}

As a test case, consider a clamped beam of length $L_b$ with a square cross section of width $W$, deformed under self-weight in three dimensions. A typical finite element mesh for such beam with $31598$ nodes and $182681$ linear tetrahedral elements and 320 subdomains (see \Cref{fig:clampedBeam}).

For the numerical experiments we have used the
nondimensionalized and scaled system parameters suggested in FEniCS tutorial~\cite{logg2012FEniCS}. The parameters are made dimensionless by using beam length and \ $\bar{\mathbfcal{U}} = \mathbfcal{U}/\mathbfcal{U}_m$. The scaling is achieved by choosing $\mathbfcal{U}_m$ equal to the maximum deflection of a clamped beam 
(refer to FEniCS tutorial~\cite{logg2012FEniCS} for further details).
The advantage of using scaled problem is to reduce the need for setting the physical parameters and also, the obtained dimensionless numbers can be used to understand the competition of parameters and physical effects.
Furthermore, the dimensionless parameters simplify the computational model and make it ideal for numerical experiments.
The following values are used for numerical simulations (as suggested in FEniCS tutorial~\cite{logg2012FEniCS}); Lam\'e constants are fixed to $\mu=1$ and $\lambda = 1.25$.
The density of the beam is set to $\rho=1$ and the acceleration due to gravity is set to $g = 0.4(W/L_b)^{2}$.
The beam is deformed under its weight with $F=(0,-\rho g, 0)$, where $F$ is the body force per unit volume 
and the boundary $\Gamma_{1}$ is traction free, i.e., $b_{T}=0$.

\begin{figure}[htbp]
 \centering
 \includegraphics[width=0.7\textwidth,height=0.35\textheight]{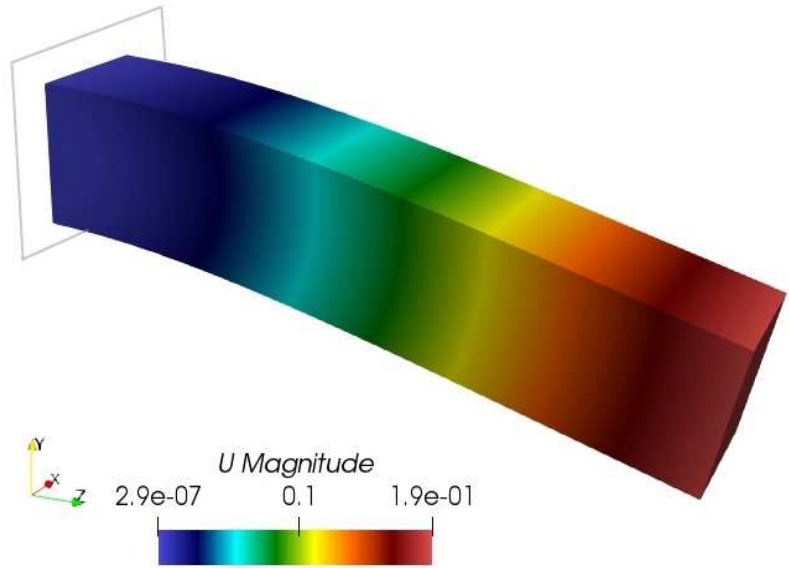}
 \caption{Mean magnitude of the beam deflection subjected to self-weight.}
 \label{fig:deflactedBeam2}
\end{figure}

\begin{figure}[htbp]
 \centering
  \subfloat[Mean: magnitude\label{subfig:pe21}]{%
 \includegraphics[width=0.35\textwidth,height=0.2\textheight]{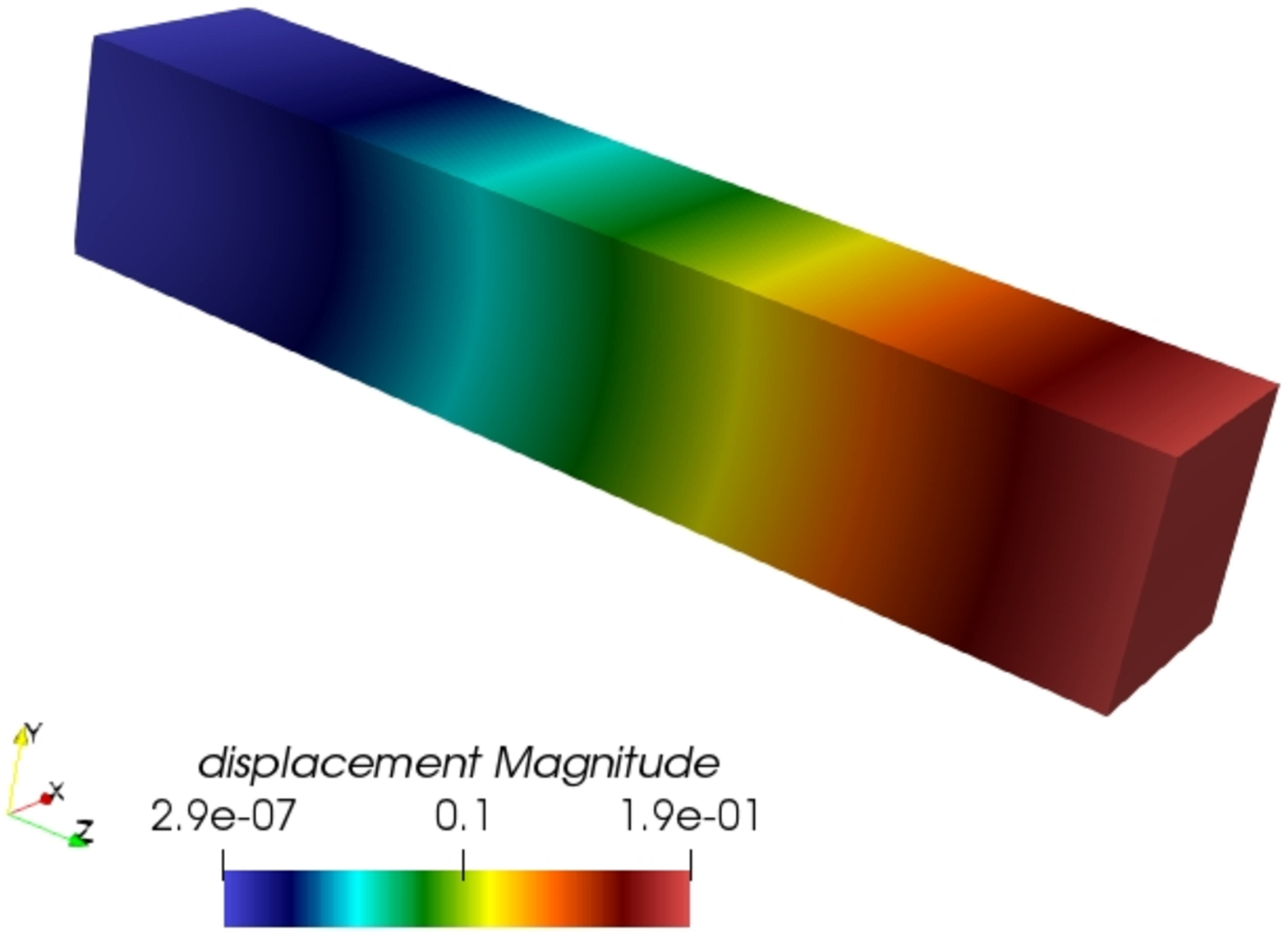}}
  \subfloat[SD: magnitude \label{subfig:pe22}]{%
 \includegraphics[width=0.35\textwidth,height=0.2\textheight]{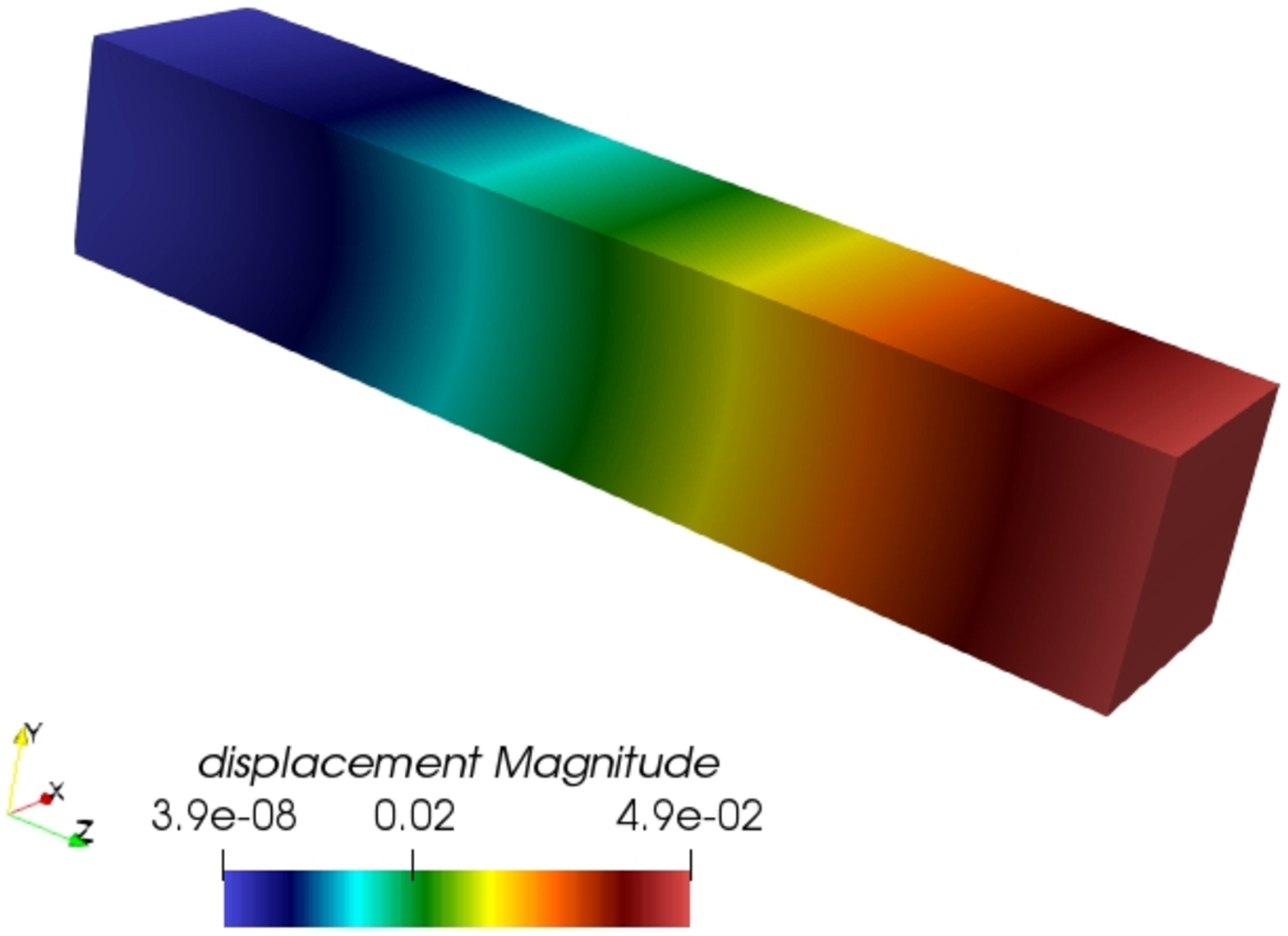}}

  \subfloat[Mean: $x$ \label{subfig:pe23}]{%
 \includegraphics[width=0.35\textwidth,height=0.2\textheight]{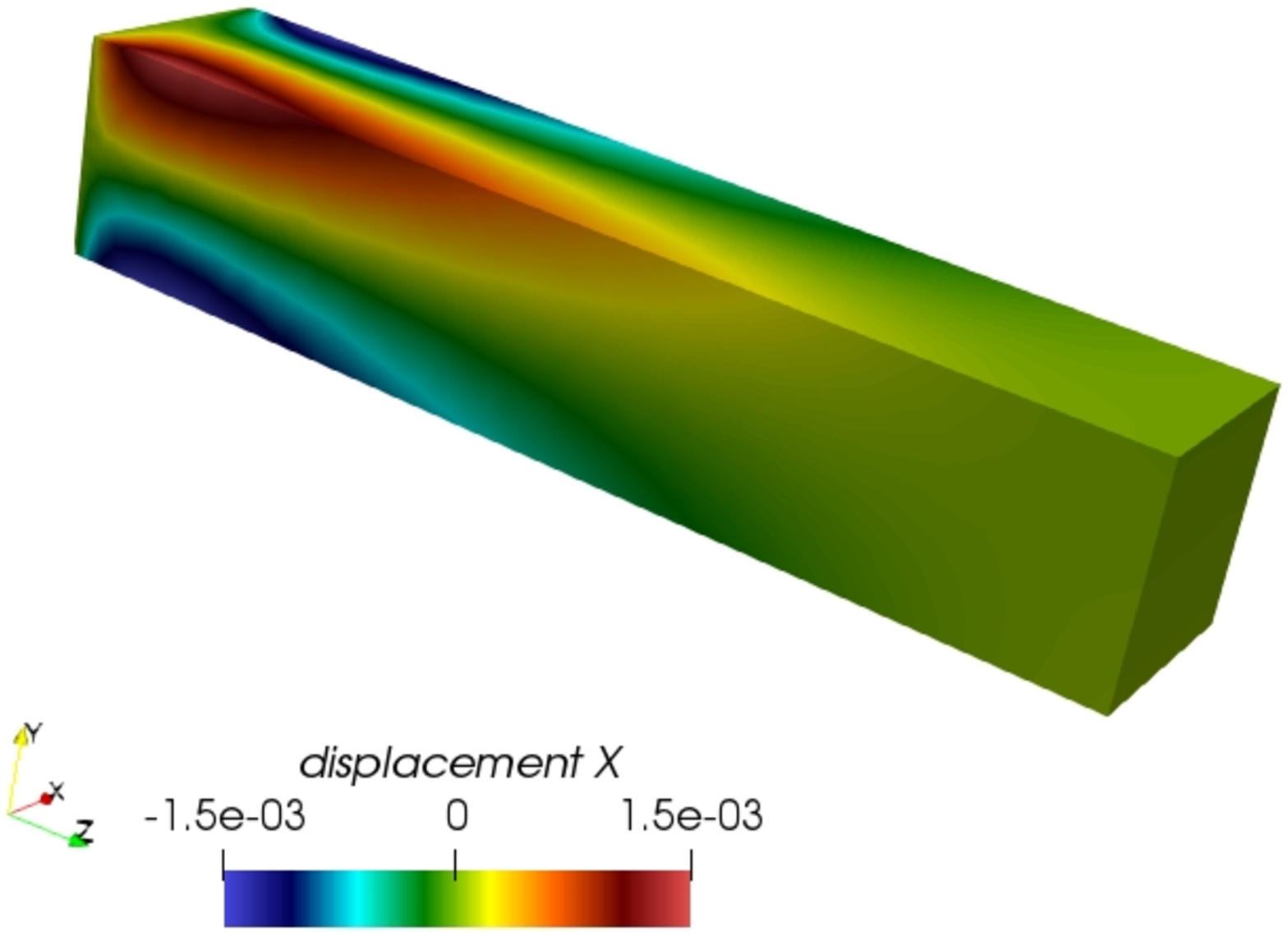}}
  \subfloat[SD: $x$ \label{subfig:pe24}]{%
 \includegraphics[width=0.35\textwidth,height=0.2\textheight]{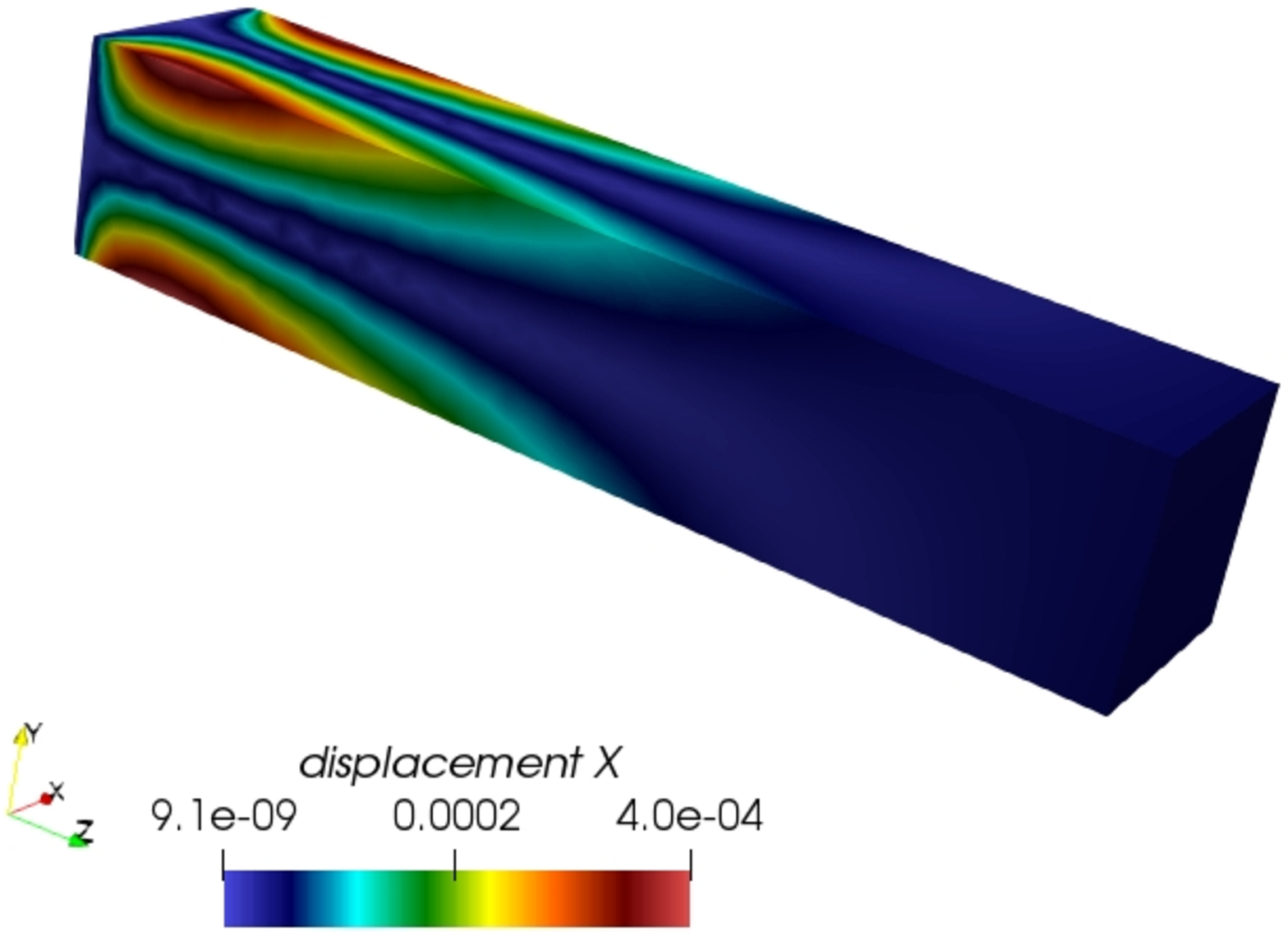}}

  \subfloat[Mean: $y$ \label{subfig:pe25}]{%
 \includegraphics[width=0.35\textwidth,height=0.2\textheight]{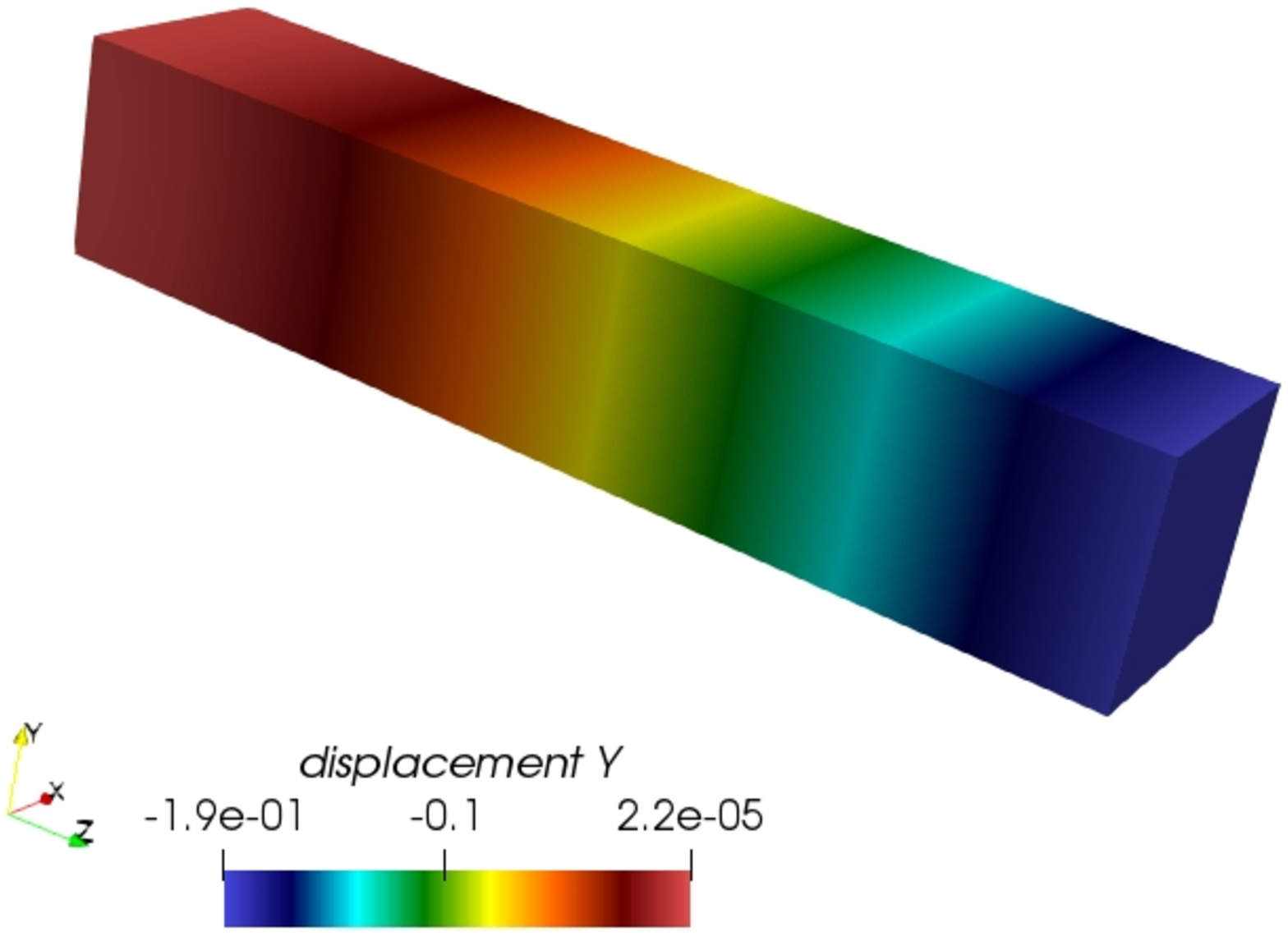}}
 \subfloat[SD: $y$ \label{subfig:pe26}]{%
 \includegraphics[width=0.35\textwidth,height=0.2\textheight]{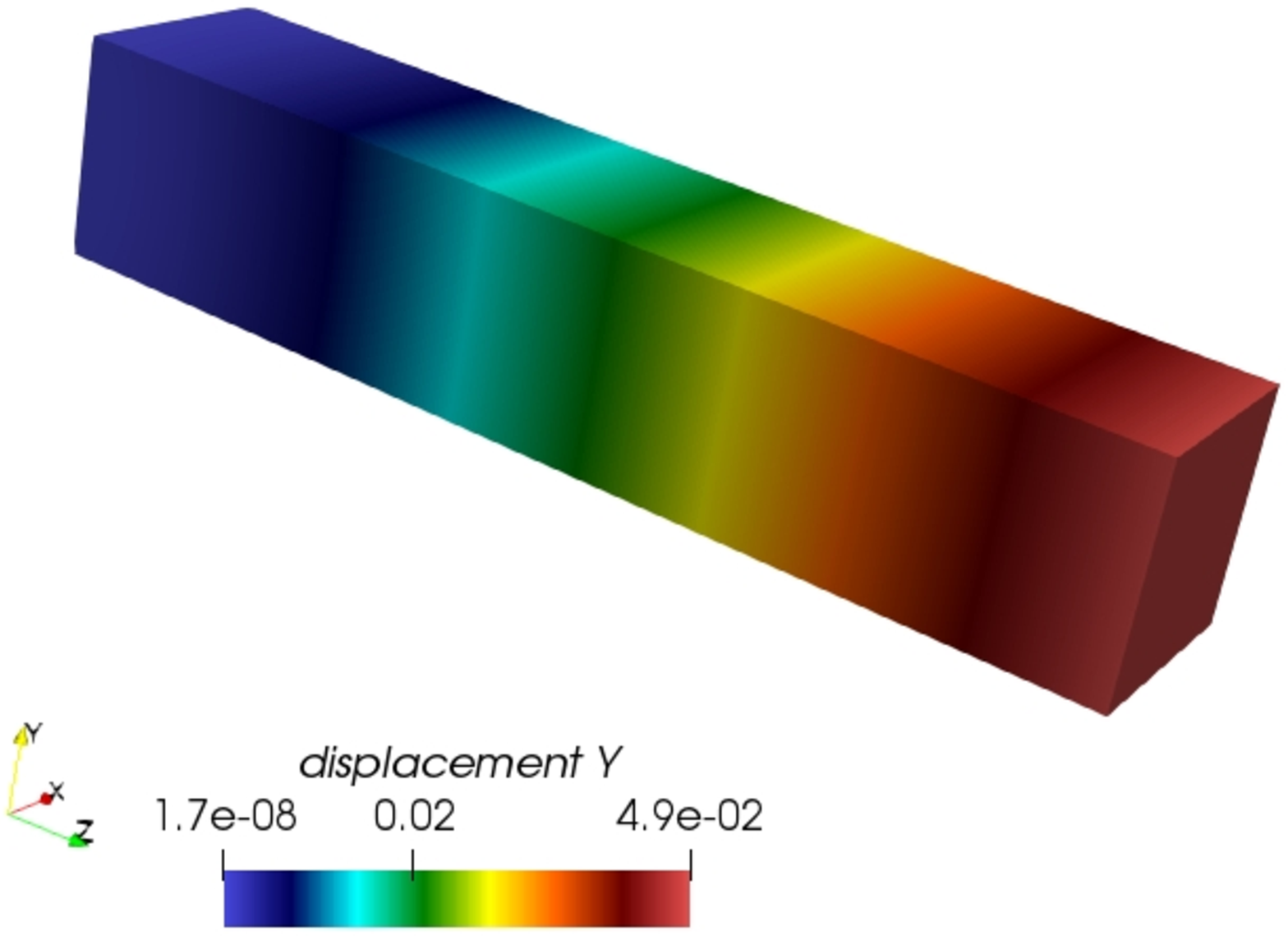}}

  \subfloat[Mean: $z$ \label{subfig:pe27}]{%
 \includegraphics[width=0.35\textwidth,height=0.2\textheight]{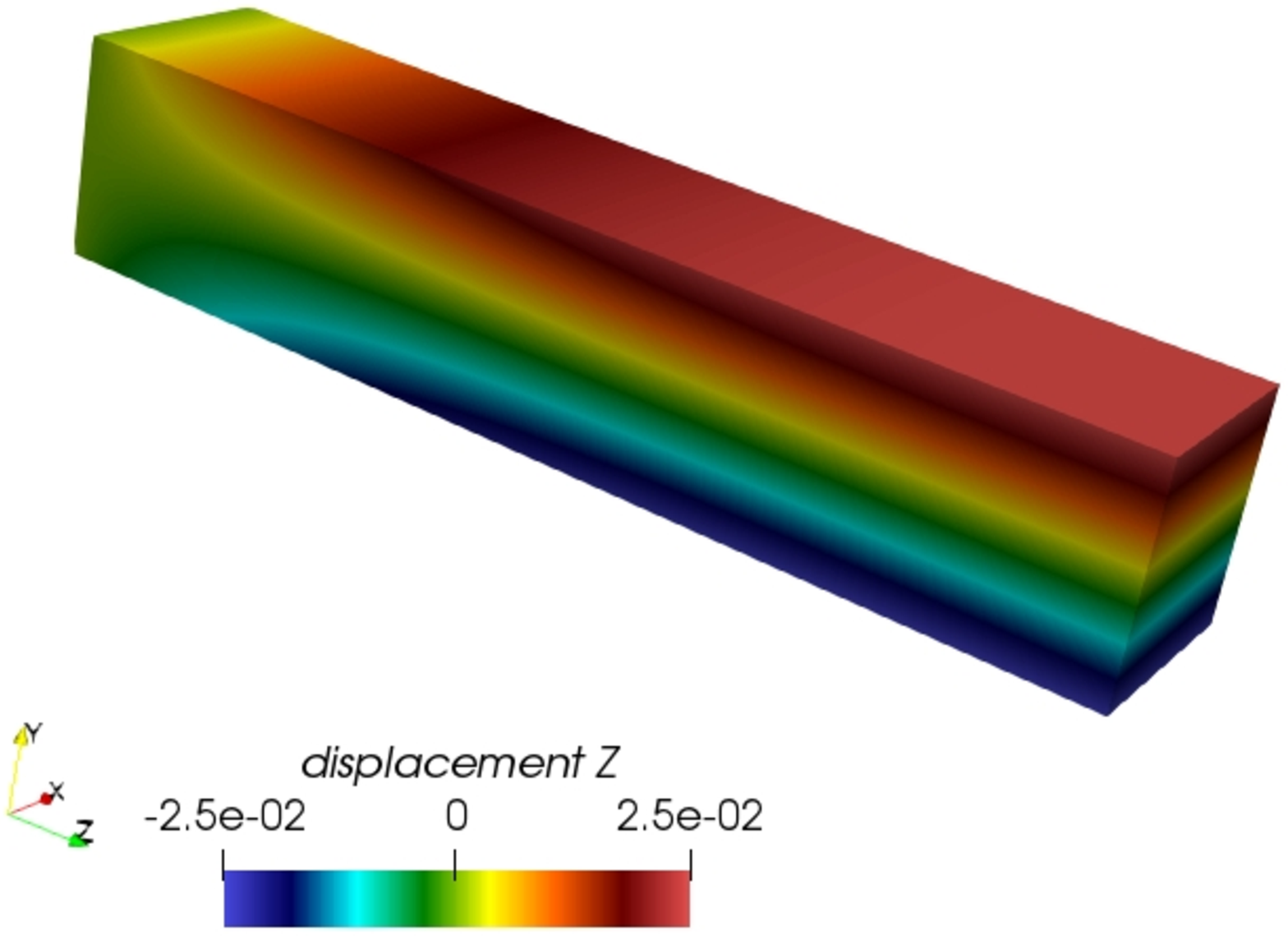}}
  \subfloat[SD: $z$ \label{subfig:pe28}]{%
 \includegraphics[width=0.35\textwidth,height=0.2\textheight]{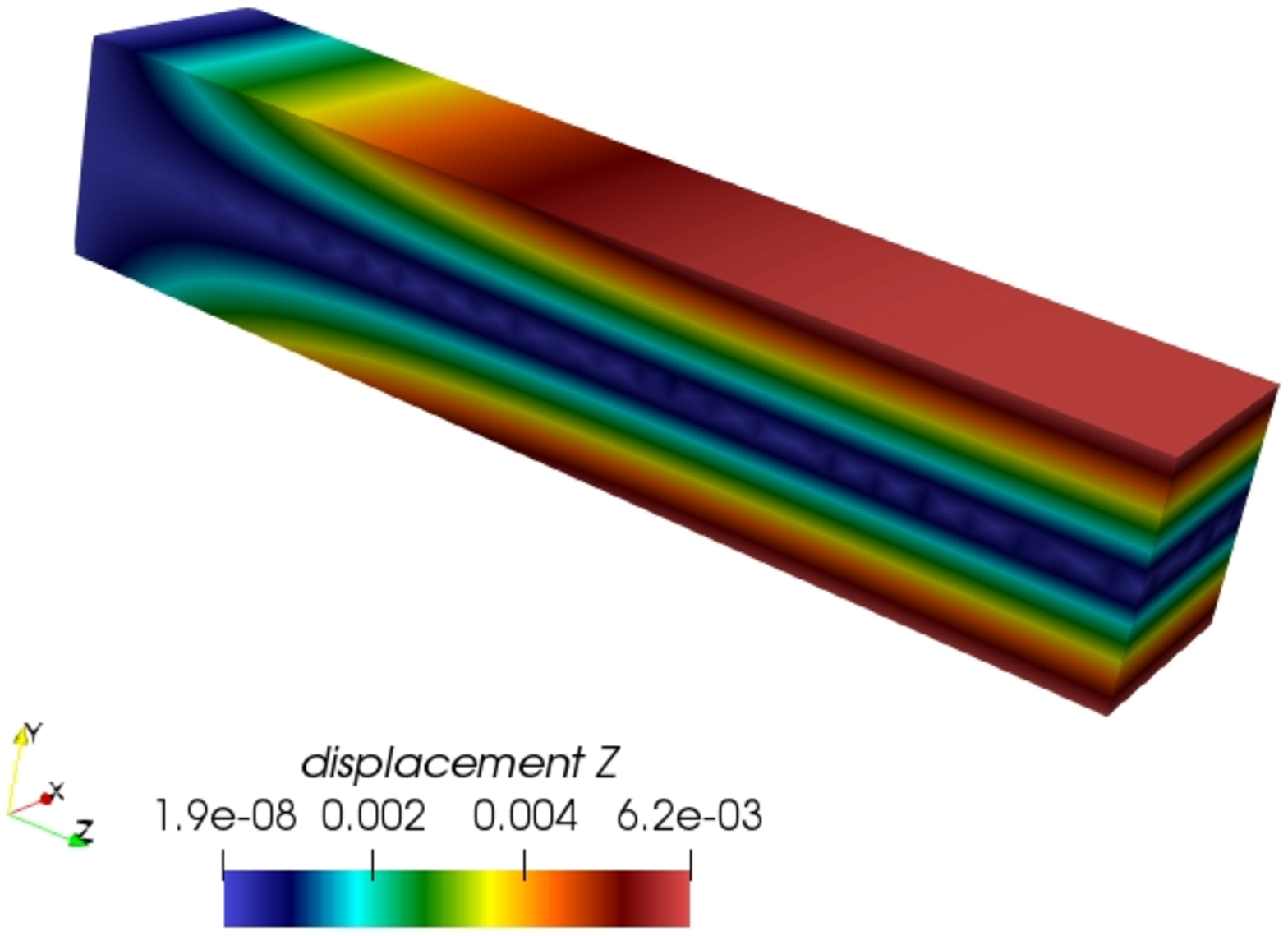}}
 \caption{Magnitude and $x,y$ and $z$ components of the mean and standard deviation of the solution process $\mathbfcal{U}$.}
 \label{fig:component}
\end{figure}

For simplicity, we assume a constant  Poisson's ratio  $\nu = 0.2778$ and model Young's modulus $E$ to be a non-Gaussian  process representing spatially varying uncertainty as shown in \Cref{eq:stoE}. The $E_{0}=2.556$ is calculated from \Cref{eq:lameparams} using $\mu=1$ and $\lambda = 1.25$. Young's modulus, represented as a lognormal stochastic process, is derived from the underlying Gaussian process with zero mean, 0.3 standard deviation and exponential covariance kernel with the correlation lengths $b_{x} = b_{y} = b_{z} = 1$.
The PCE of the input stochastic process is represented using three random variables ($L=3$) and the second order expansion ($p_{a} = 2$).
The PCE of the solution process with $L=3$ and $p_{u} = 3$ is used.
For the selected $L$ and $p_{\alpha}$, we need 10 PCE terms for the input and 20 PCE terms for the output~\cite{subber2012PhDTh}.
The mean value of the magnitude of the deflected beam is shown in~\Cref{fig:deflactedBeam2}. As expected, the maximum deflection occurs at the tip. 
The maximum coefficient of variation in the displacement magnitude  is about $26\%$, highlighting the effect of input uncertainty.
The mean and standard deviation (SD) of the magnitude of $\mathbfcal{U}$ and its components $\mathbfcal{U}_x,\mathbfcal{U}_y$ and $\mathbfcal{U}_z$ are shown in \Cref{fig:component}. The magnitude of the mean and SD have the same trend but the components of the displacement vector exhibit more complicated features.
The mean of the magnitude of deflection at the mid-span and the free end of the beam approximately matches with the analytical solution obtained at the mean values of the system parameters.

\begin{figure}[htbp]
 \centering
 \subfloat[$u_{1}$: magnitude \label{subfig:pe29}]{%
 \includegraphics[width=0.35\textwidth,height=0.2\textheight]{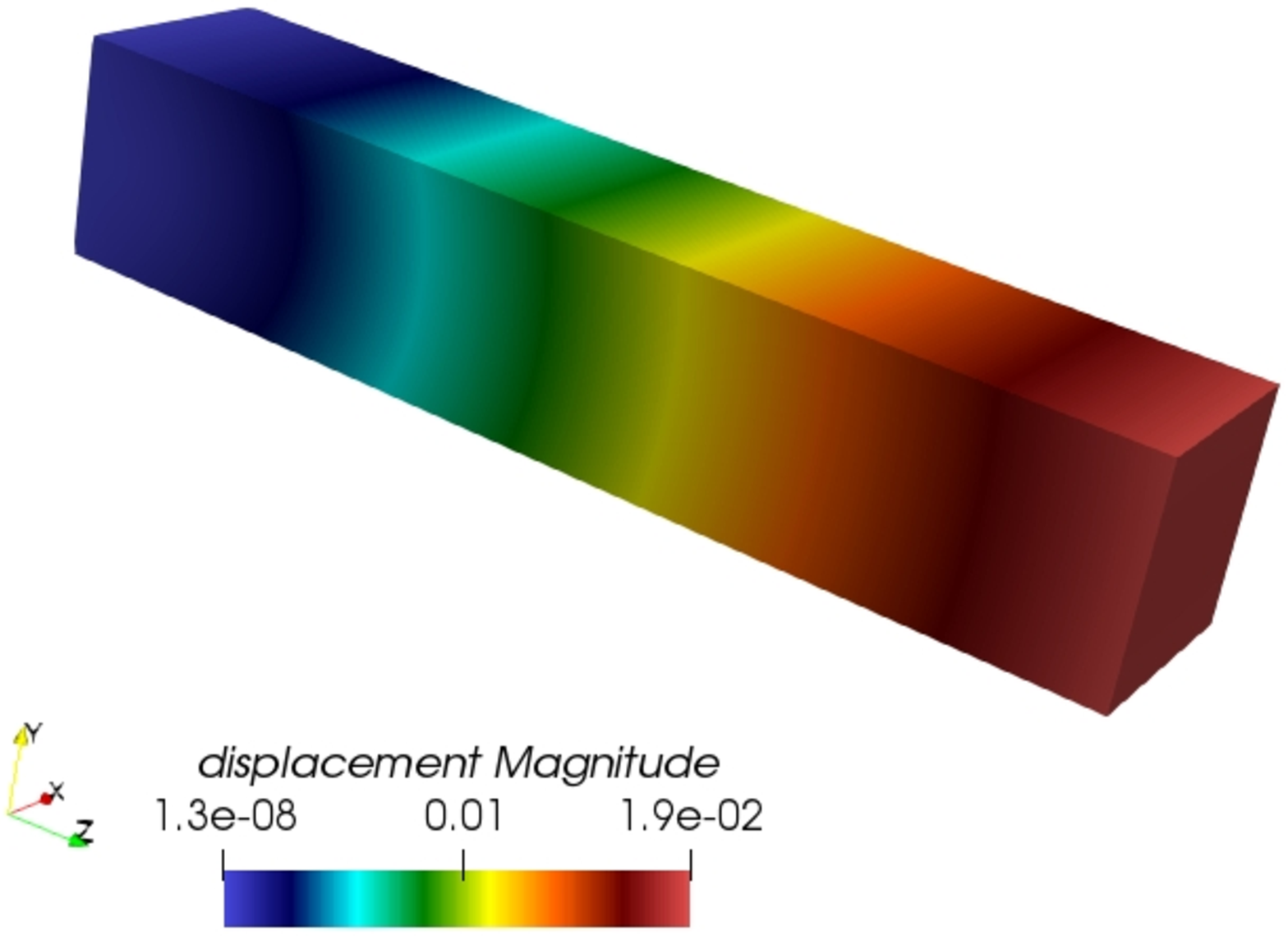}}
 \subfloat[$u_{2}$: magnitude\label{subfig:pe30}]{%
 \includegraphics[width=0.35\textwidth,height=0.2\textheight]{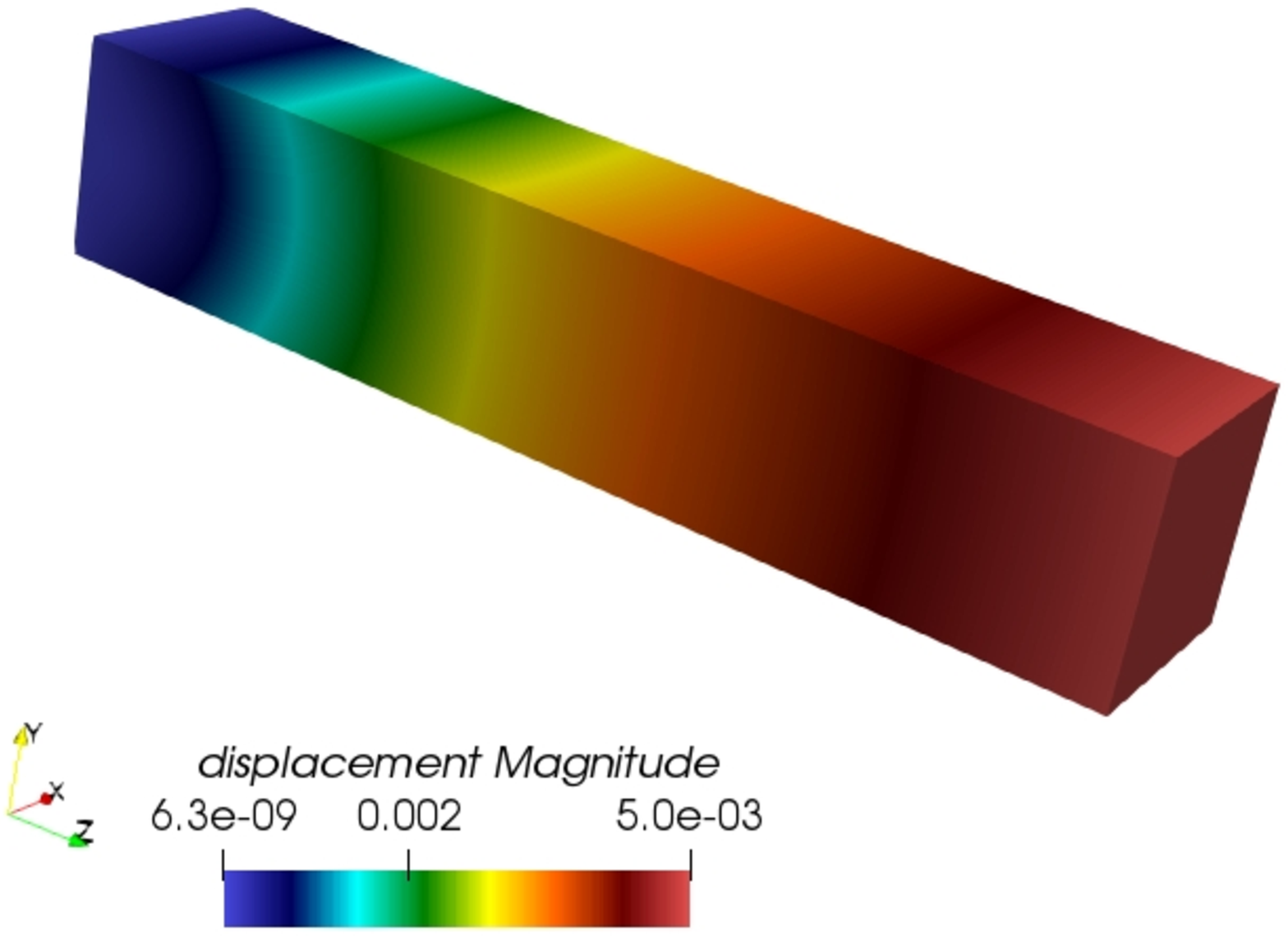}}

 \subfloat[$u_1$: $x$\label{subfig:pe31}]{%
 \includegraphics[width=0.35\textwidth,height=0.2\textheight]{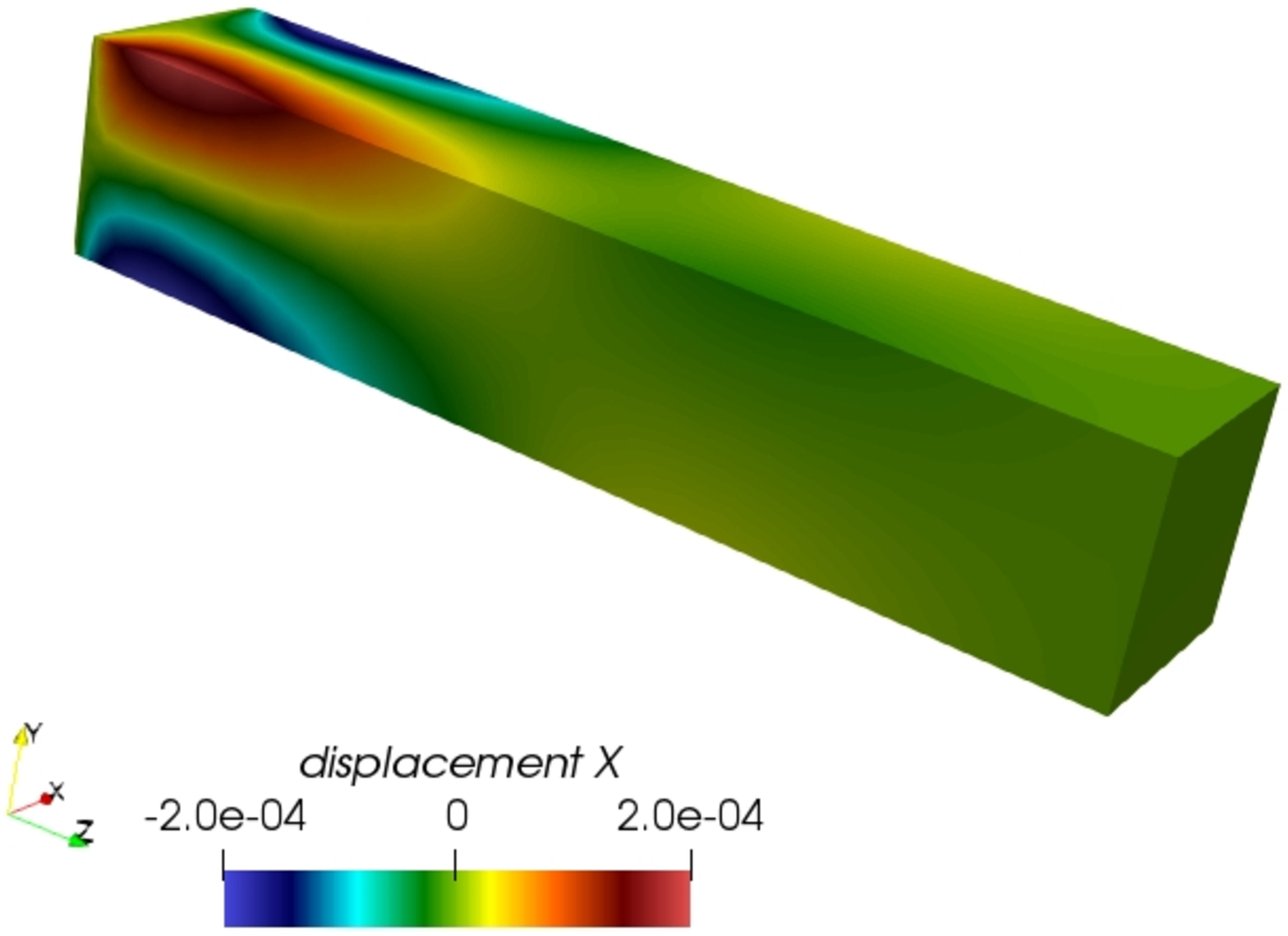}}
 \subfloat[$u_2$: $x$\label{subfig:pe32}]{%
 \includegraphics[width=0.35\textwidth,height=0.2\textheight]{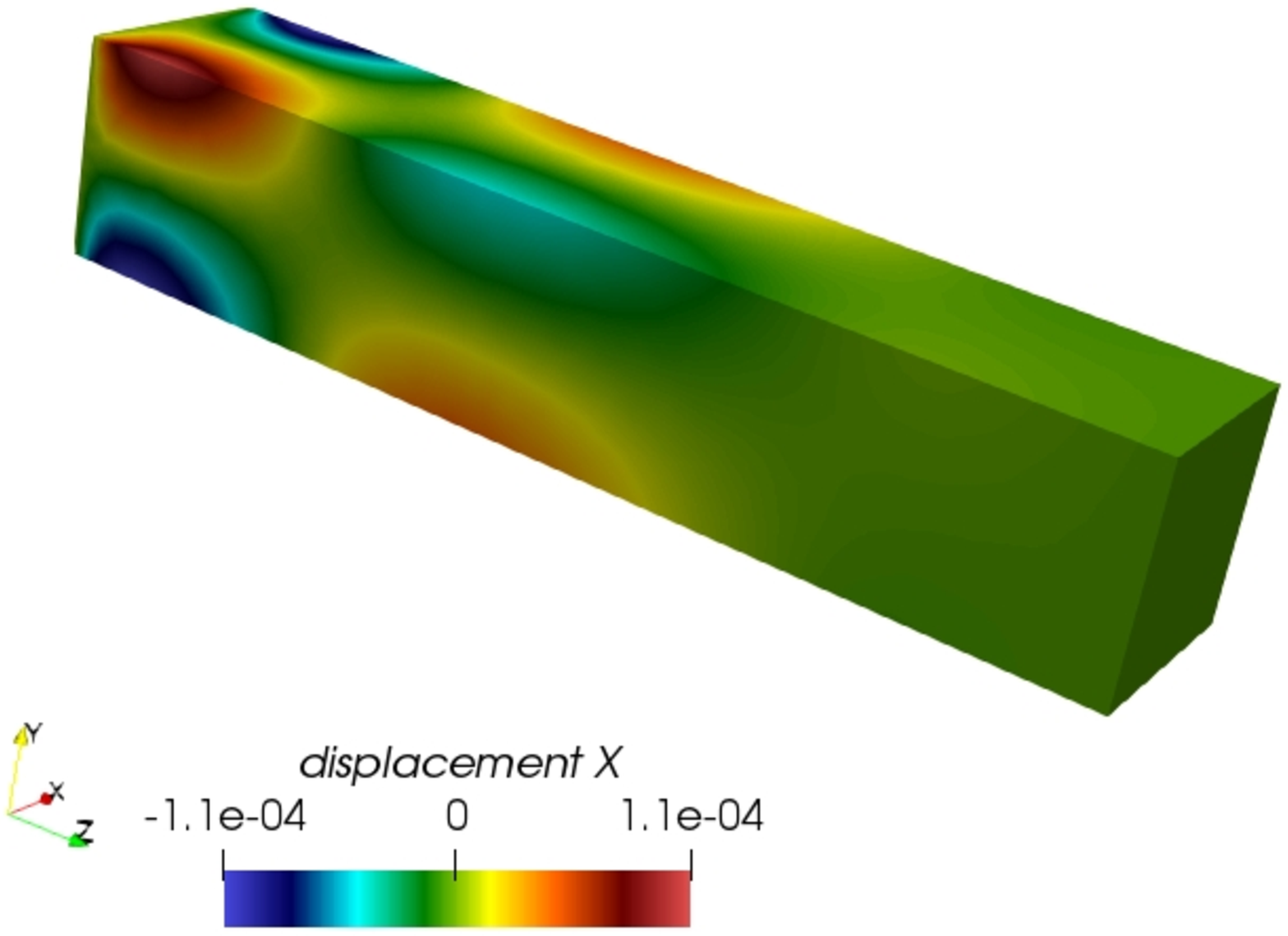}}

 \subfloat[$u_1$: $y$\label{subfig:pe33}]{%
 \includegraphics[width=0.35\textwidth,height=0.2\textheight]{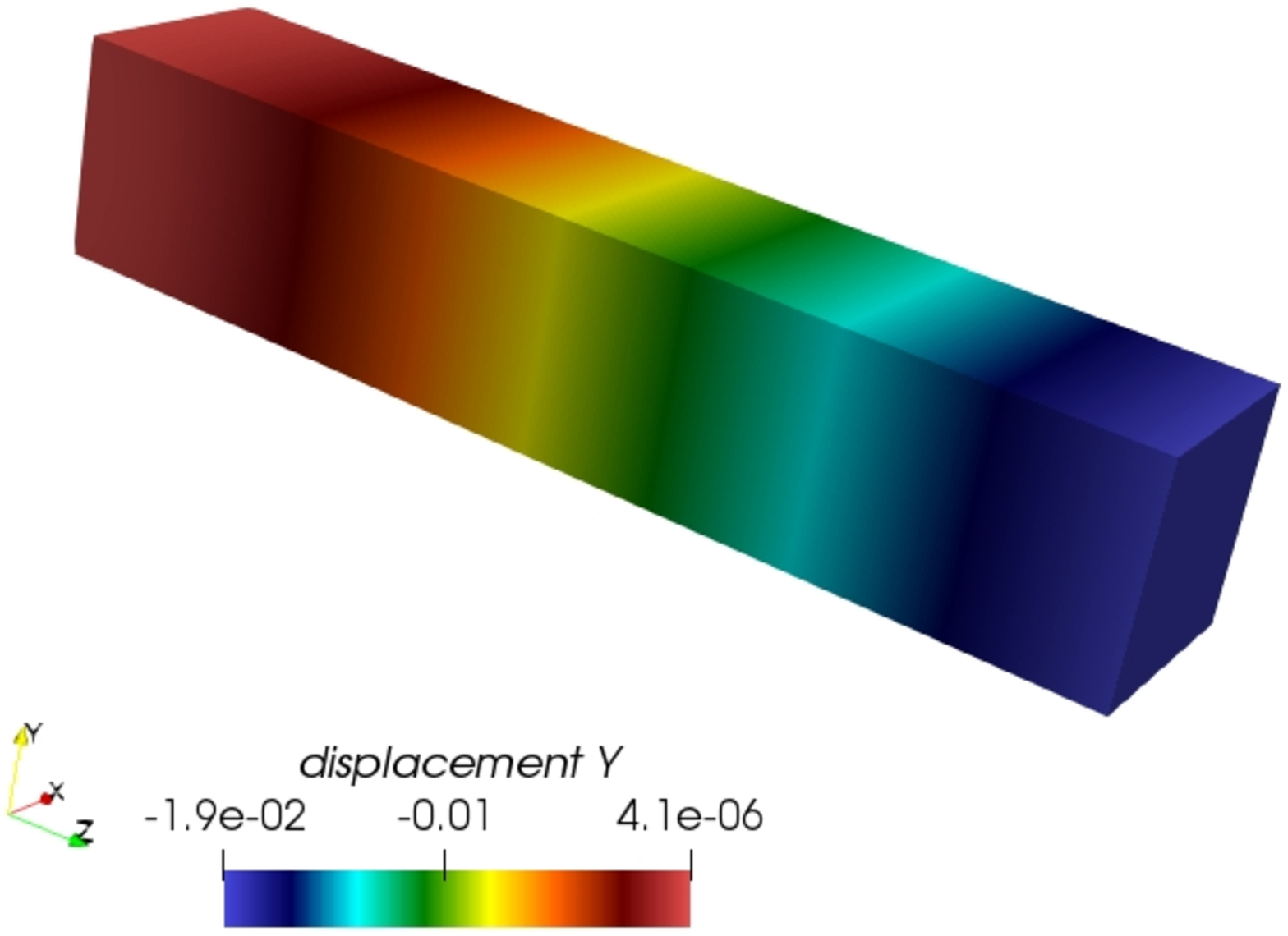}}
 \subfloat[$u_2$: $y$\label{subfig:pe34}]{%
 \includegraphics[width=0.35\textwidth,height=0.2\textheight]{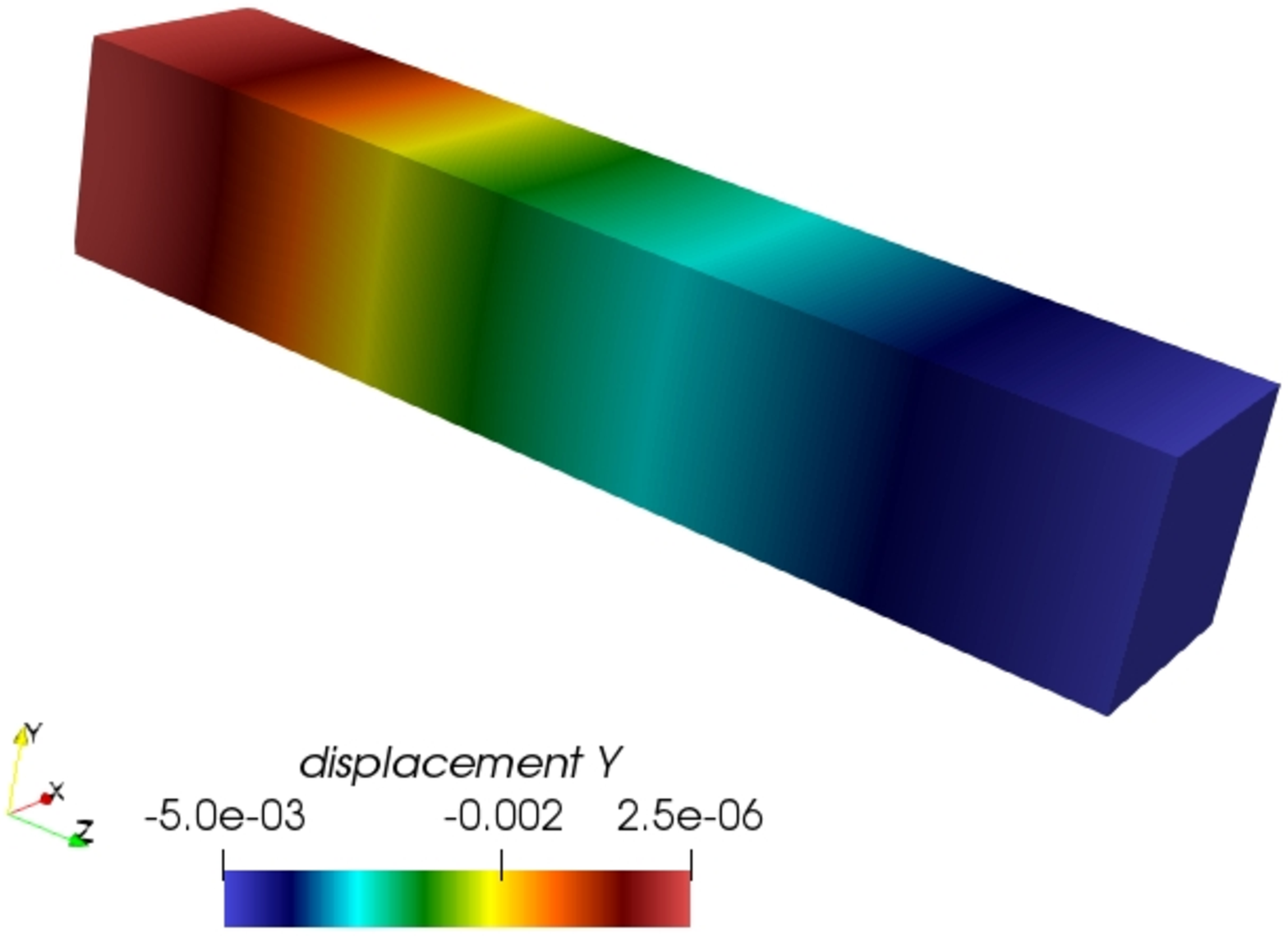}}

 \subfloat[$u_1$: $z$\label{subfig:pe35}]{%
 \includegraphics[width=0.35\textwidth,height=0.2\textheight]{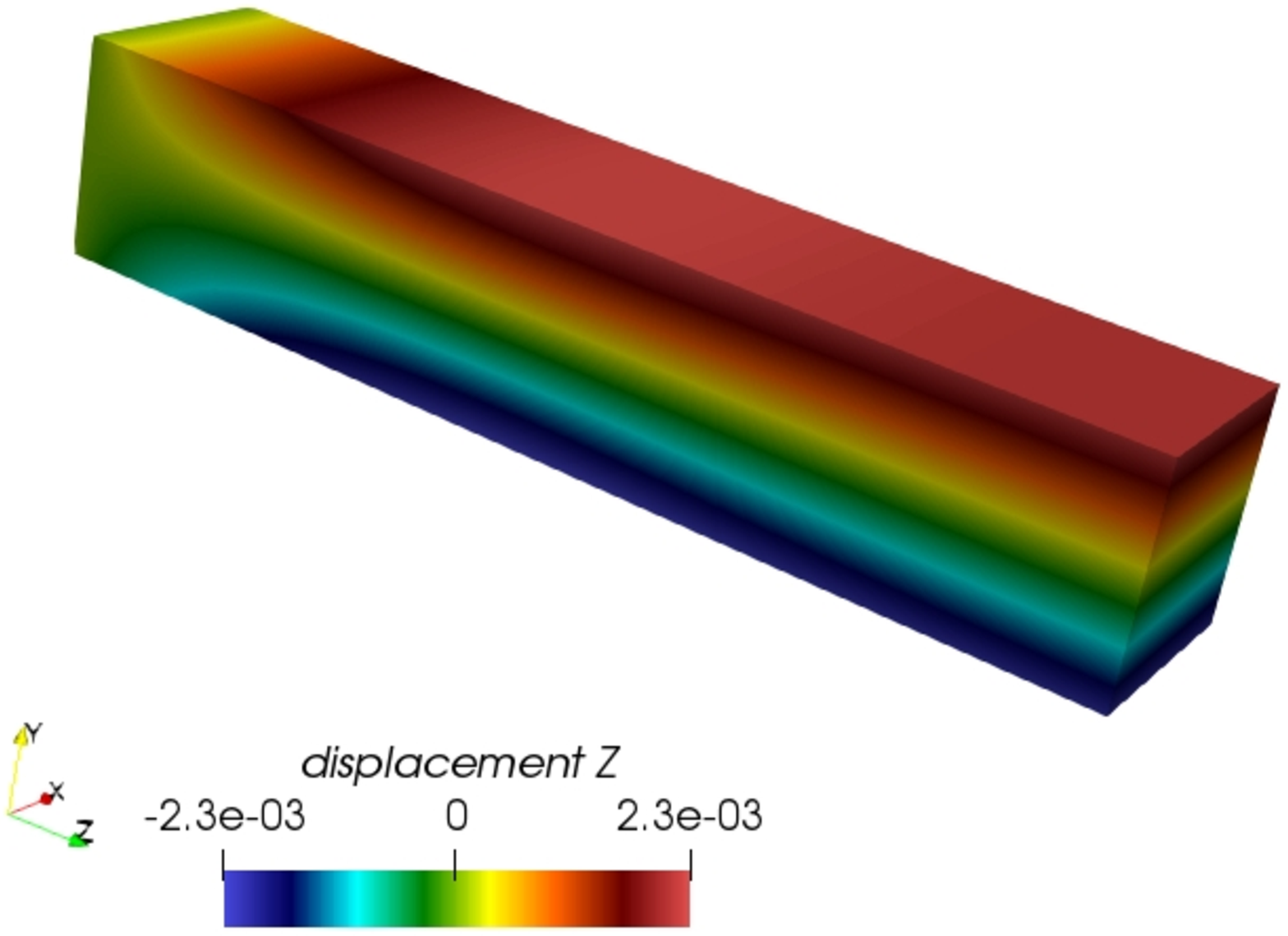}}
 \subfloat[$u_2$: $z$\label{subfig:pe37}]{%
 \includegraphics[width=0.35\textwidth,height=0.2\textheight]{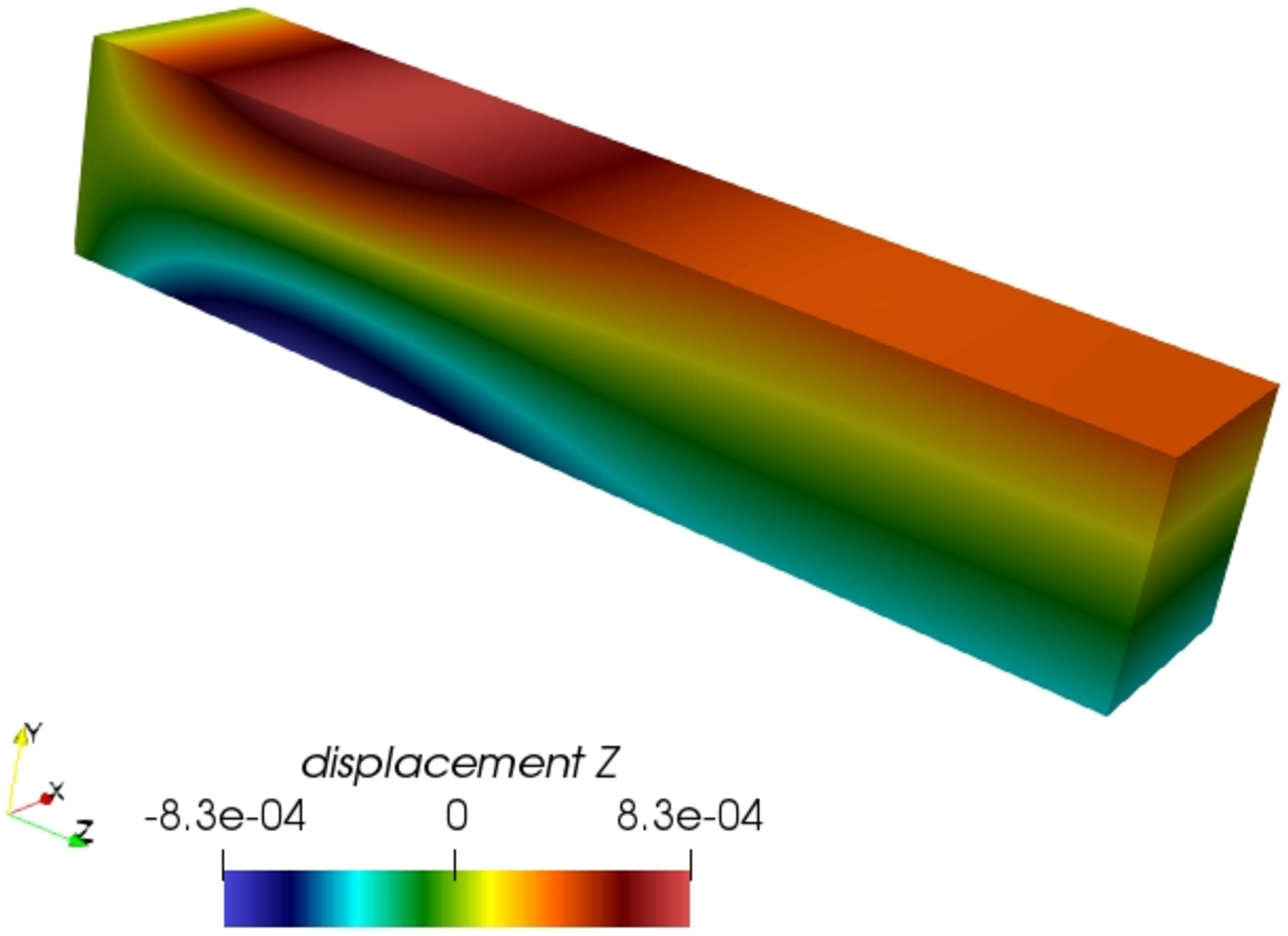}}
 \caption{Magnitude and $x,y$ and $z$ components of the selected PCE coefficients of the solution process $\mathbfcal{U}$.}
 \label{fig:pce3DE1}
\end{figure}

\begin{figure}[htbp]
 \centering
 \subfloat[$u_4$: magnitude\label{subfig:pe38}]{%
 \includegraphics[width=0.35\textwidth,height=0.2\textheight]{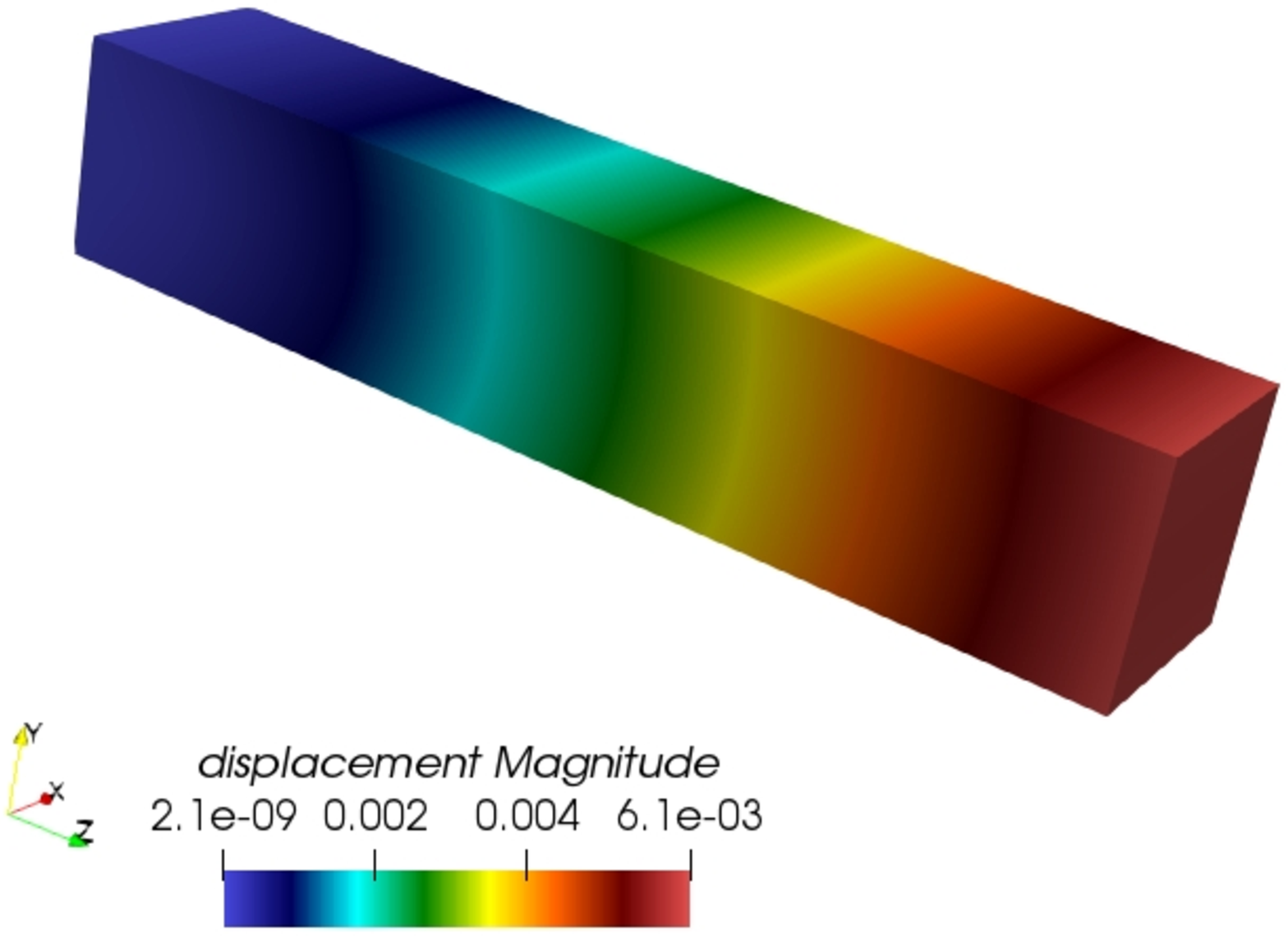}}
 \subfloat[$u_6$: magnitude\label{subfig:pe39}]{%
 \includegraphics[width=0.35\textwidth,height=0.2\textheight]{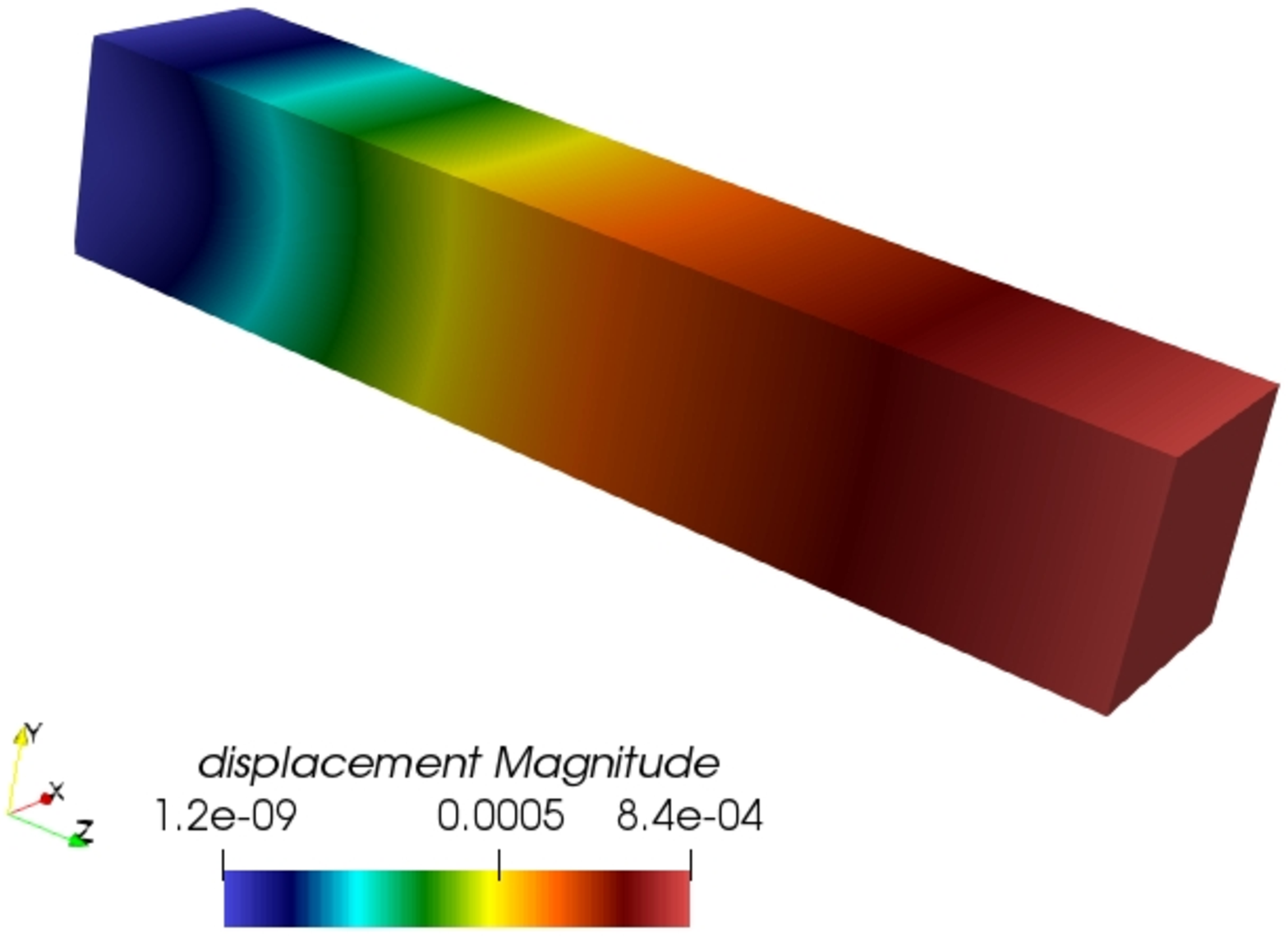}}

 \subfloat[$u_4$: $x$\label{subfig:pe40}]{%
 \includegraphics[width=0.35\textwidth,height=0.2\textheight]{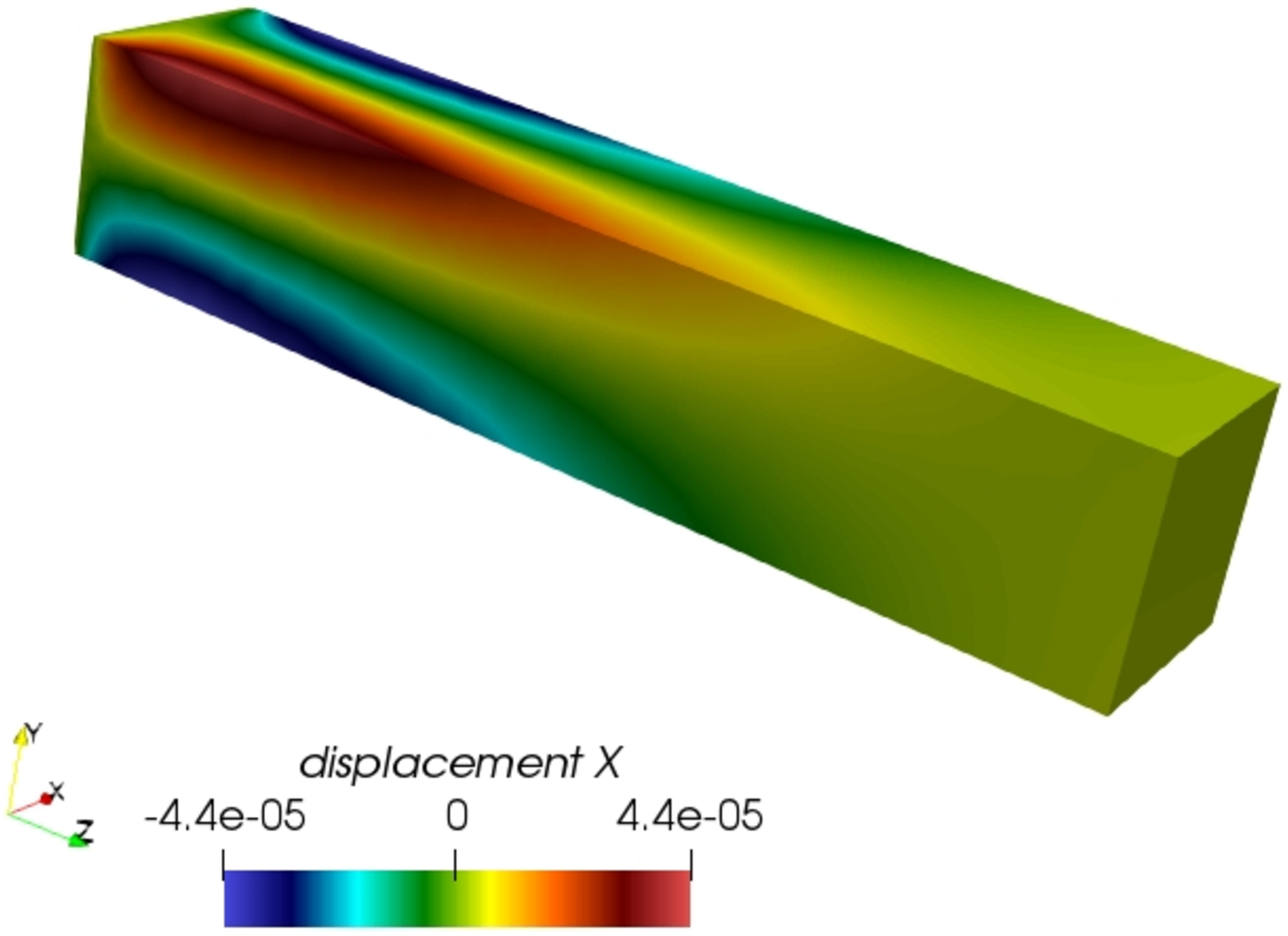}}
 \subfloat[$u_6$: $x$\label{subfig:pe41}]{%
 \includegraphics[width=0.35\textwidth,height=0.2\textheight]{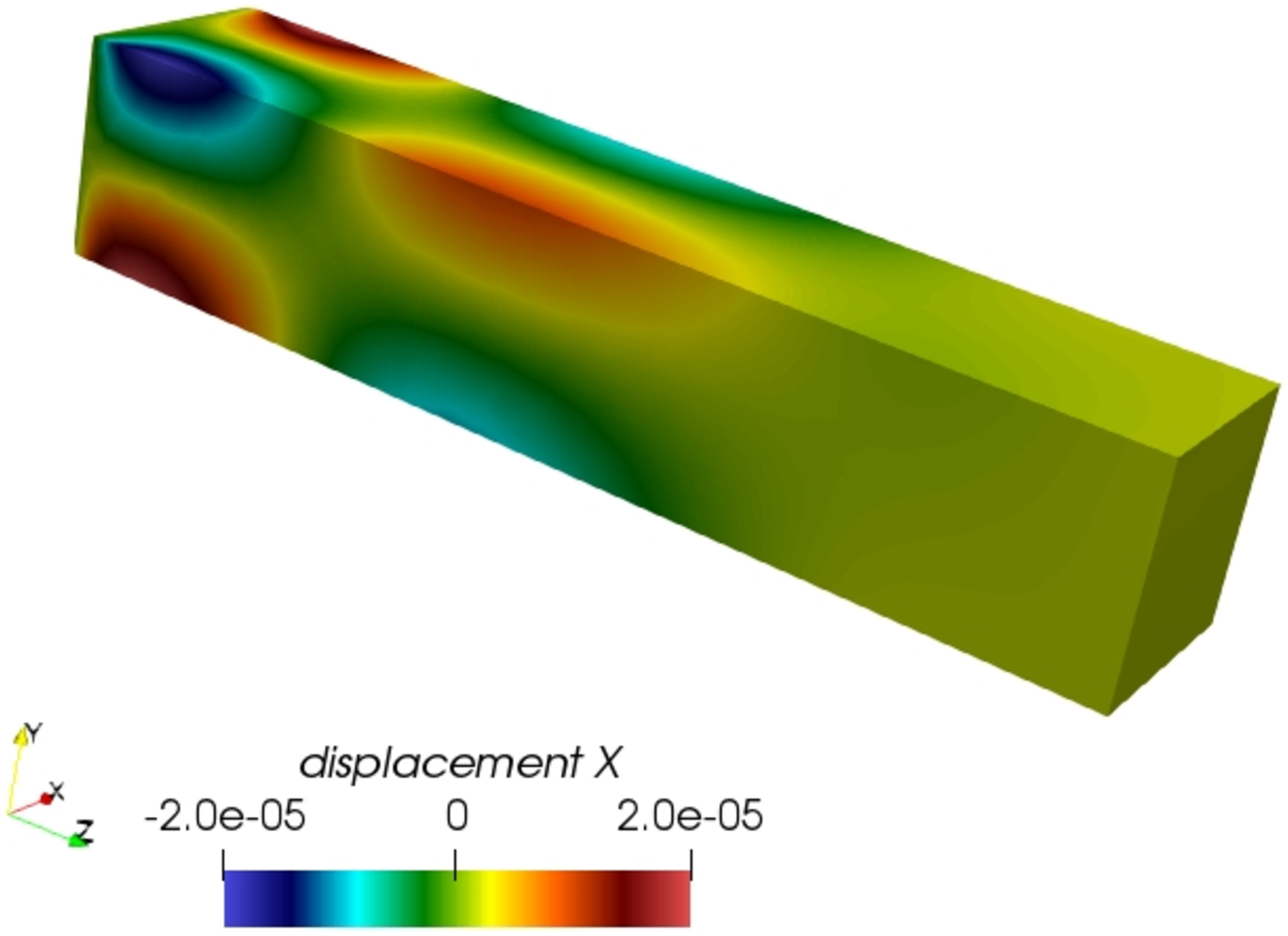}}

 \subfloat[$u_4$: $y$\label{subfig:pe42}]{%
 \includegraphics[width=0.35\textwidth,height=0.2\textheight]{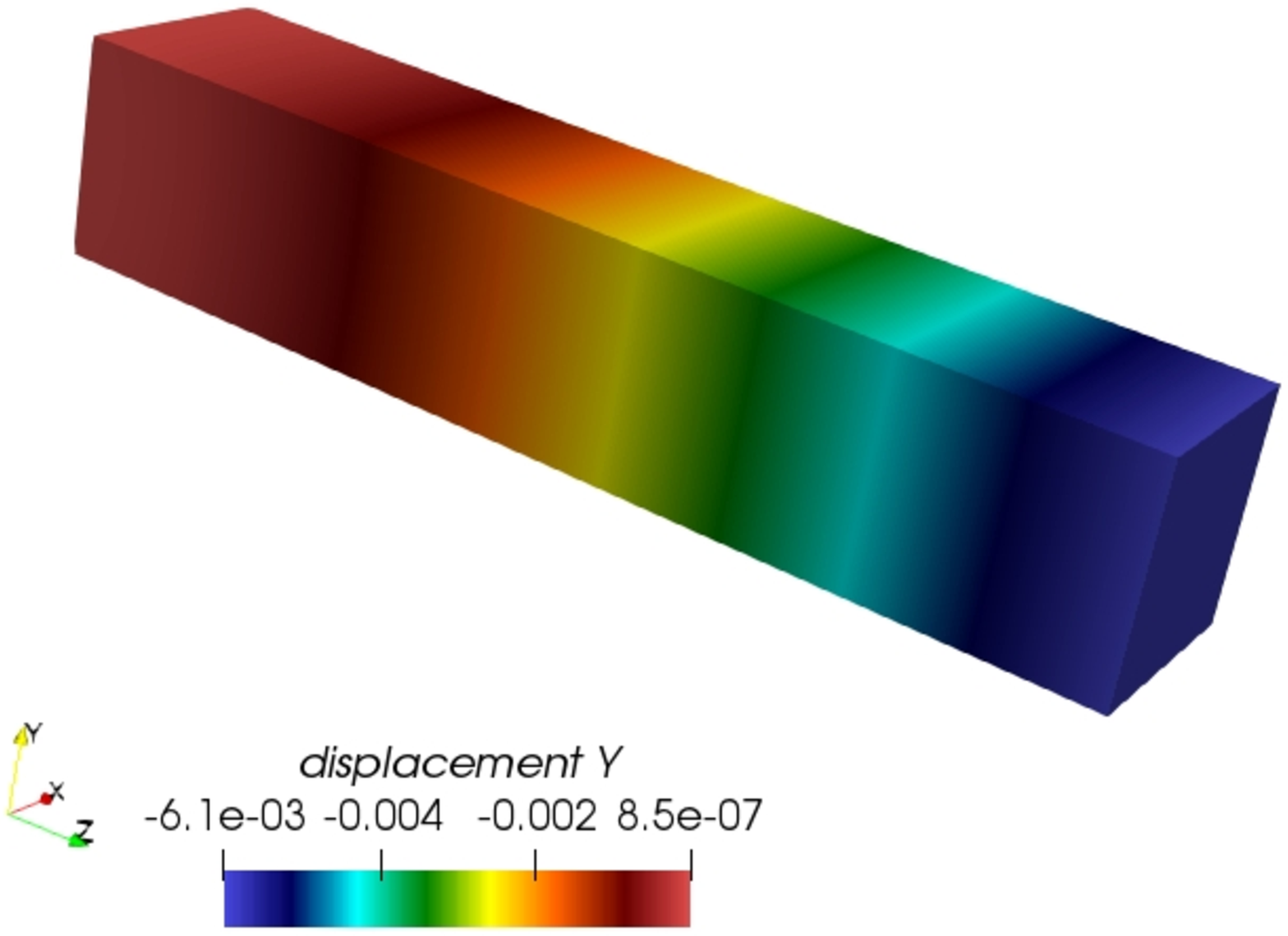}}
 \subfloat[$u_6$: $y$\label{subfig:pe43}]{%
 \includegraphics[width=0.35\textwidth,height=0.2\textheight]{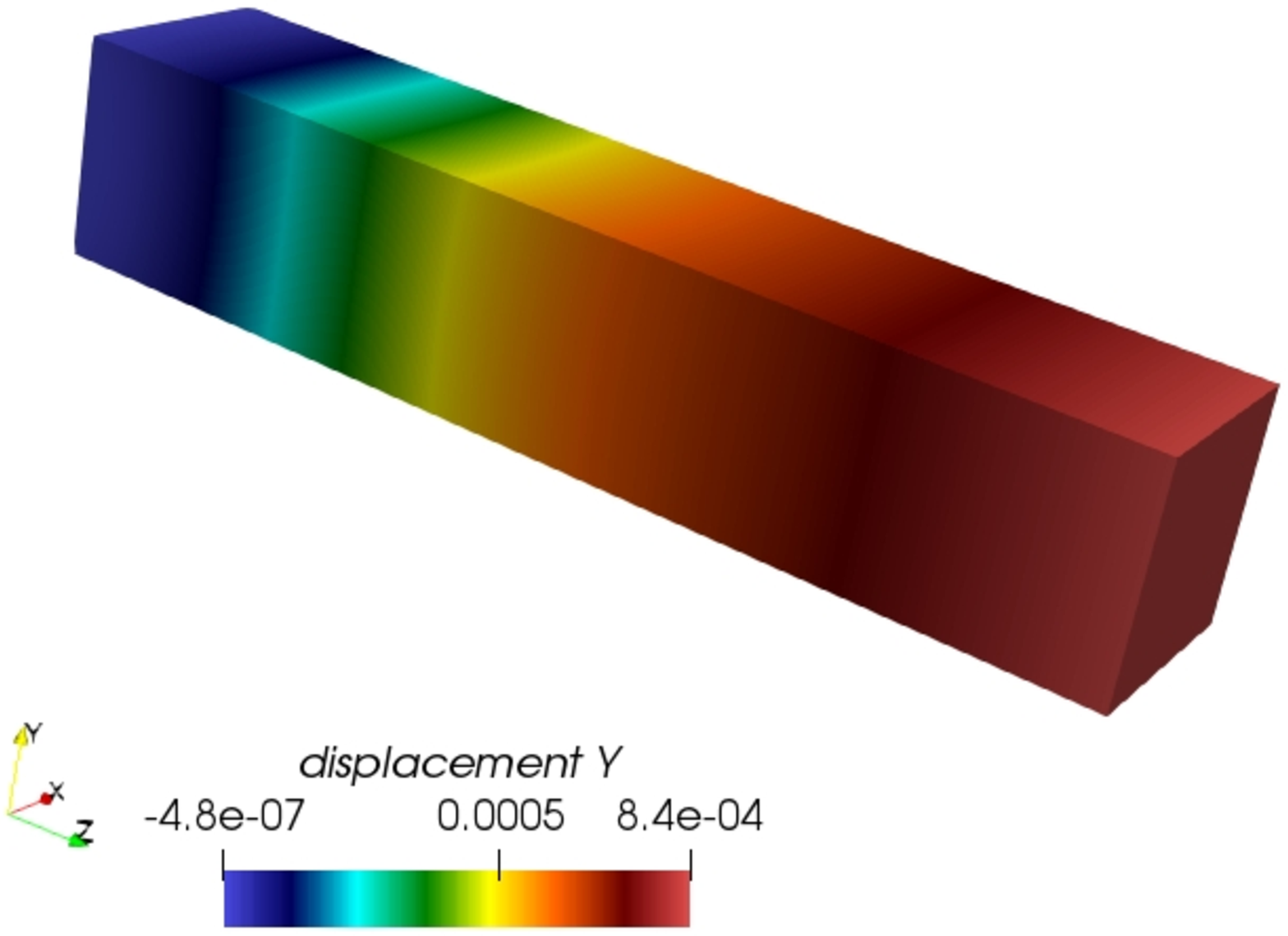}}

 \subfloat[$u_4$: $z$\label{subfig:pe44}]{%
 \includegraphics[width=0.35\textwidth,height=0.2\textheight]{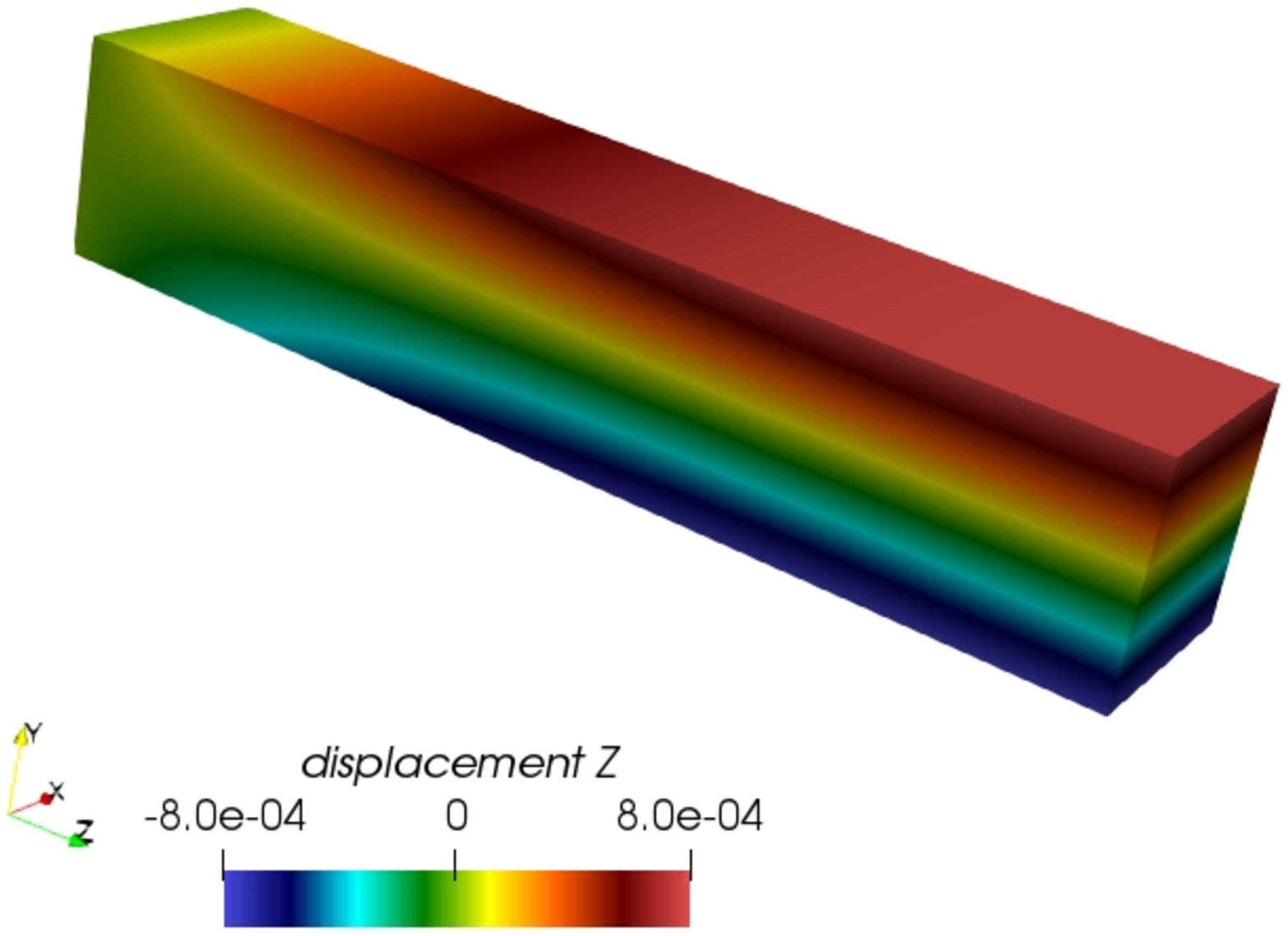}}
 \subfloat[$u_6$: $z$\label{subfig:pe45}]{%
 \includegraphics[width=0.35\textwidth,height=0.2\textheight]{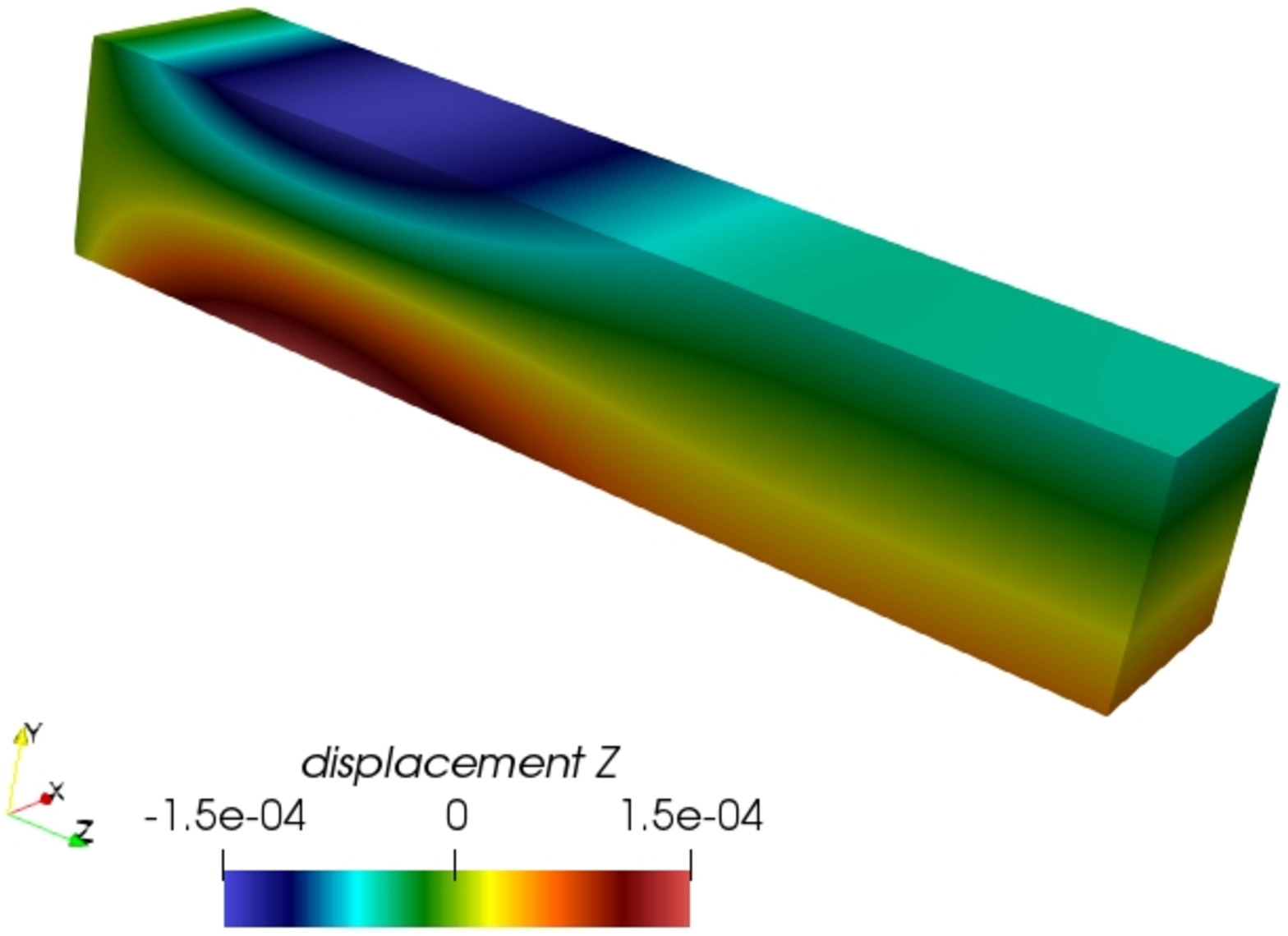}}
 \caption{Magnitude and $x,y$ and $z$ components of the selected PCE coefficients of the solution process $\mathbfcal{U}$.}
 \label{fig:pce3DE2}
\end{figure}

\begin{figure}[htbp]
 \centering
 \subfloat[$u_{10}$: magnitude\label{subfig:pe46}]{%
 \includegraphics[width=0.35\textwidth,height=0.2\textheight]{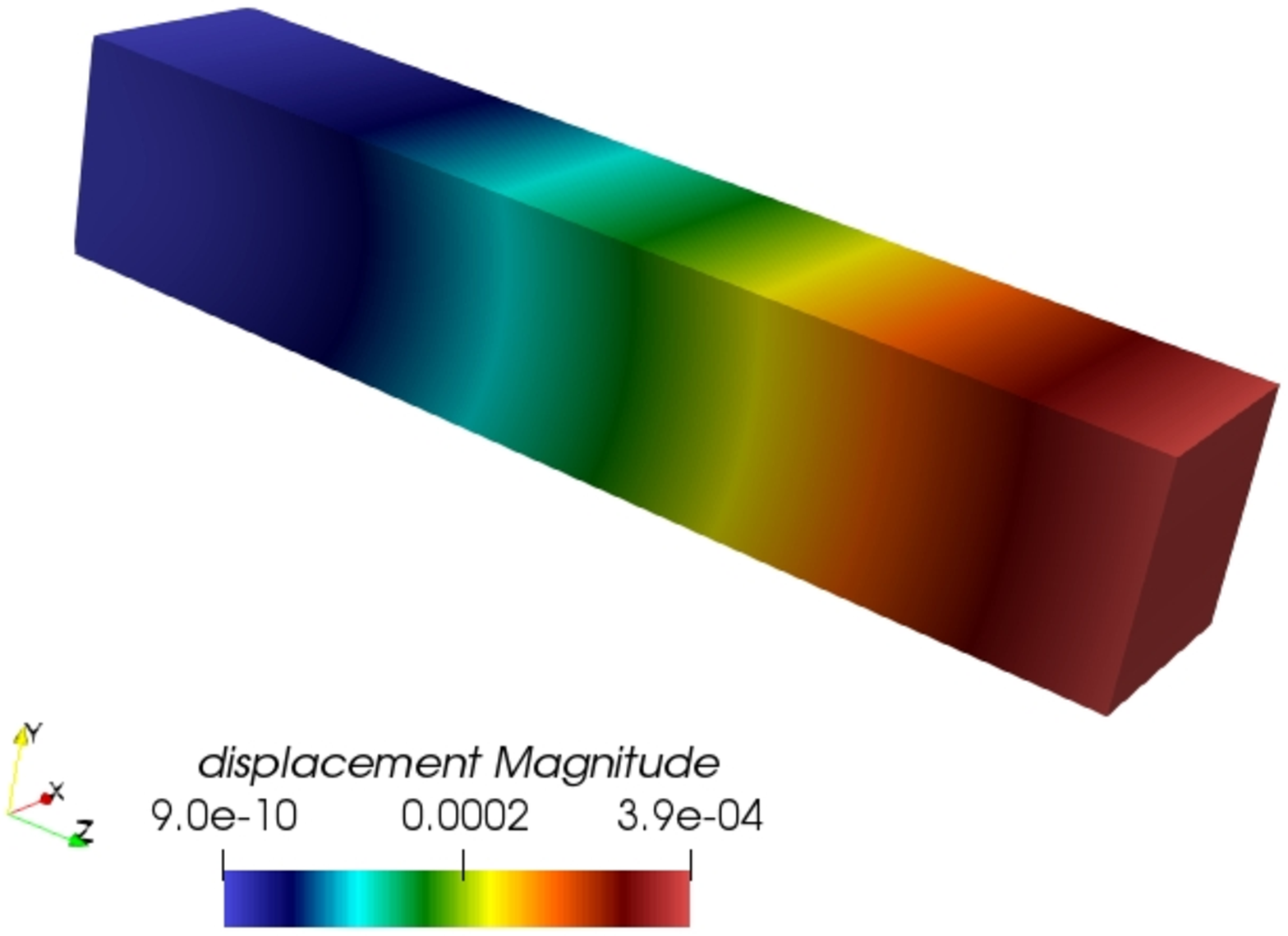}}
 \subfloat[$u_{12}$: magnitude\label{subfig:pe47}]{%
 \includegraphics[width=0.35\textwidth,height=0.2\textheight]{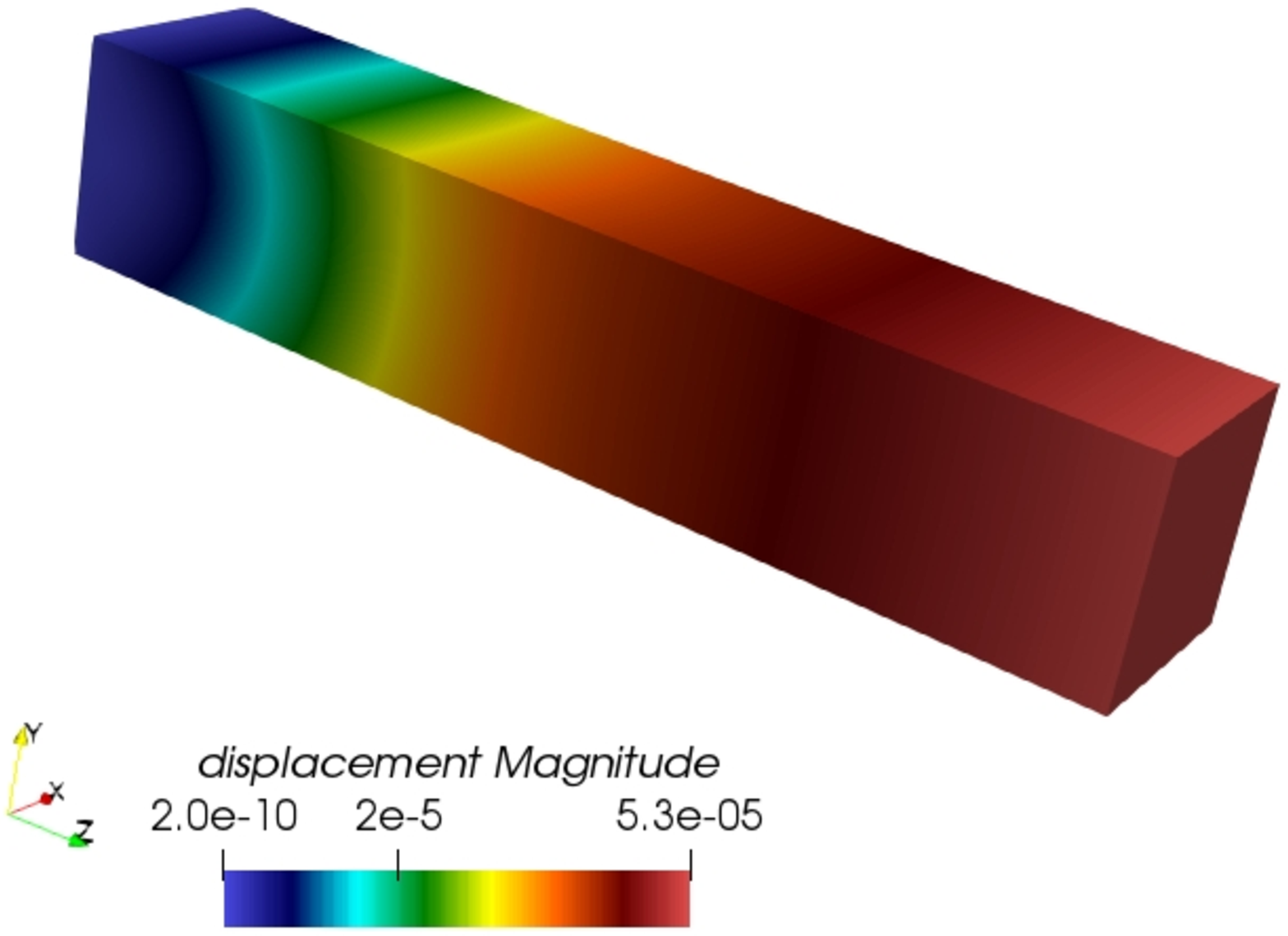}}

 \subfloat[$u_{10}$: $x$\label{subfig:pe48}]{%
 \includegraphics[width=0.35\textwidth,height=0.2\textheight]{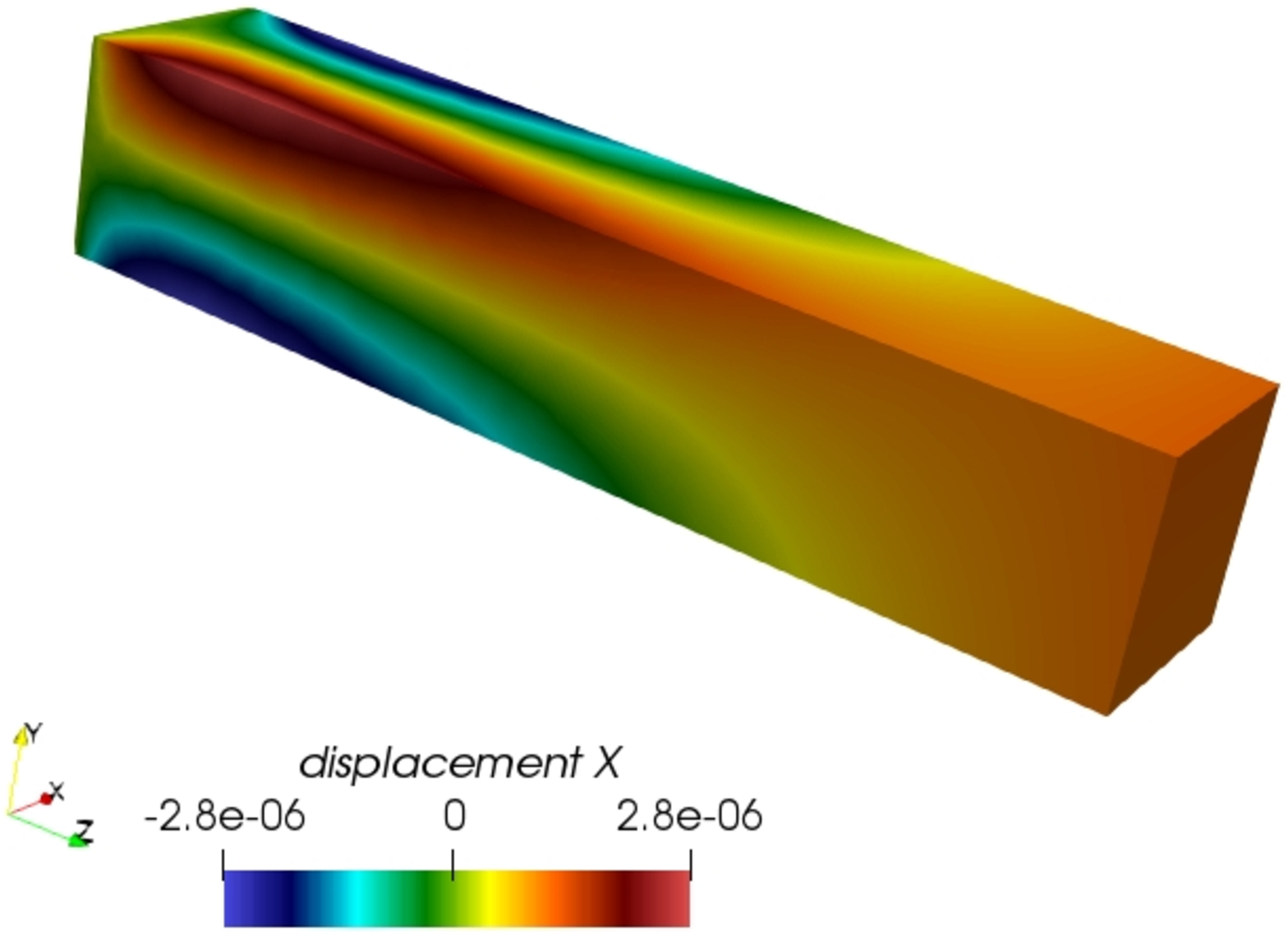}}
 \subfloat[$u_{12}$: $x$\label{subfig:pe49}]{%
 \includegraphics[width=0.35\textwidth,height=0.2\textheight]{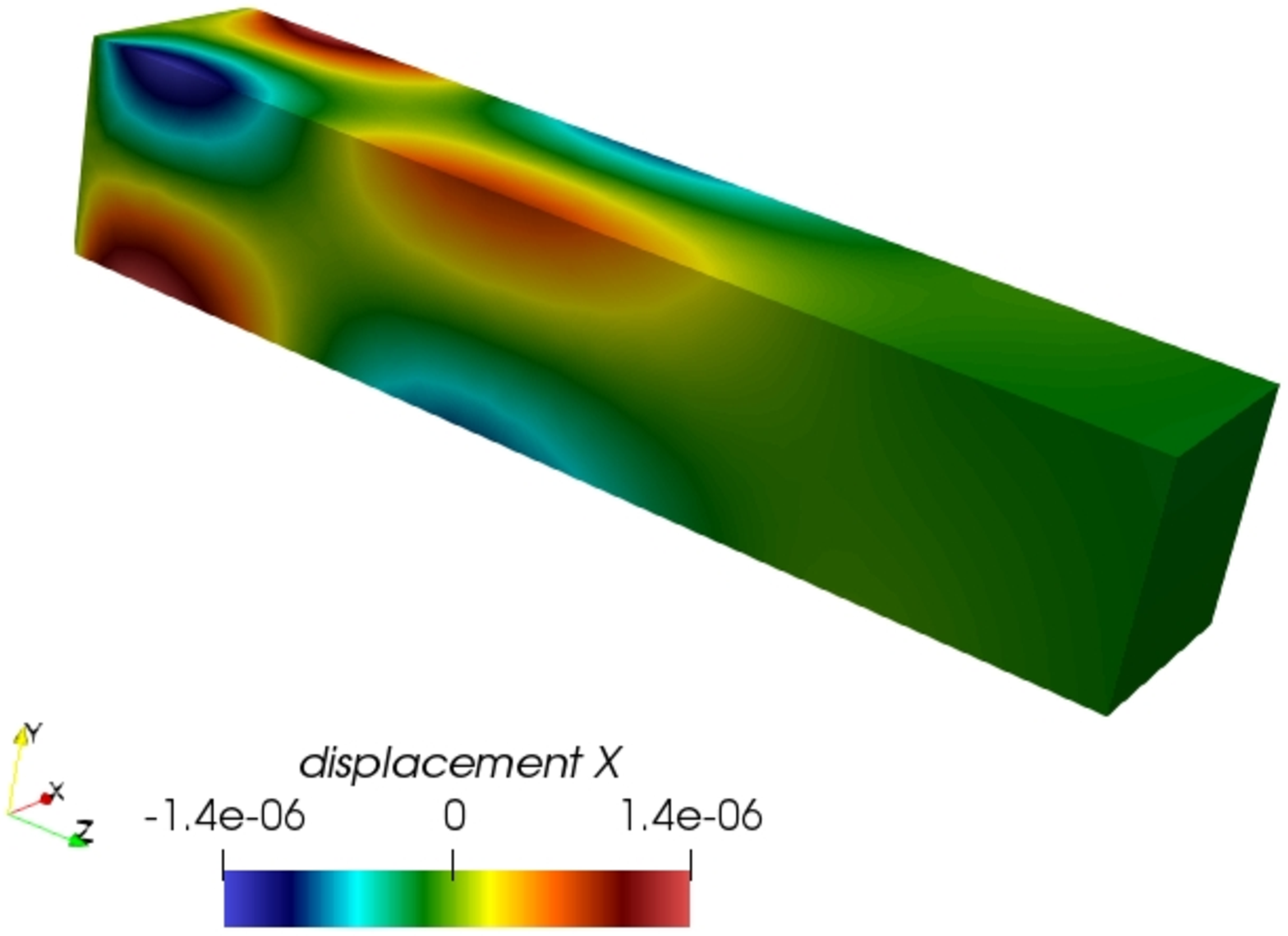}}

 \subfloat[$u_{10}$: $y$\label{subfig:pe50}]{%
 \includegraphics[width=0.35\textwidth,height=0.2\textheight]{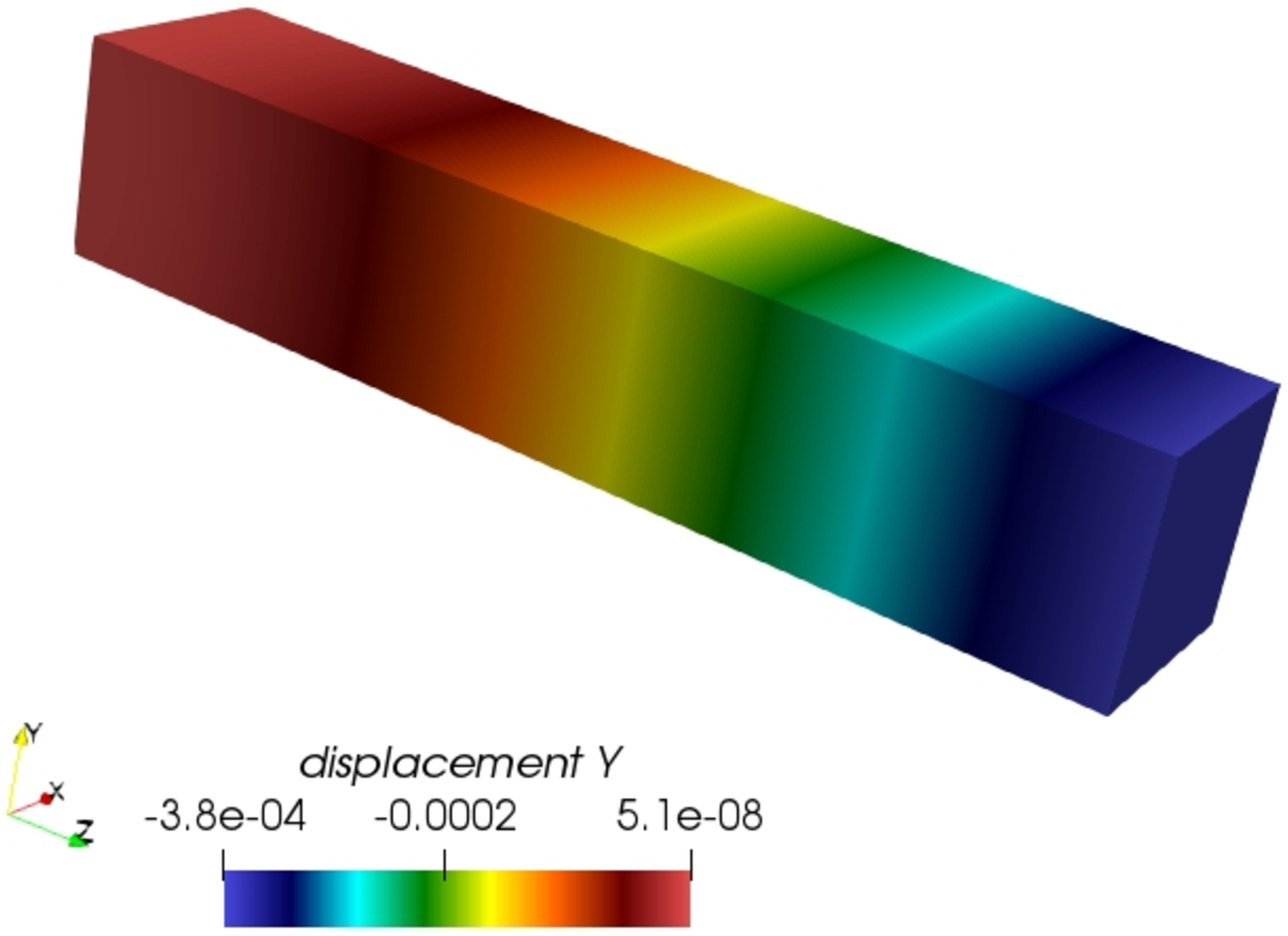}}
 \subfloat[$u_{12}$: $y$\label{subfig:pe51}]{%
 \includegraphics[width=0.35\textwidth,height=0.2\textheight]{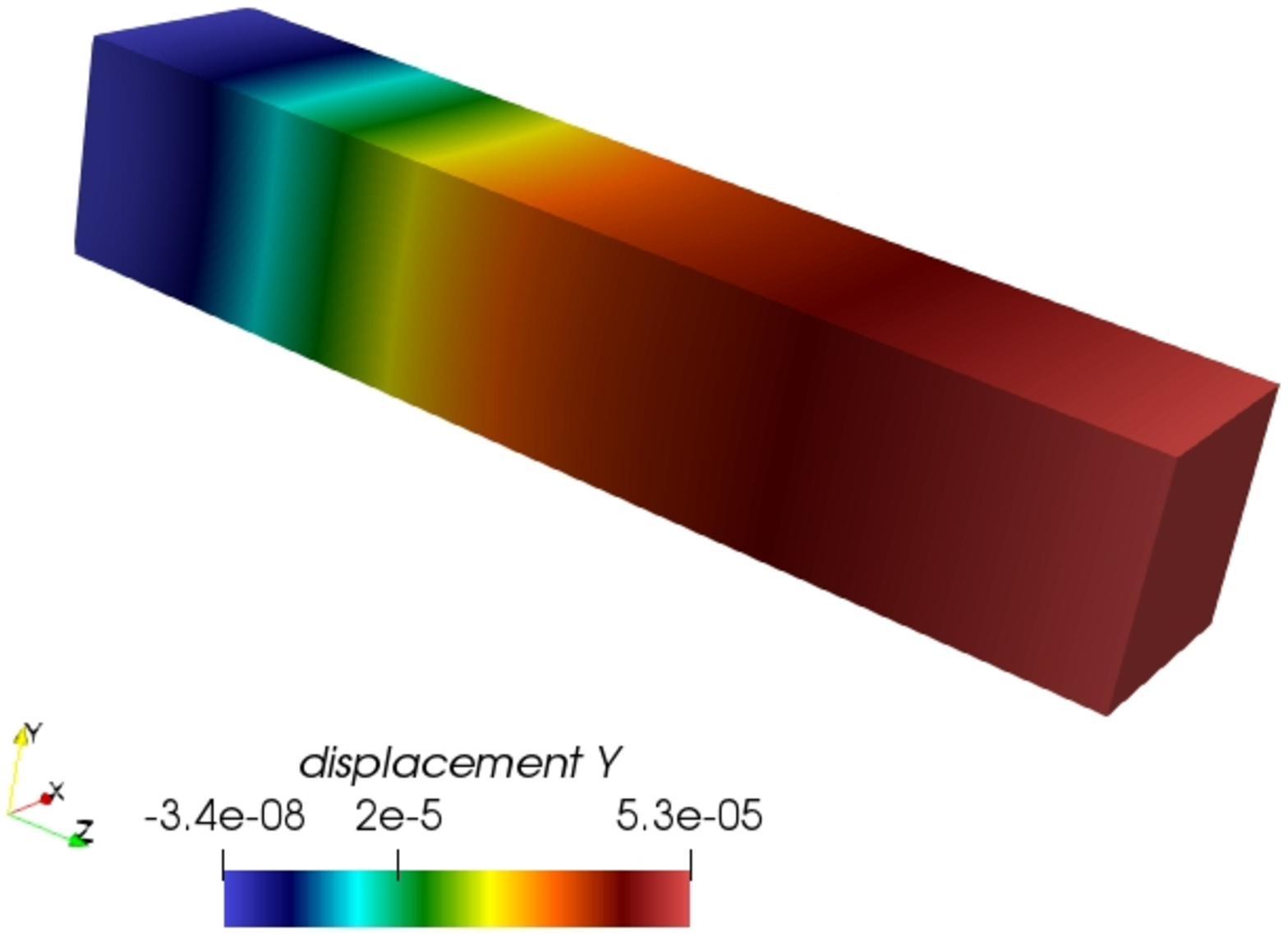}}

 \subfloat[$u_{10}$: $z$\label{subfig:pe52}]{%
 \includegraphics[width=0.35\textwidth,height=0.2\textheight]{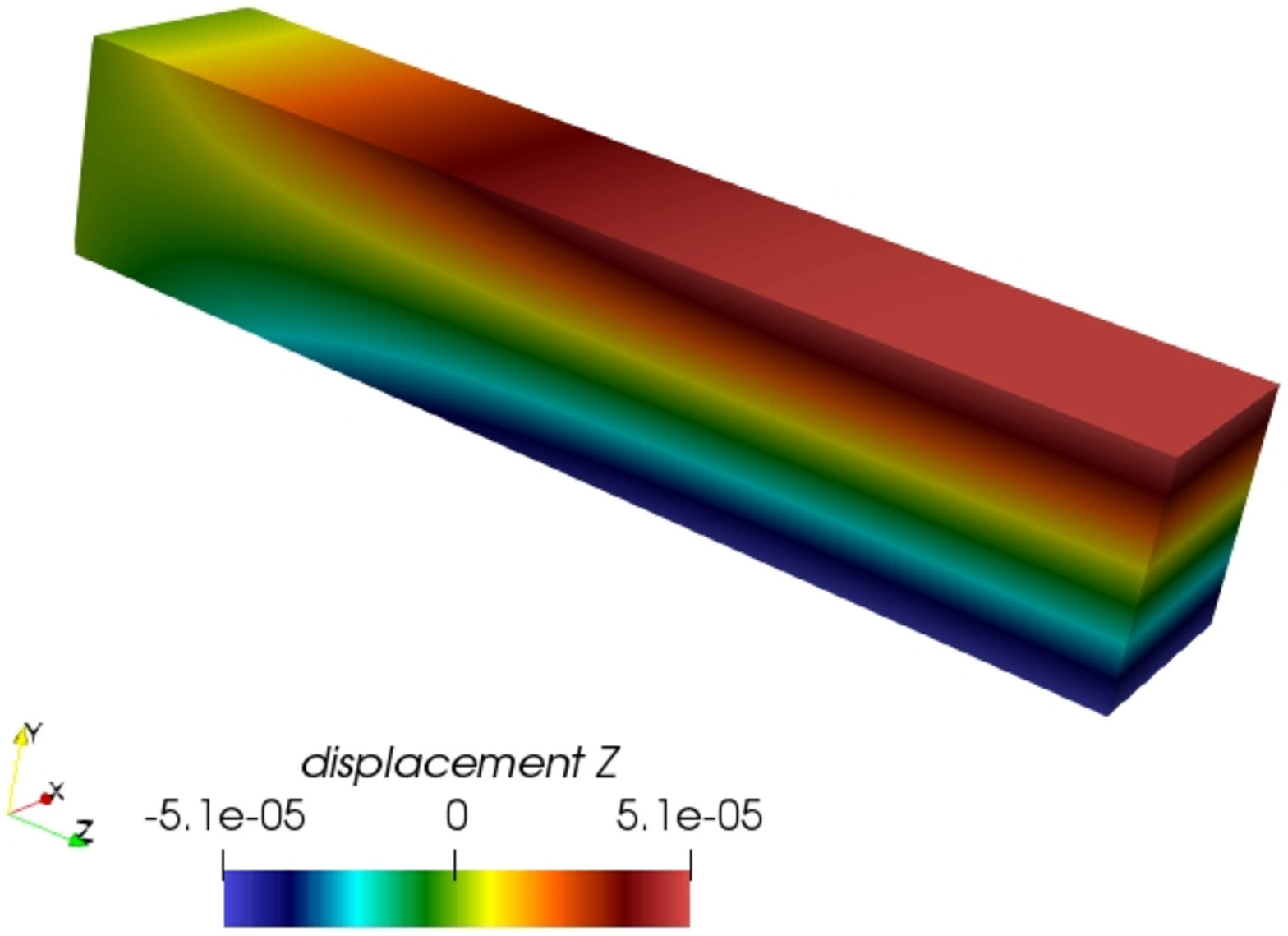}}
 \subfloat[$u_{12}$: $z$\label{subfig:pe53}]{%
 \includegraphics[width=0.35\textwidth,height=0.2\textheight]{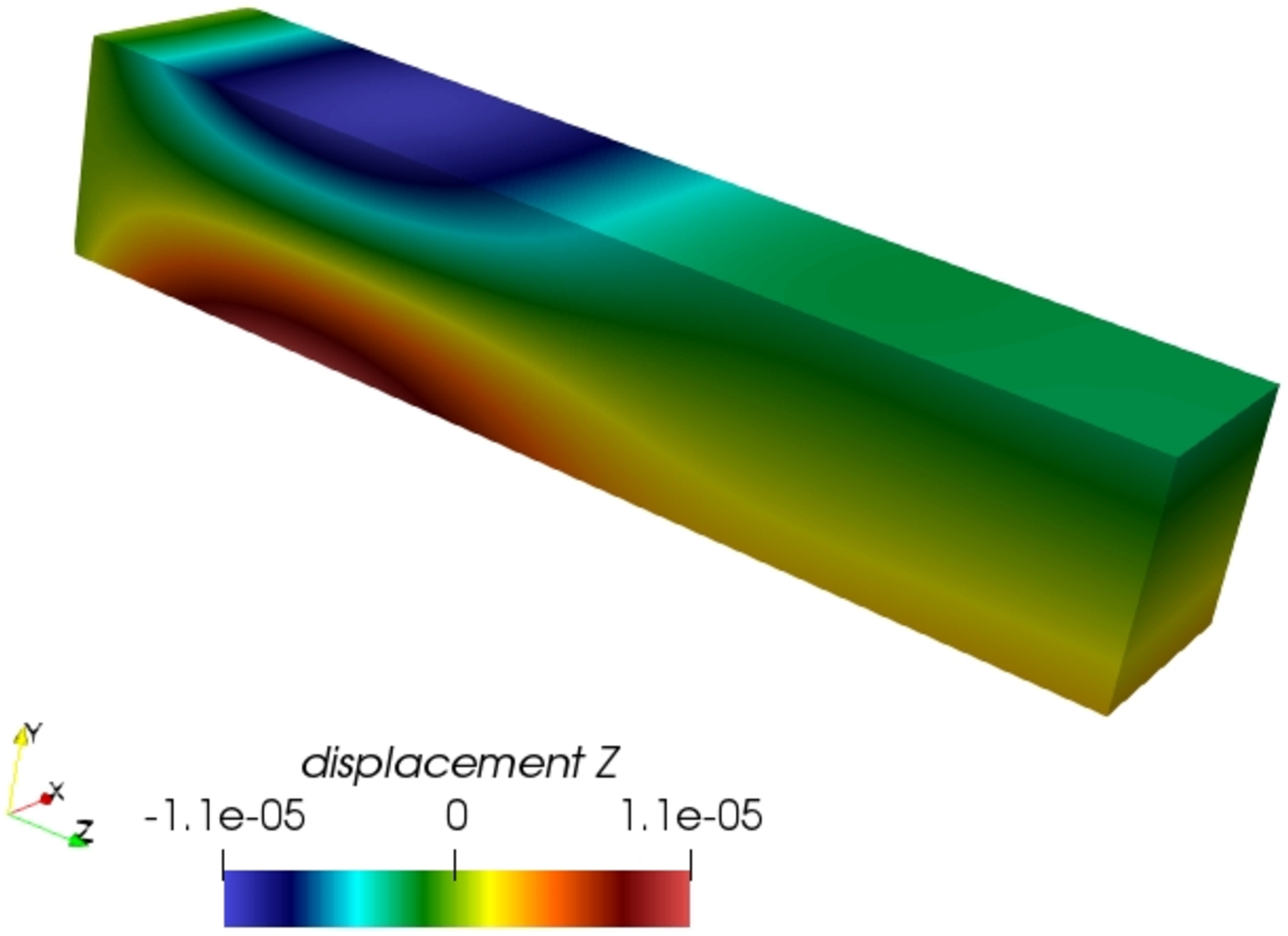}}
 \caption{Magnitude and $x,y$ and $z$ components of the selected PCE coefficients of the solution process $\mathbfcal{U}$.}
 \label{fig:pce3DE3}
\end{figure}

\begin{figure}[htbp]
 \centering
 \subfloat[$u_{15}$: magnitude\label{subfig:pe54}]{%
 \includegraphics[width=0.35\textwidth,height=0.2\textheight]{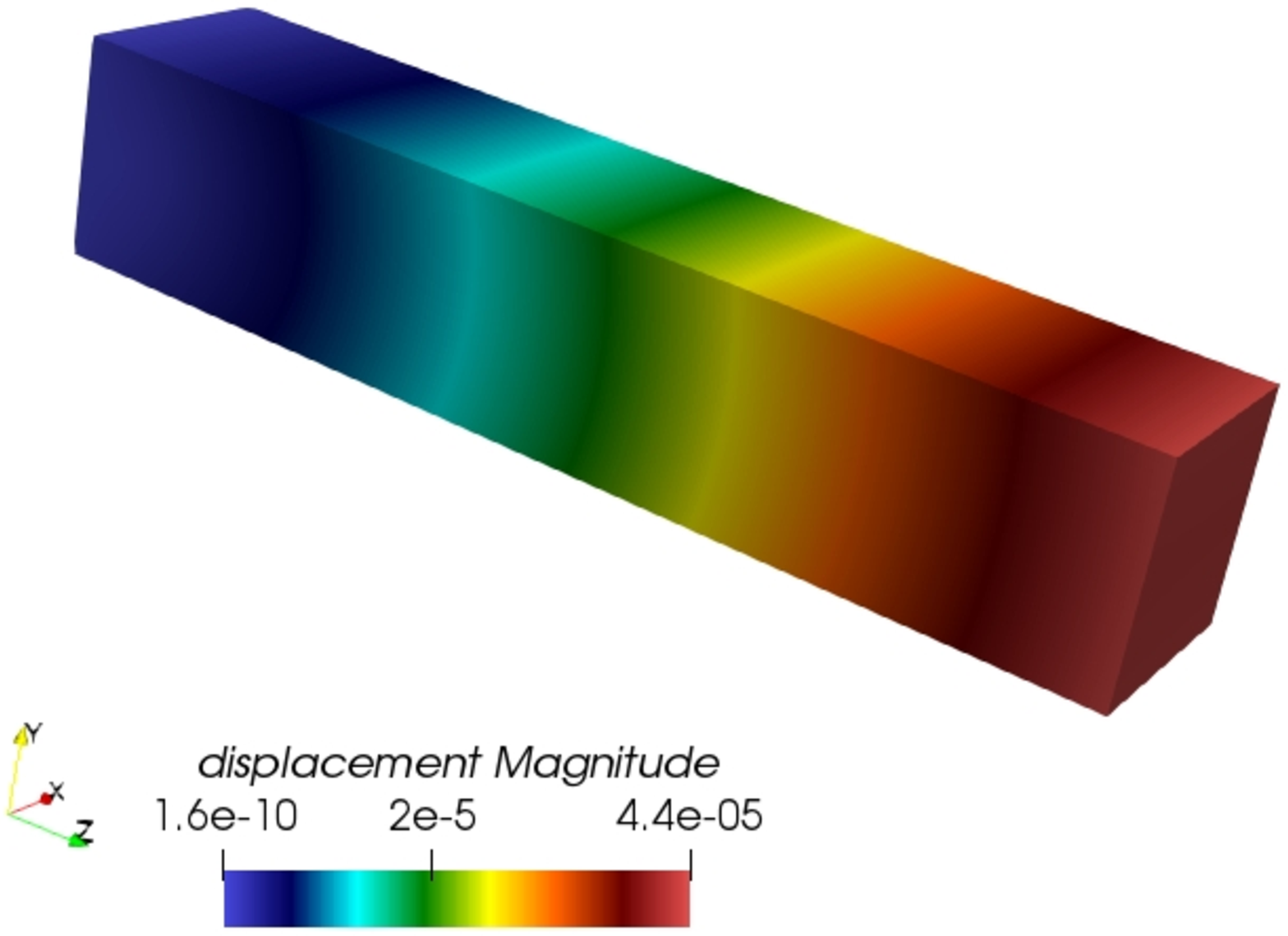}}
 \subfloat[$u_{19}$: magnitude\label{subfig:pe55}]{%
 \includegraphics[width=0.35\textwidth,height=0.2\textheight]{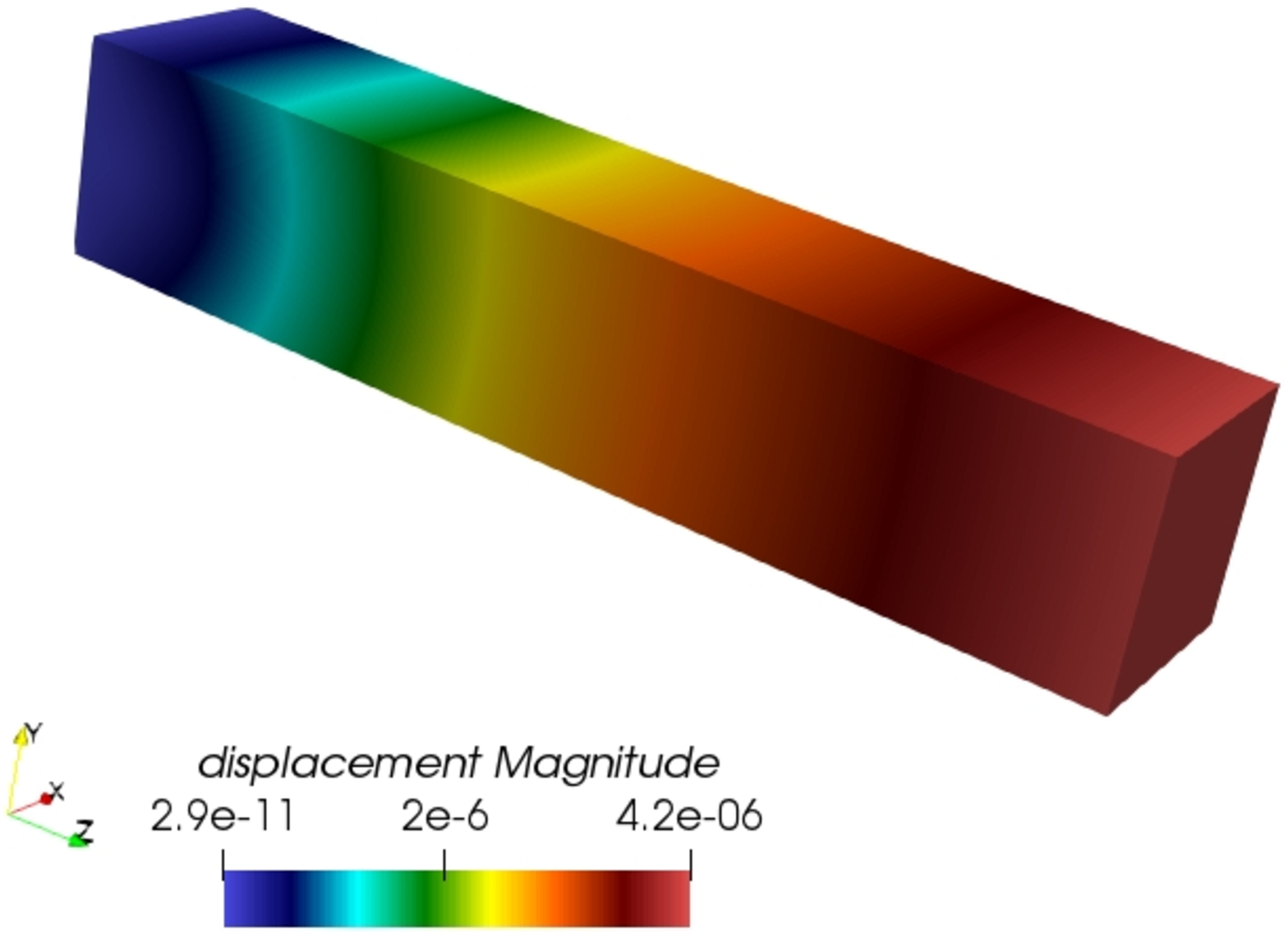}}

 \subfloat[$u_{15}$: $x$\label{subfig:pe56}]{%
 \includegraphics[width=0.35\textwidth,height=0.2\textheight]{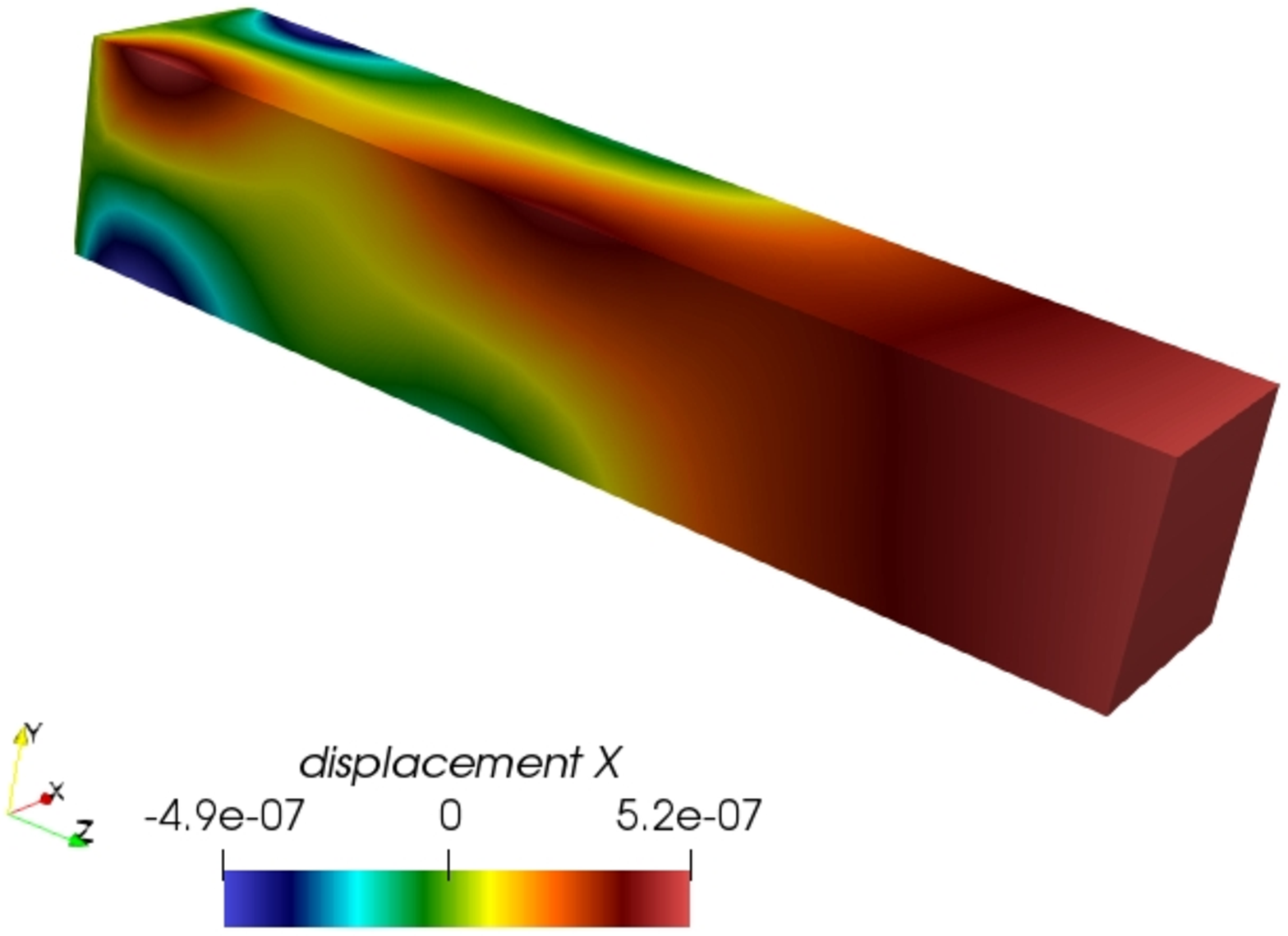}}
 \subfloat[$u_{19}$: $x$\label{subfig:pe57}]{%
 \includegraphics[width=0.35\textwidth,height=0.2\textheight]{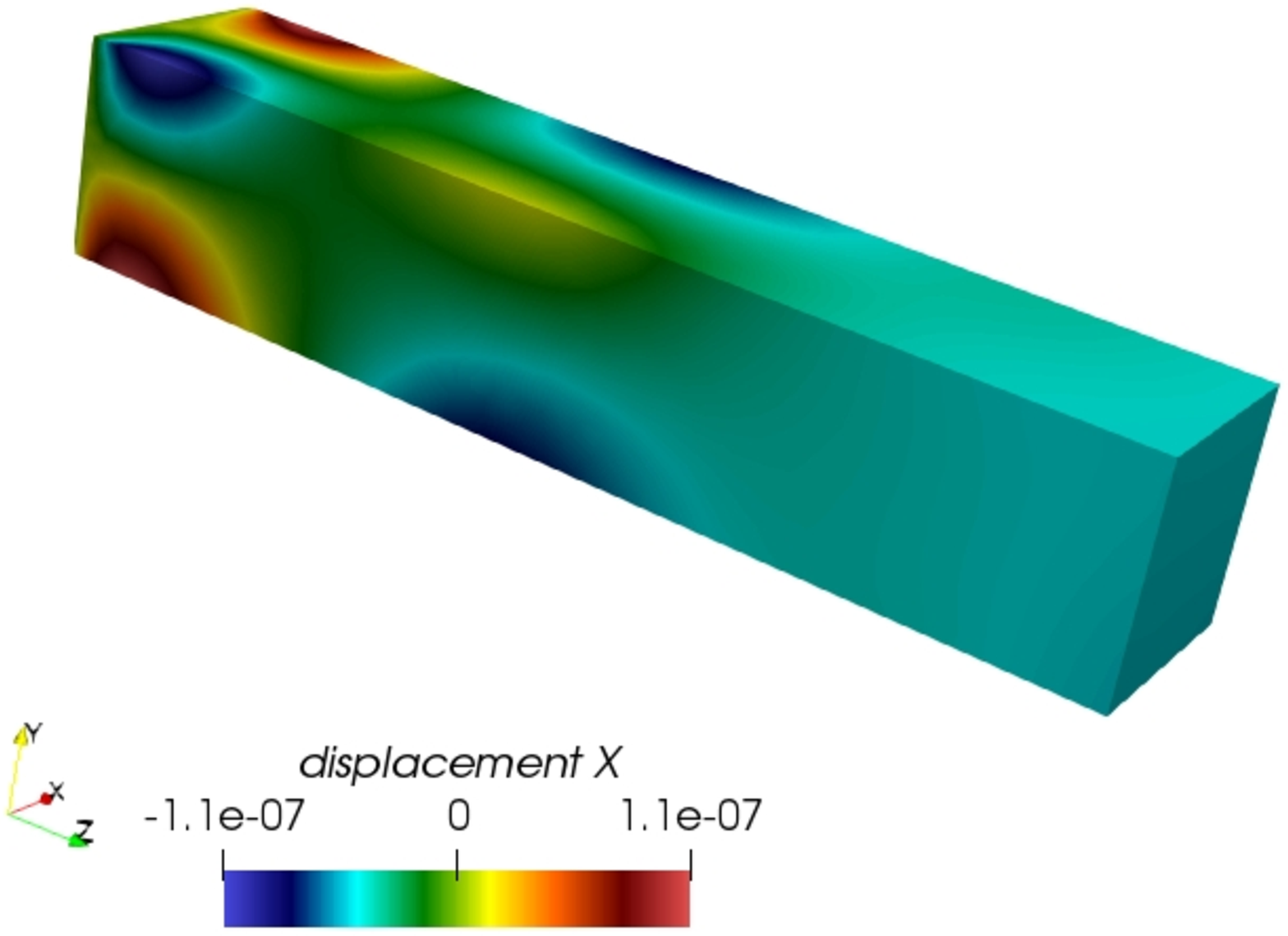}}

 \subfloat[$u_{15}$: $y$\label{subfig:pe58}]{%
 \includegraphics[width=0.35\textwidth,height=0.2\textheight]{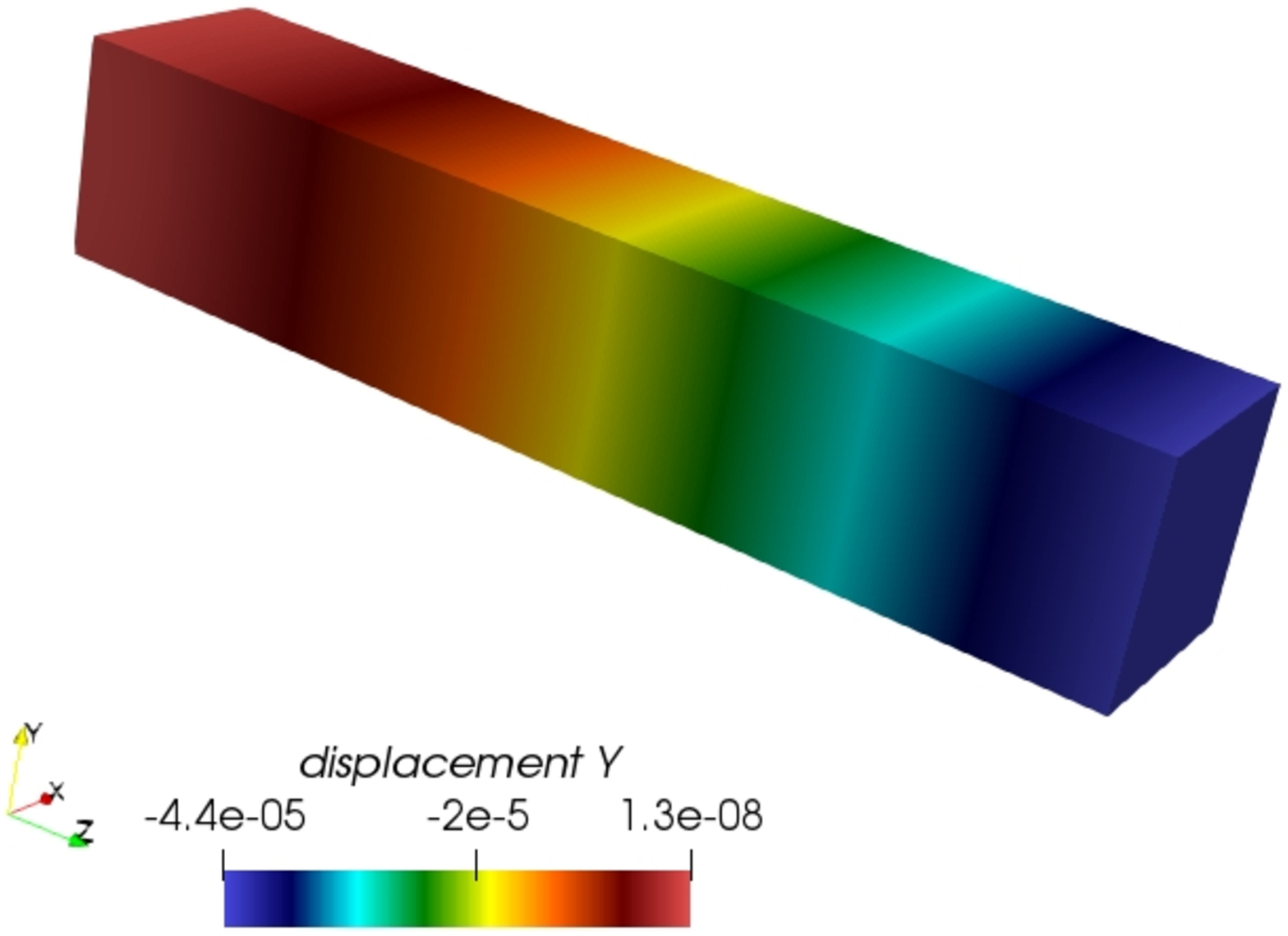}}
 \subfloat[$u_{19}$: $y$\label{subfig:pe59}]{%
 \includegraphics[width=0.35\textwidth,height=0.2\textheight]{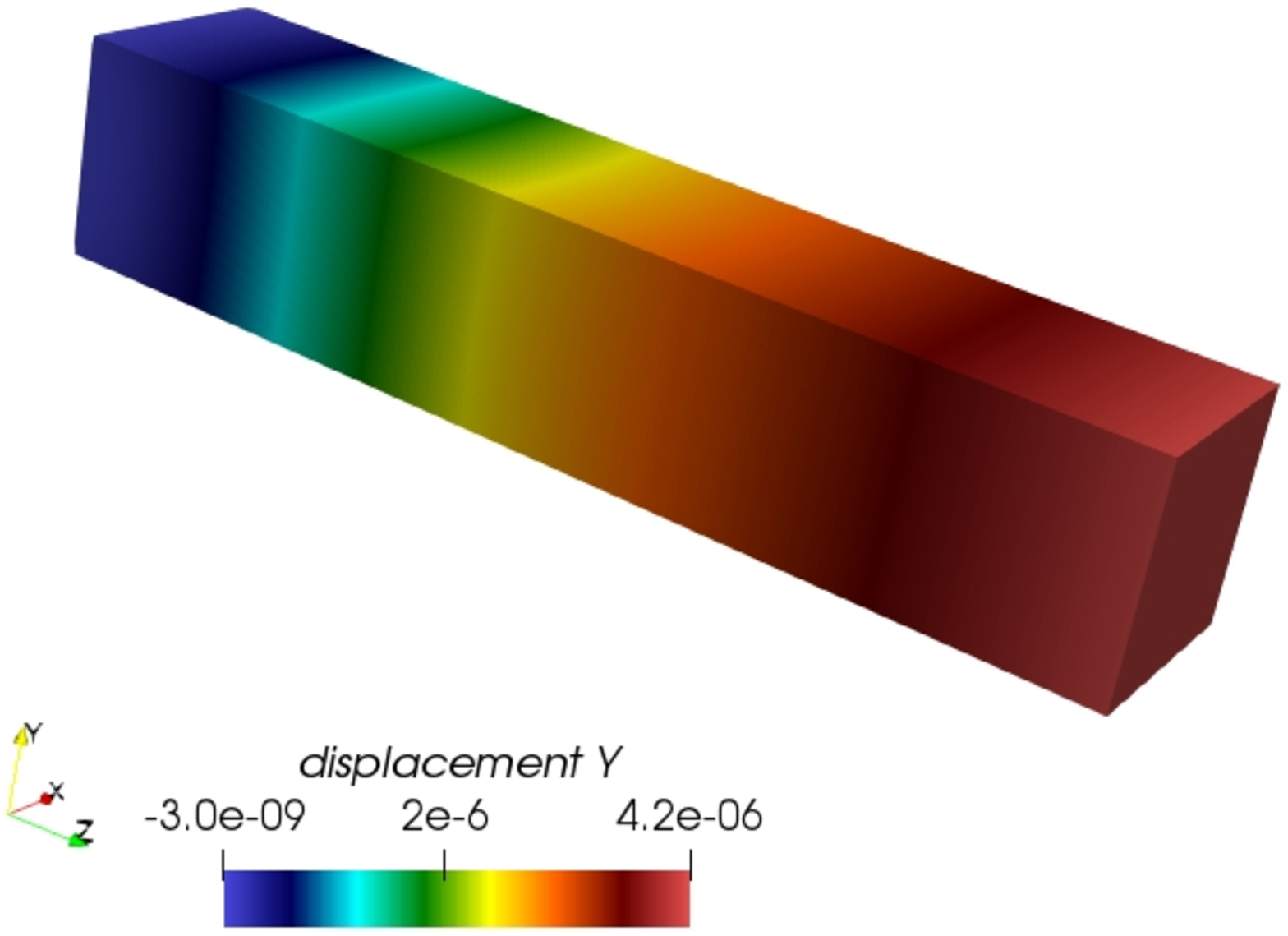}}

 \subfloat[$u_{15}$: $z$\label{subfig:pe60}]{%
 \includegraphics[width=0.35\textwidth,height=0.2\textheight]{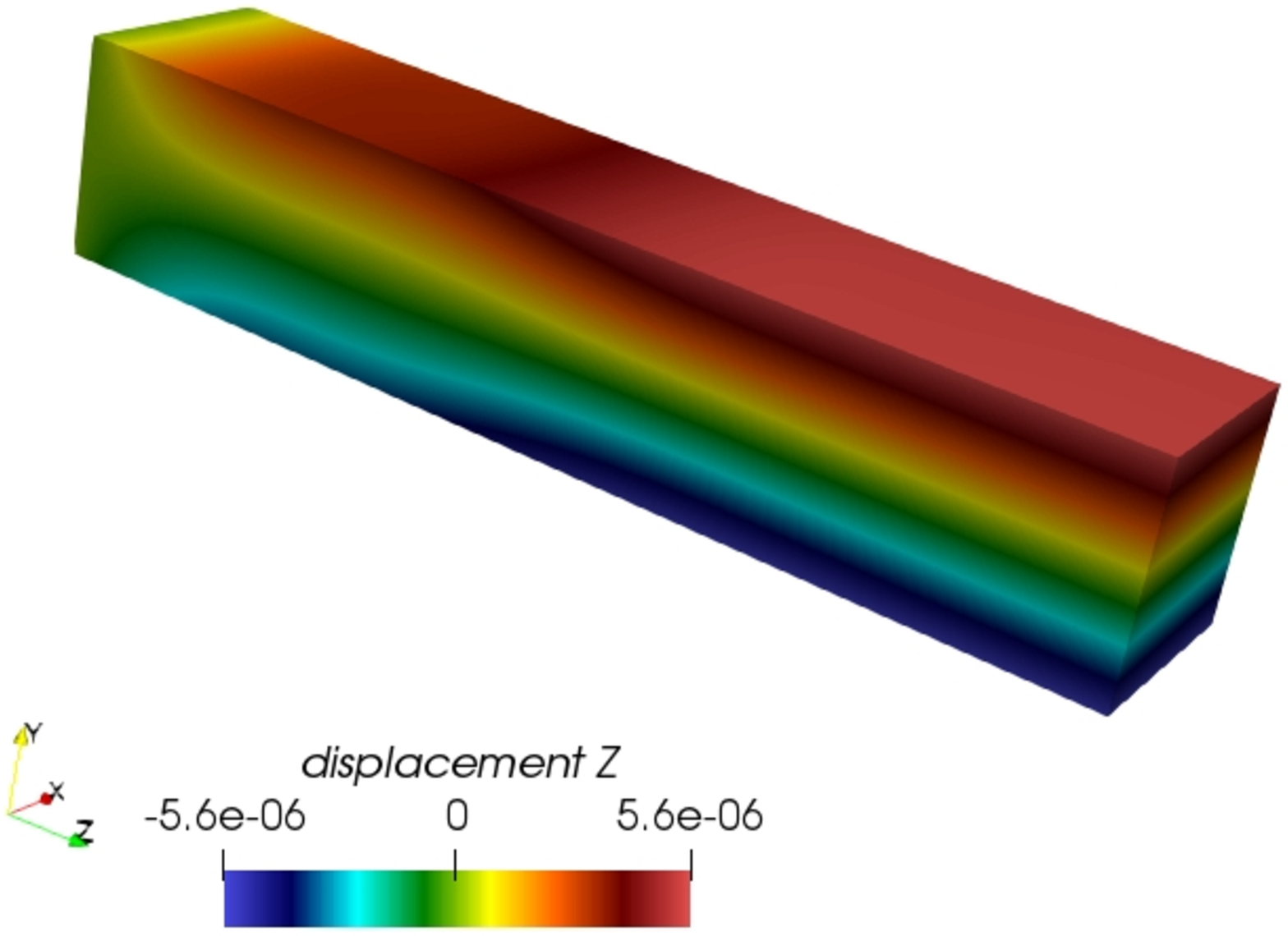}}
 \subfloat[$u_{19}$: $z$\label{subfig:pe61}]{%
 \includegraphics[width=0.35\textwidth,height=0.2\textheight]{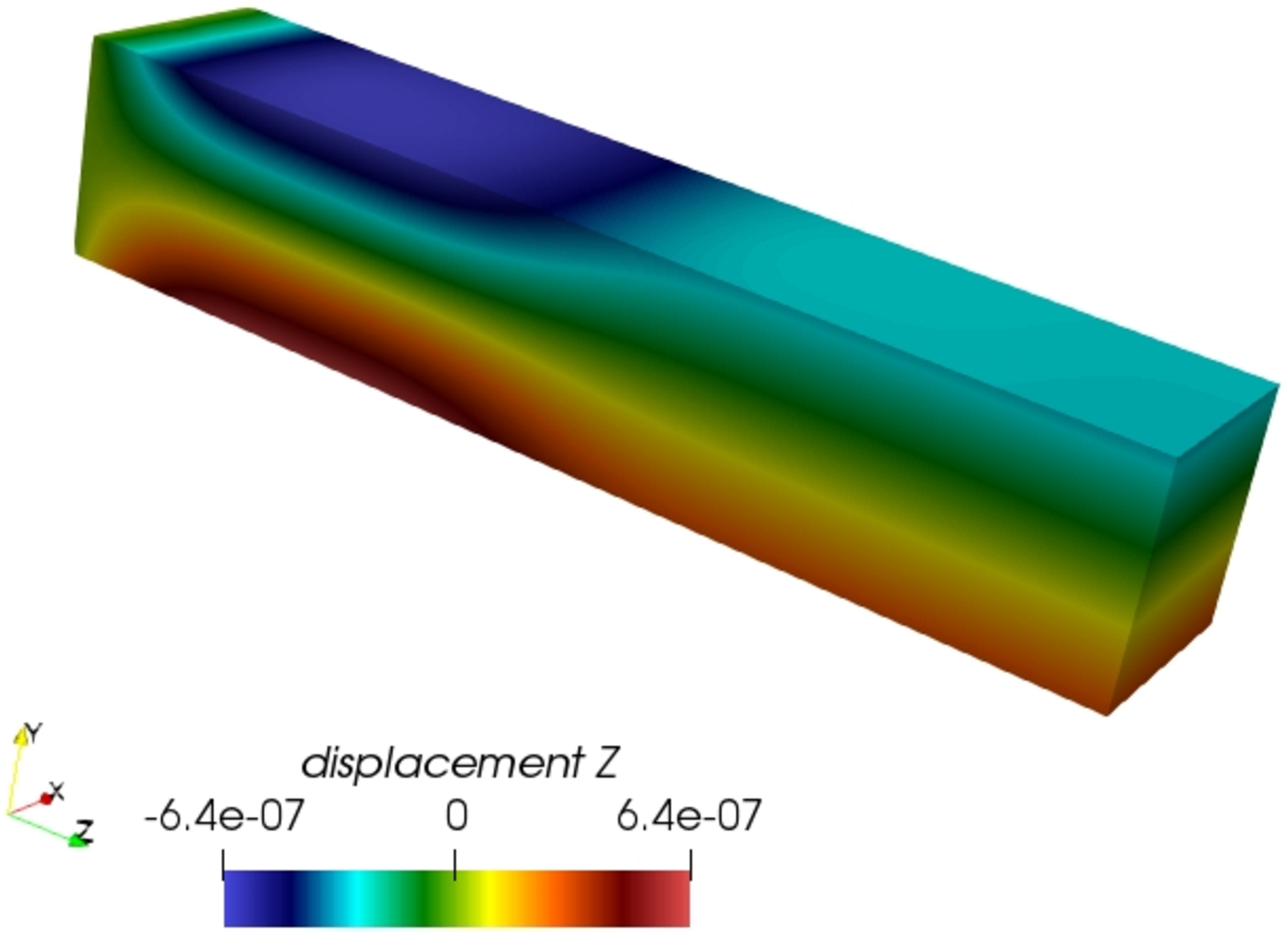}}
 \caption{Magnitude and $x,y$ and $z$ components of the selected PCE coefficients of the solution process $\mathbfcal{U}$.}
 \label{fig:pce3DE4}
\end{figure}

To get further insights into the stochastic aspects of the solution process, the first few PCE coefficients showing the magnitude and the respective displacement components of the solution process are plotted in \Crefrange{fig:pce3DE1}{fig:pce3DE4}.
Similar to the earlier observation, the contribution of the chaos coefficients $u_{j}$ to the solution process decreases with increasing PCE index $j$.
Among these coefficients, the first order coefficients contain Gaussian contributions and the higher order coefficients contain the non-Gaussian effects.
When the PCE terms increase, the solution  mean and standard deviation converge as shown in~\Cref{fig:component}.

\subsection{Comparison of Extended Wirebasket with Vertex Coarse Grid}\label{sec:3DE_compare}
The performance of the extended wirebasket-based coarse grid preconditioner is compared to the extended vertex-based coarse grid to tackle the stochastic PDE system in linear elasticity.
We focus on the numerical scalabilities of the solvers, i.e., the PCGM iteration count for  convergence with a tolerance: $$\frac{\| \mathbfcal{U}_{{\it \Gamma}_{i+1}} - \mathbfcal{U}_{{\it \Gamma}_{i}} \|_2}{\| \mathbfcal{U}_{{\it \Gamma}_{i}} \|_2} \le tol = 10^{-5},$$
where the subscript denotes iteration number.

\begin{figure}[htbp]
\centering
\includegraphics[width=0.63\textwidth]{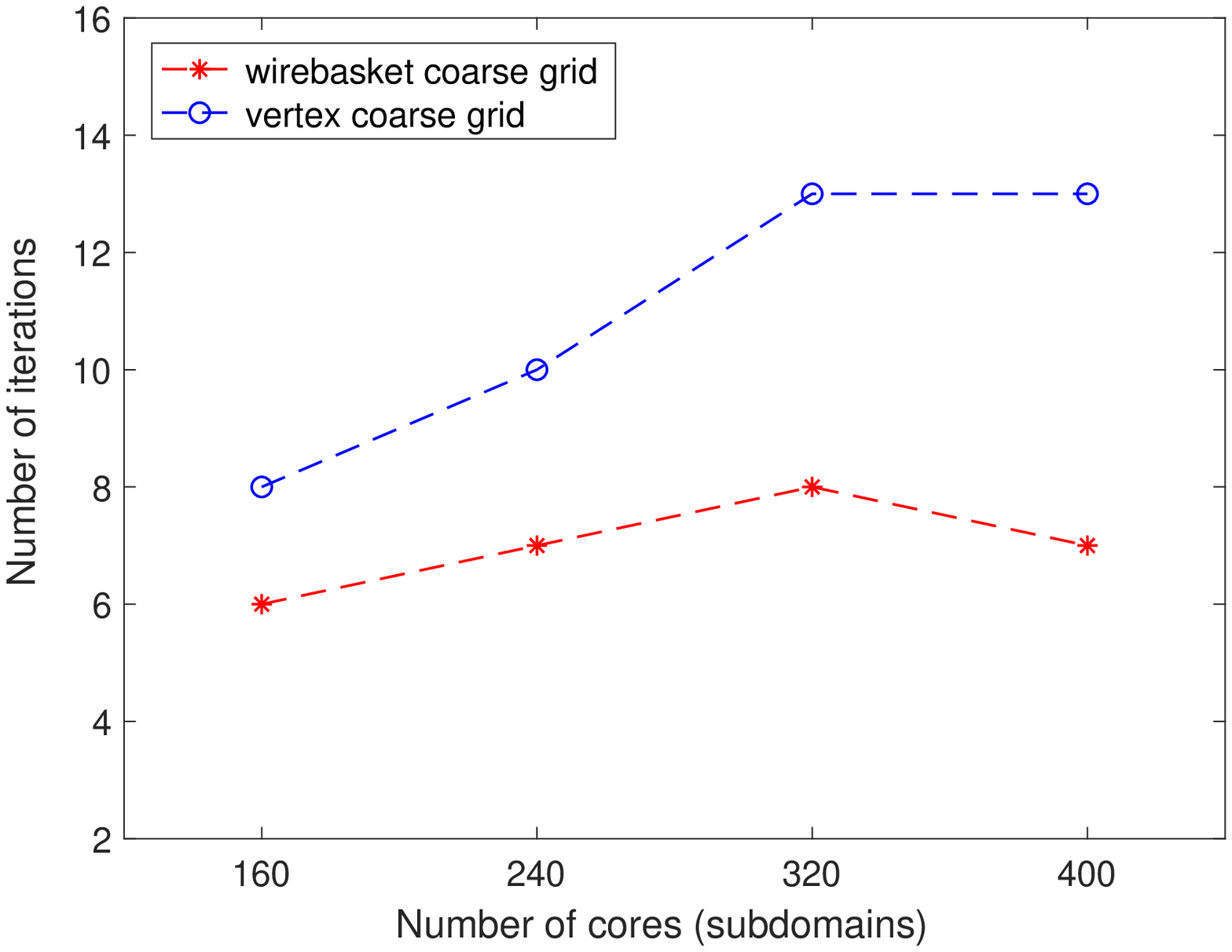}
\caption{Iteration count versus number of subdomains for the fixed number of PCE terms with fixed mesh resolution.}
\label{fig:3DE_comp_nIterVnProc_stoStrong}
\end{figure}

For a fixed mesh resolution, i.e., $31598$ nodes and $182681$ (four node) linear tetrahedral elements, with a fixed number of PCE terms ($P_u=56$ for 5 RVs), the PCGM iteration count grows quickly with the number of subdomains for the vertex-based coarse grid. The growth of the iteration count with the wirebasket-based coarse grid is smaller compared to that of the vertex-based coarse grid. \Cref{fig:3DE_comp_nIterVnProc_stoStrong} suggests that the BDDC/NNC solver with wirebasket-based coarse grid showed superior numerical scaling with respect to the number of subdomains compared to the vertex-based coarse grid.
For a fixed mesh resolution ($20753$ nodes and $119179$ linear tetrahedral element), the global problem size grows with increasing number of PCEs by changing the number of RVs $L=2, 3, 5$ and $7$. 

\begin{figure}[htbp]
\centering
\includegraphics[width=0.60\textwidth]{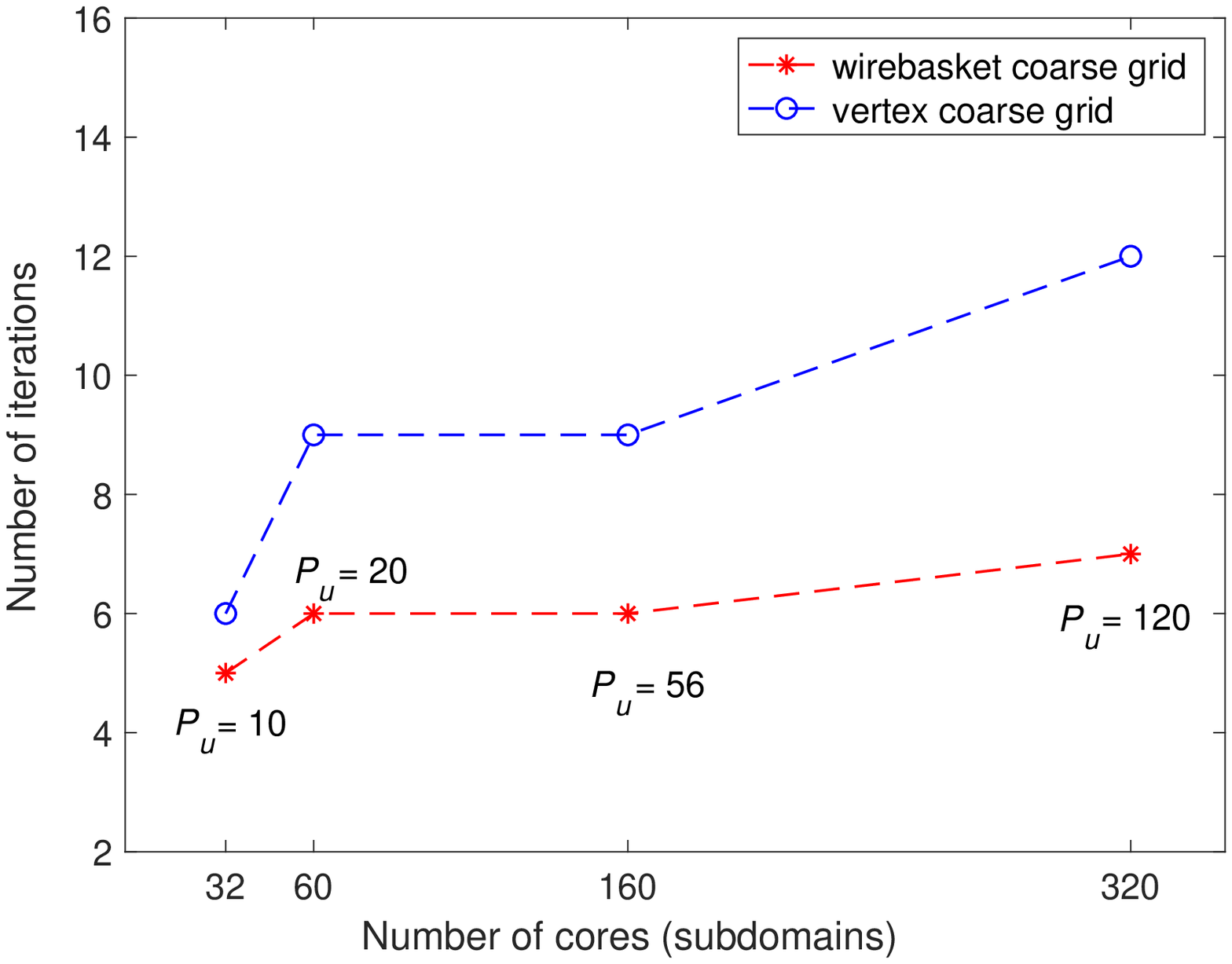}
\caption{Iteration count versus number of subdomains for the fixed mesh resolution and fixed problem size per subdomain with increasing number of PCEs.}
\label{fig:3DE_comp_nIterVnProc_stoWeak}
\end{figure}

The resulting larger system is tackled by adding more cores, maintaining the problem size per core $\approx 22500$. The iteration count for the wirebasket-based  solver  increases slightly with the subdomain numbers as shown in \Cref{fig:3DE_comp_nIterVnProc_stoWeak}. On the other hand, for the same case, the number of iterations grows quickly with the vertex-based coarse grid.
Therefore, the BDDC/NNC solver with the wirebasket-based coarse grid is superior to the vertex-based coarse grid regarding the numerical scalability against the PCE terms.
\begin{figure}[htbp]
\centering
\includegraphics[width=0.60\textwidth,height=0.30\textheight]{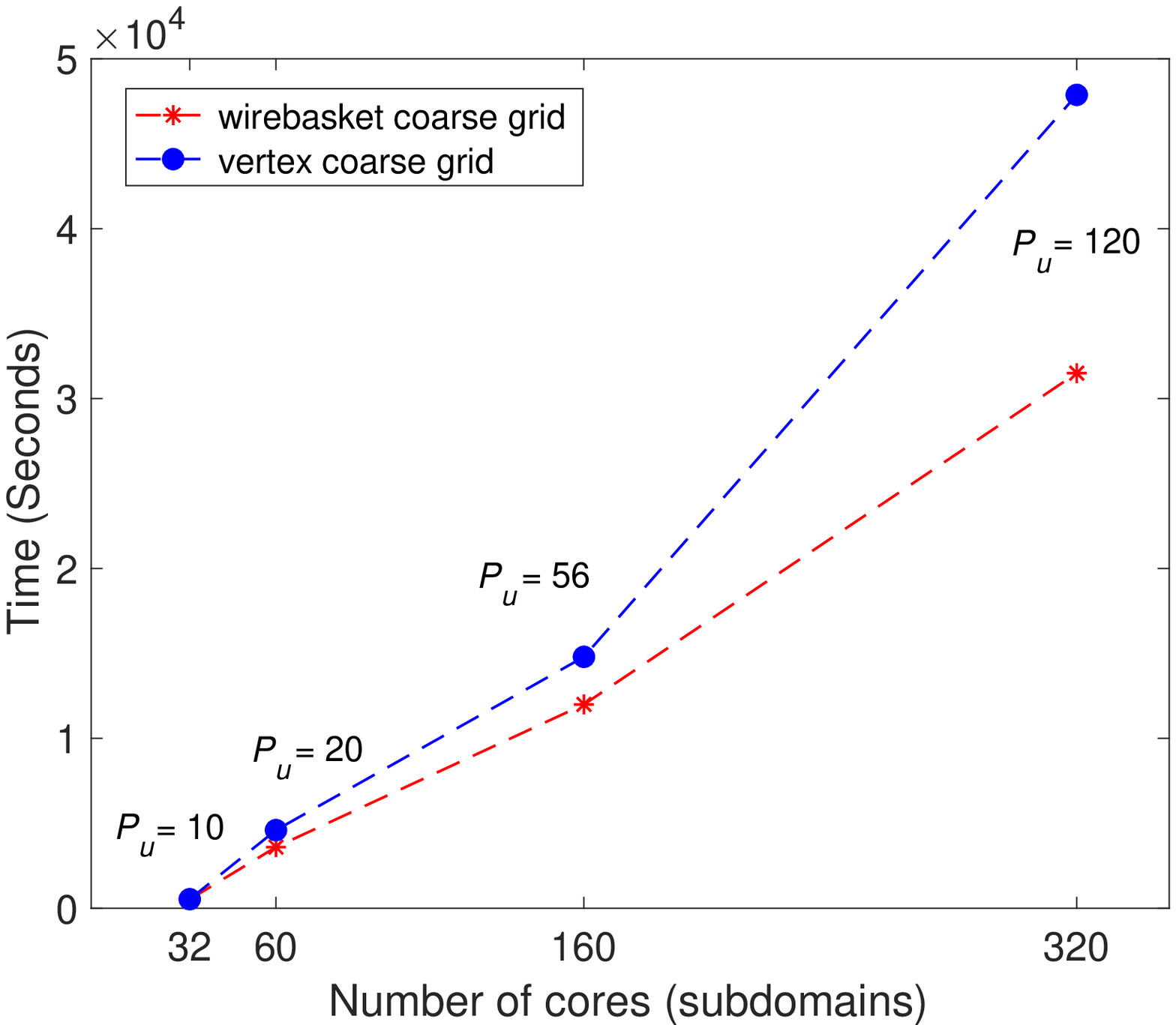}
\caption{Execution time versus number of subdomains for the fixed mesh resolution and fixed problem size per subdomain with increasing number of PCEs (fixed mesh resolution).}
\label{fig:3DE_comp_weak_timeVnProcs}
\end{figure}

For the same case as in \Cref{fig:3DE_comp_nIterVnProc_stoWeak}, the run time for the wirebasket-based  solver and vertex-based solver against the  subdomain counts (with the fixed problem size/core) are compared in~\Cref{fig:3DE_comp_weak_timeVnProcs}.
The execution time for the vertex-based solver increases faster than  the wirebasket-based solver. This is primarily due to more PCGM iteration required with the vertex-based coarse grid.

\begin{figure}[htbp]
\centering
\includegraphics[width=0.65\textwidth]{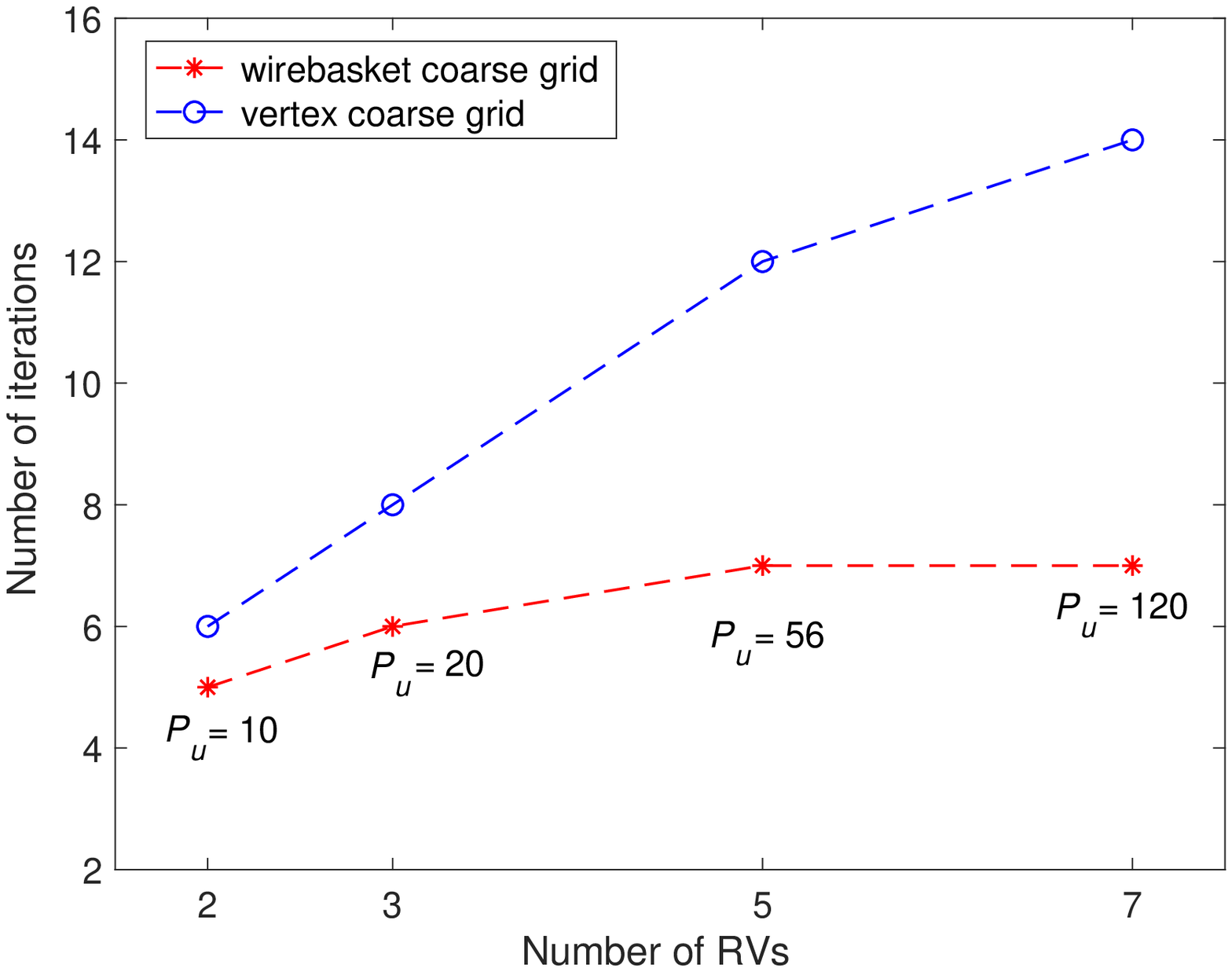}
\caption{Iteration count versus number of PCE terms for the fixed mesh having fixed number of subdomains.}
\label{fig:3DE_comp_nIterVnRVs}
\end{figure}

For a fixed mesh ($31598$ nodes and $182681$ linear tetrahedral element) and a fixed subdomain ($n_s=400$), increasing PCE terms by adding more random variables (while maintaining the third-order expansion), results in a small increase in the PCGM iteration counts for the case with the wirebasket-based coarse grid, as shown in \Cref{fig:3DE_comp_nIterVnRVs}. Conversely, for the same case, the PCGM iteration count increases faster with the vertex-based coarse grid. Therefore, the BDDC/NNC solver with the extended wirebasket-based coarse grid is superior to  vertex-based coarse grid for numerical scaling against the output PCE terms (due to increasing input RVs). 

\subsection{Scalabilities Studies for Stochastic Simulations}\label{sec:3DE_scalabilites}
Next we study the performance of the wirebasket-based BDDC/NNC solver in
high dimensional stochastic space.
Therefore, for most of the simulations, a fixed mesh (with  $48563$
nodes and  $283886$ elements) is used. For some cases, the mesh size is increased up to $89657$ nodes and $523450$ elements.
The performance plots include the number of random variables ranging from 2 to 9.  The relative contribution of the KLE eigenvalues to the total energy (variance) of the signal guides the selection of these cases (see \Cref{fig:relativeEnergy3D}).

The orders of expansion $p_{\textsc{a}} = 2$ and $p_u = 3$ are used for the input and output PCE representations, respectively.
Note that the orders of expansion, $p_{\textsc{a}}$ and $p_u$ are kept constant for most of the experiments.
For some cases $p_u$ is varied between $2$ to $4$;
resulting in 20 to 220  PCE output terms~\cite{ghanemSFEM1991}.
The maximum linear system size handled is about $60$ million, i.e., using finite element mesh with approximately $90,000$  node points and $220$ PCE terms (9-RVs).

First, we study the numerical scalability of the solver using PCGM iteration counts (with the tolerance value $tol = 10^{-5}$).
Next,  the strong and weak parallel scalabilities are investigated using the execution time of the solvers against the number of subdomains. Finally, we present the numerical and parallel scalability plots against the number of random variables and the order of PC expansion.

\subsubsection{Numerical Scalability}
\begin{figure}[htbp]
\centering
\includegraphics[width=0.6\textwidth]{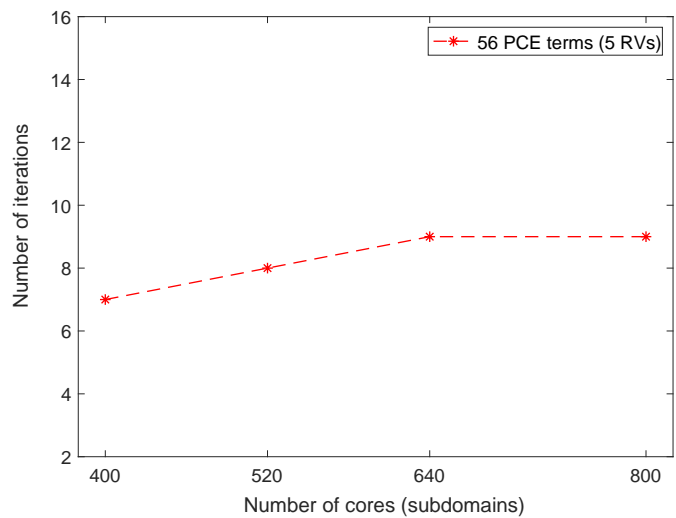}
\caption{Iteration count versus number of subdomains for the fixed number of PCE terms with fixed mesh resolution.}
\label{fig:3DE_nIterVnProc}
\end{figure}

For a fixed  PCE terms ($P_u=56$) and fixed problem size (48563 nodes and 283886 elements),  the PCGM iteration count increases slightly with increasing subdomains, indicating numerical scalability of  the wirebasket-based BDDC/NNC solver (see \Cref{fig:3DE_nIterVnProc}).

\begin{figure}[htbp]
\centering
\includegraphics[width=0.6\textwidth]{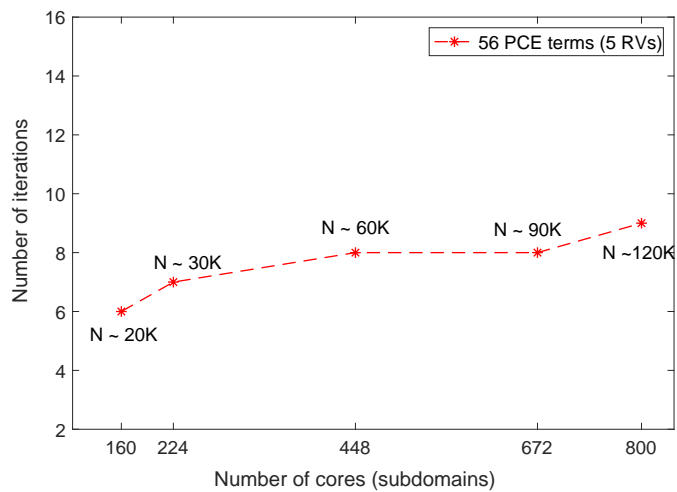}
\caption{Iteration count versus number of subdomains for the fixed number of PCE terms and the fixed problem size per core with increasing mesh resolution.}
\label{fig:3DE_weak_IterVnProc}
\end{figure}
Next, for the fixed PCE terms $(P_u=56)$, the global problem size is increased by increasing mesh resolution while keeping a fixed problem size per subdomain $(\approx 22500)$. As shown in~\Cref{fig:3DE_weak_IterVnProc}, the PCGM iteration count increases slowly. \Cref{fig:3DE_weak_IterVnProc} demonstrates a reasonable numerical scaling of the wirebasket-based  solver against mesh resolution and fixed problem size per subdomains.

\subsubsection{Parallel Scalabilities}
\begin{figure}[htbp]
\centering
\includegraphics[width=0.65\textwidth]{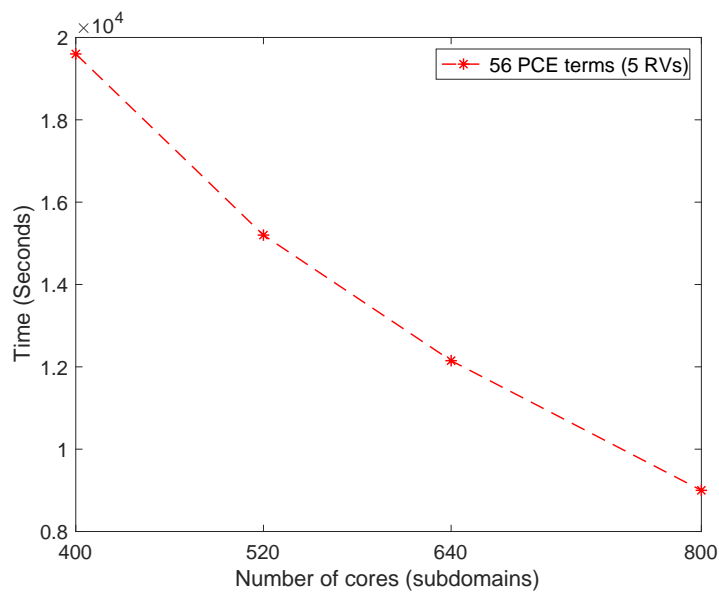}
\caption{Execution time versus number of subdomains with the number of PCE terms and the fixed mesh resolution.}
\label{fig:3DE_timeVnProcs}
\end{figure}

For a fixed  mesh (48563 nodes and 283886 elements) and fixed PCE terms ($P_u$ = 56),  the execution time reduces with increasing cores as shown in \Cref{fig:3DE_timeVnProcs}. This fact indicates an excellent strong scaling of the BDDC/NNC solver with the wirebasket based coarse grid.

\begin{figure}[htbp]
\centering
\includegraphics[width=0.65\textwidth]{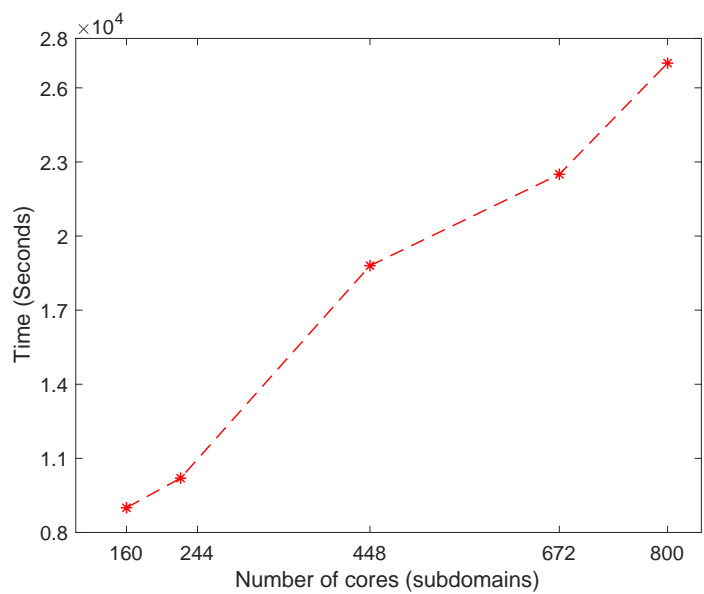}
\caption{Execution time versus number of subdomains for the fixed number of PCE terms and the fixed problem size per core with increasing mesh resolution.}
\label{fig:3DE_weak_timeVnProcs}
\end{figure}

 For a fixed PCE terms $(P_u=56)$, the global problem size is increased by increasing mesh resolution while keeping a fixed problem size per subdomain $(\approx 22500)$; and the  corresponding results are shown in \Cref{fig:3DE_weak_timeVnProcs}.
The suboptimal weak scaling is evident in \Cref{fig:3DE_weak_timeVnProcs} as the execution time grows with core counts.
This is mainly because of the increased parallel overhead with the number of cores caused by the MPI collective communication~\cite{desai2017scalable,subber2012PhDTh}.
Also, in the current case, the global coarse problem size grows quickly with subdomains due to vector-valued solution process. Consequently, the resulting coupled system of equations is computationally challenging. Therefore, it demands significantly more computational efforts compared to that of scalar-valued stochastic PDEs.

\subsubsection{Scalability with respect to Stochastic Parameters}
In this section, the numerical and parallel scalabilities against stochastic parameters such as the number PCE terms and the order of expansions are presented for a mesh having 48563 nodes and 283886 elements.

\begin{figure}[htbp]
\centering
\includegraphics[width=0.65\textwidth]{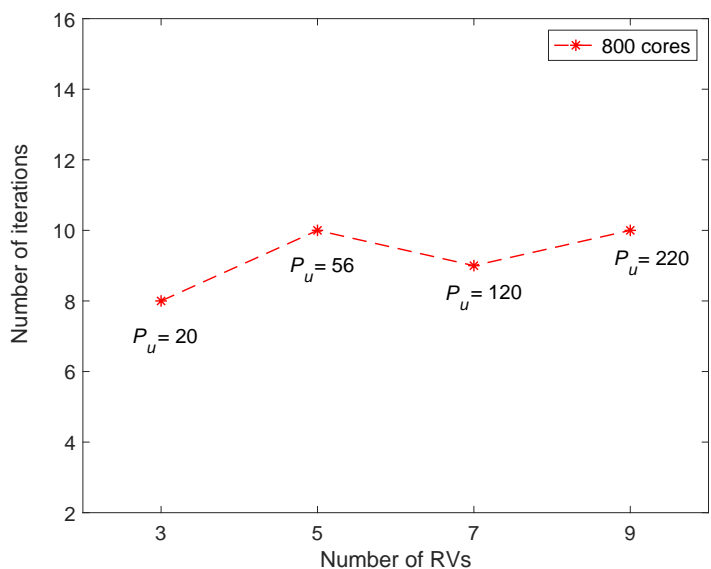}
\caption{Iteration count versus number of PCE terms for the fixed mesh resolution with fixed number of subdomains.}
\label{fig:3DE_nIterVnPCE}
\end{figure}

\begin{figure}[htbp]
\centering
\includegraphics[width=0.65\textwidth]{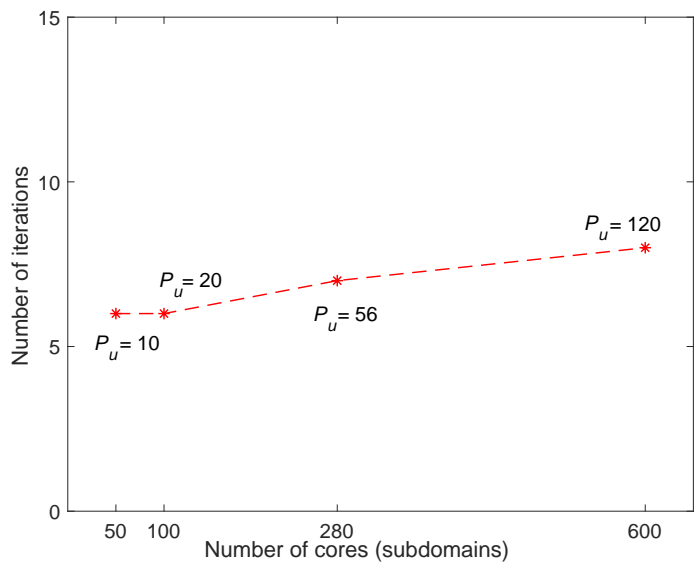}
\caption{Iteration count versus number of subdomains for increasing number of PCE terms (fixed mesh resolution with the fixed problem size per core).}
\label{fig:3DE_StoWeak_IterVnProc}
\end{figure}

The four points in \Cref{fig:3DE_nIterVnPCE} and \ref{fig:3DE_StoWeak_IterVnProc} represent  PCE terms 20, 56, 120 and 220 associated with random variables 3, 5, 7 and 9 respectively.
For a fixed mesh with $n_s=800$ subdomains, increasing  PCE terms associated with  increasing random variables (with the third-order output PC expansion) lead to a slight variation in the PCGM iteration counts (see \Cref{fig:3DE_nIterVnPCE}).
Hence the wirebasket-based  solver demonstrates good numerical scaling with increasing random variables. 

\begin{figure}[htbp]
\centering
\includegraphics[width=0.65\textwidth]{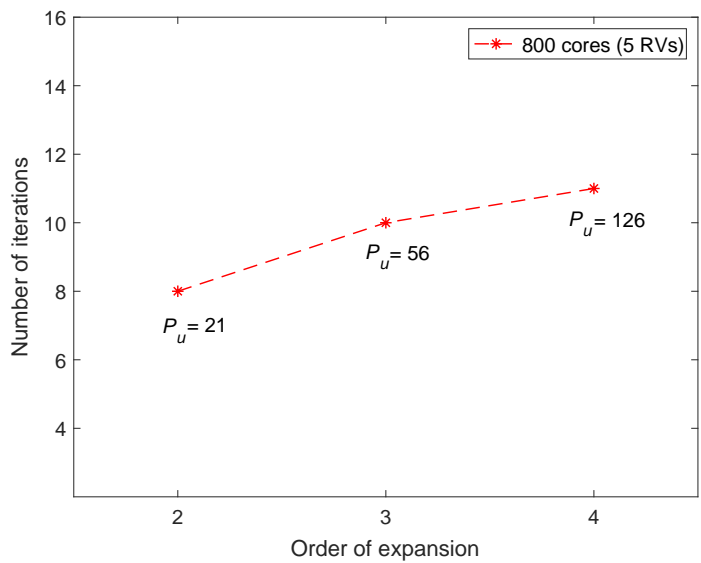}
\caption{Iteration count versus order of expansion for the fixed mesh resolution with fixed number of subdomains (fixed number of RVs).}
\label{fig:3DE_nIterVnOrd}
\end{figure}

For a fixed mesh, the global problem size grows with increasing PCE terms. The resulting larger system is tackled using additional cores (by fixing the problem size per core $\approx 18000$). The iteration count of the wirebasket based  solver against subdomain numbers remains nearly constant (\Cref{fig:3DE_StoWeak_IterVnProc}). Hence the wirebasket-based BDDC/NNC-PCGM solver showed acceptable numerical scaling with respect to  PCE terms.

For a fixed mesh with   $n_s=800$ subdomains, PCGM iterations grow
moderately with increasing  PCE orders $p_u$ (with random variables $L=5$)   as shown in \Cref{fig:3DE_nIterVnOrd}.
Hence the wirebasket-based  solver exhibits satisfactory numerical scaling with respect to the order of PC expansion.

\section{Conclusion}\label{sec:conclusion}
In summary,  extending the DD preconditioner with a wirebasket-based coarse grid~\cite{bramble1989construction,smith1991domain,dryja1990some,mandel1990two} for deterministic PDEs, we formulate a  wirebasket-based two-level preconditioner for the stochastic PDEs. The BDDC/NNC solver with the extended wirebasket-based coarse grid is employed to tackle three-dimensional (scalar and vector-valued) stochastic PDEs in high-dimensional stochastic spaces.
The novel domain decomposition preconditioner based on the extended wirebasket-based coarse grid is shown to outperform vertex-based coarse grid for stochastic PDEs in three dimensions.
The stochastic system matrix assembly procedure can utilize existing deterministic FEM packages and thereby reduces the implementational burden of DD-based intrusive SSFEM.
Various sparse data structures, routines, iterative solvers and preconditioners are leveraged to tackle the subdomain-level systems.
These implementational advances cut execution time and memory usage, and
potentially paves the way to tackle extreme-scale uncertainty quantification problems in stochastic PDEs using high performance computing.
Numerous simulations are performed to thoroughly investigate the performance of the extended wirebasket-based BDDC/NNC solver to tackle large-scale high-dimensional stochastic PDEs. The  numerical and parallel scalings (against subdomain numbers and  PCE terms) are essential to understand convergence behavior and utility of domain decomposition solvers for stochastic PDEs.

Although the extended wirebasket based DD preconditioner showed better numerical scaling for the stochastic PDEs illustrated here, the size of the extended coarse problem arising in this setting grows rapidly with (a) the mesh resolution, (b) PCE terms and (c) subdomain numbers. Therefore,  the coarse problem solver becomes computational bottleneck in regards to memory requirement and floating point operations. This problem, for instance, can be alleviated by extending the recently proposed algorithms in the deterministic setting which enforces only average constraints over selected edges~\cite{klawonn2006dual,klawonn2006parallel} and may  reduce
the cost of the coarse problem solver for stochastic PDEs. These aspects are under current investigation.

\section*{Acknowledgments}
The first author acknowledges  the support of  an Ontario Trillium scholarship.
The fourth author acknowledges  the support of the Department of National Defence, Canada and a Discovery Grant from Natural Sciences and Engineering Research Council of Canada.   The fifth author acknowledges the support of a  Discovery Grant from Natural Sciences and Engineering Research Council of Canada. The computing infrastructure is supported by the Canada Foundation for Innovation (CFI), the Ontario Innovation Trust (OIT),  Compute Canada and Calcul Qu\'ebec.

{\it Sandia National Laboratories is a multimission laboratory managed and operated by National Technology \& Engineering Solutions of Sandia, LLC, a wholly owned subsidiary of Honeywell International Inc., for the U.S. Department of Energy’s National Nuclear Security Administration under contract DE-NA0003525. This paper describes objective technical results and analysis. Any subjective views or opinions that might be expressed in the paper do not necessarily represent the views of the U.S. Department of Energy or the United States Government.}

% \clearpage
 %\section*{References}
%\bibliographystyle{elsarticle-num}
%\bibliography{references}

%\newpage
\appendix

\section{Parallel Implementational Details of  Domain Decomposition Solvers for SPDEs}\label{sec:ddmImplementation}
The distributed implementation of two-level domain decomposition solvers 
uses PCGM algorithm~\cite{subber2012PhDTh,desai2019scalable} with MPI and PETSc libraries~\cite{petsc2016}, as  described in~\Cref{alg:pcgm}~\cite{subber2012PhDTh,desai2019scalable}. 
The procedure (e.g. ~\cite{subber2012PhDTh,desai2019scalable}) is illustrated for the preconditioned extended Schur complement system below:

\begin{equation}\label{eq:3DE_Schur2}
\mathbfcal{S} {\mathbfcal{U}}_{\it\Gamma} = \textbf{\textit{g}}_{\it\Gamma},
\end{equation}

%%***************************************************************************************
\alglanguage{pseudocode}
\begin{algorithm}
\caption{: Parallel PCGM Algorithm}
\label{alg:pcgm}
\begin{algorithmic}[1]
\State {\bf{Input :}} $\mathbfcal{U}_{\it\Gamma_0}$
\State Initial Residual : $r_{\it{\Gamma}_0}  = \textbf{\textit{g}}_{\it\Gamma} - \mathbfcal{S} \mathbfcal{U}_{\it{\Gamma}_0}$
\State Preconditioned Residual : $z_0 = \mathbfcal{M}^{-1} \ r_{\it{\Gamma}_0}$
\State Compute : $p_0 = z_0$
\State Compute : ${\delta}_0 = (r_{\it\Gamma_0} , z_0) $
\For{each iteration, $i = 0, 1, 2, ... \ , $}
\State \textcolor{blue}{\bf{Parallel Mat-Vec Product Procedure}} : $q_i = \mathbfcal{S} p_i$
\State Compute : ${\gamma}_i = (q_i  \ , \ p_i)$
\State Compute : ${\alpha}_i = {\delta}_i / {\gamma}_i$
\State Update : $u_{\it\Gamma_{i+1}} =  u_{\it\Gamma_i} +  {\alpha}_i p_i $
\State Update : $r_{{\it\Gamma}_{i+1}} =  r_{{\it\Gamma}_i} -  {\alpha}_i q_i $
\State {\bf{{If}}} {\it{iteration has converged,}} {\bf{Exit}}
\State  \textcolor{blue}{\bf{Parallel Preconditioning Procedure}} : $z_{i+1} = \mathbfcal{M}^{-1} \ r_{{\it\Gamma}_{i+1}} $
\State Compute : ${\delta}_{i+1} = (r_{{\it\Gamma}_{i+1}} , z_{i+1})$
\State Compute : ${\beta}_{i} = {\delta}_{i+1} / {\delta}_i $
\State Update : $p_{i+1} =  z_{i+1} + {\beta}_{i}  p_i $
\EndFor
\State {\bf{Output :}}  $\mathbfcal{U}_{\it\Gamma}$
\end{algorithmic}
\end{algorithm}
%%***************************************************************************************

 The steps 7 and 13  (in blue) are computationally intensive; hence they need
careful implementation. The matrix-vector product and effect of preconditioner on the residual are performed concurrently on each subdomain.
Explicit construction of the preconditioner $(M^{-1})$ and Schur complement matrix $(S)$ are avoided, and Parallel Preconditioning (PP) procedure and Parallel Mat-Vec Product (PMVP) procedure~(refer to ~\cite{desai2019scalable} for more details) are leveraged instead. For large-scale systems (both due to spatial and stochastic resolutions), steps 7 and 13 are critical for minimizing  memory requirement and floating-point operations.
 
The following algorithm outlines the stochastic finite element assembly 
procedure of  a typical block of the subdmain-level matrix defined as:

\begin{align}
\left[{\mathbfcal{A}_{\gamma\delta}^s}\right]_{jk} &= \sum_{i=0}^{P_{\textsc{A}}}\left< \psi_i \psi_j \psi_k \right> \bar{\textbf{\textit{A}}}_{\gamma\delta,i}^s, \label{eq:3d_Stoassembly2} 
\end{align}

%%***************************************************************************************
\alglanguage{pseudocode}
\begin{algorithm}
\caption{: {Stochastic Finite Element Assembly for DD Blocks}}
\label{alg:smap}
\begin{algorithmic}[1]
\State {\bf{Input :}} non-zero : $ijk$ and $C_{ijk}$
\State {Create :} \textcolor{red}{PETSc-$MatCreateSeqAIJ$} ($[{\mathbfcal{A}_{\gamma\delta}^s}]$)
\For{$i = 1, 2, ..., P_{\textsc{a}}$}
\State  {Initialize :} $[\bar{\textbf{\textit{A}}}_{\gamma\delta,i}^s]_i$
\State  {Call :}  \textcolor{magenta}{Deterministic Finite Element Matrix Assembly} ($[\bar{\textbf{\textit{A}}}_{\gamma\delta,i}^s]_i$)
\For{$j = 1, 2, ... P_{u}$}
\For{$k = 1, 2, ... P_{u}$}
\If {$( \ i,j,k == nonZero(ijk) \ ) $}
\State  {Insert :}  \textbf{call} \textcolor{red}{PETSc-$MatSetValues$}($\{C_{ijk} \times [\bar{\textbf{\textit{A}}}_{\gamma\delta,i}^s]_i\} => [{\mathbfcal{A}_{\gamma\delta}^s}]$)
%\EndIf
 %\EndFor
 %\EndFor
\EndIf
 \EndFor
 \EndFor
 \State  {Destroy :} $[\bar{\textbf{\textit{A}}}_{\gamma\delta,i}^s]_i$
 \EndFor
 \State {\bf{Output :}}  $[{\mathbfcal{A}_{\gamma\delta}^s}]$
\end{algorithmic}
\end{algorithm}
%%*************************

Clearly the memory requirement to assemble and solve the  subdmain-level
system limits the maximum size of the global problem.
For subdomain-level (local) systems,  sparse PETSc~\cite{petsc2016} data structures and solvers are used for effective memory usage in assembly, storage and matrix-vector operations.
The local sparse PETSc matrices are stored in compressed row format using
 PETSc {\it{MatCreateSeqAIJ}} routine ~\cite{petsc2016}).

%START HERE! (ithenticate page 52 and paper pdf 52)!!!\\
\Cref{alg:smap} outlines the assembly procedure of a typical block of the 
subdomain-level stiffness matrix, corresponding to \Cref{eq:3d_Stoassembly2} in which stochastic finite element matrix blocks are constructed with deterministic finite element blocks.
The PETSc {\it{Mat}} sparse data structure is used for efficient memory
usage. For effective memory reusage (e.g. in handling a large number of random variables), the deterministic matrix blocks are destroyed after completing the stochastic finite element matrix assembly.

%Bibliography
\bibliographystyle{unsrt}  
\bibliography{references}  

\begin{thebibliography}{10}

\bibitem{sarkarIJNME2009}
Abhijit Sarkar, Nabil Benabbou, and Roger Ghanem.
\newblock Domain decomposition of stochastic {PDE}s: theoretical formulations.
\newblock {\em International Journal for Numerical Methods in Engineering},
  77(5):689--701, 2009.

\bibitem{subberJCP2014}
Waad Subber and Abhijit Sarkar.
\newblock A domain decomposition method of stochastic {PDE}s: An iterative
  solution techniques using a two-level scalable preconditioner.
\newblock {\em Journal of Computational Physics}, 257:298--317, 2014.

\bibitem{subberCMAME2013}
Waad Subber and Abhijit Sarkar.
\newblock Dual-primal domain decomposition method for uncertainty
  quantification.
\newblock {\em Computer Methods in Applied Mechanics and Engineering},
  266:112--124, 2013.

\bibitem{subber2012PhDTh}
Waad Subber.
\newblock {\em Domain decomposition methods for uncertainty quantification}.
\newblock PhD thesis, Carleton University Ottawa, 2012.

\bibitem{desai2017scalable}
Ajit Desai, Mohammad Khalil, Chris Pettit, Dominique Poirel, and Abhijit
  Sarkar.
\newblock Scalable domain decomposition solvers for stochastic {PDE}s in high
  performance computing.
\newblock {\em Computer Methods in Applied Mechanics and Engineering},
  335:194--222, 2017.

\bibitem{desai2019scalable}
Ajit Desai.
\newblock {\em Scalable Domain Decomposition Algorithms for Uncertainty
  Quantification in High Performance Computing}.
\newblock PhD thesis, Carleton University, 2019.

\bibitem{bramble1989construction}
James~H Bramble, Joseph~E Pasciak, and Alfred~H Schatz.
\newblock The construction of preconditioners for elliptic problems by
  substructuring. iv.
\newblock {\em Mathematics of Computation}, 53(187):1--24, 1989.

\bibitem{smith1991domain}
Barry~F Smith.
\newblock A domain decomposition algorithm for elliptic problems in three
  dimensions.
\newblock {\em Numerische Mathematik}, 60(1):219--234, 1991.

\bibitem{smith2004domain}
Barry Smith, Petter Bjorstad, and William Gropp.
\newblock {\em Domain decomposition: parallel multilevel methods for elliptic
  partial differential equations}.
\newblock Cambridge University Press, 2004.

\bibitem{toselli2005domain}
Andrea Toselli and Olof Widlund.
\newblock {\em Domain decomposition methods: algorithms and theory}, volume~34.
\newblock Springer, 2005.

\bibitem{smith1992optimal}
Barry~F Smith.
\newblock An optimal domain decomposition preconditioner for the finite element
  solution of linear elasticity problems.
\newblock {\em SIAM Journal on Scientific and Statistical Computing},
  13(1):364--378, 1992.

\bibitem{ghosh2009feti}
Debraj Ghosh, Philip Avery, and Charbel Farhat.
\newblock A {FETI}-preconditioned conjugate gradient method for large-scale
  stochastic finite element problems.
\newblock {\em International Journal for Numerical Methods in Engineering},
  80(6-7):914--931, 2009.

\bibitem{dryja1990some}
Maksymilian Dryja and Olof~B Widlund.
\newblock Some domain decomposition algorithms for elliptic problems.
\newblock In {\em Iterative methods for large linear systems}, pages 273--291.
  Elsevier, 1990.

\bibitem{mandel1990two}
Jan Mandel.
\newblock Two-level domain decomposition preconditioning for the p-version
  finite element method in three dimensions.
\newblock {\em International Journal for Numerical Methods in Engineering},
  29(5):1095--1108, 1990.

\bibitem{mandel1990iterative}
Jan Mandel.
\newblock Iterative solvers by substructuring for the p-version finite element
  method.
\newblock {\em Computer Methods in Applied Mechanics and Engineering},
  80(1-3):117--128, 1990.

\bibitem{logg2012FEniCS}
Anders Logg, Andre Mardal, and Garth Wells.
\newblock {\em Automated solution of differential equations by the finite
  element method: The {FEniCS} book}, volume~84.
\newblock Springer Science \& Business Media, 2012.

\bibitem{fenics/dolfin17}
Anders Logg, Garth Wells, and Johan Hake.
\newblock {DOLFIN}: A {C}++/{P}ython finite element library.
\newblock In {\em Automated Solution of Differential Equations by the Finite
  Element Method}. Springer, 2012.

\bibitem{le2010spectral}
Olivier Le~Ma{\^\i}tre and Omar~M Knio.
\newblock {\em Spectral methods for uncertainty quantification: with
  applications to computational fluid dynamics}.
\newblock Springer Science \& Business Media, 2010.

\bibitem{eldred2009comparison}
MS~Eldred and John Burkardt.
\newblock Comparison of non-intrusive polynomial chaos and stochastic
  collocation methods for uncertainty quantification.
\newblock {\em AIAA paper}, 976:1--20, 2009.

\bibitem{reagana2003uncertainty}
Matthew Reagana, Habib Najm, Roger Ghanem, and Omar Knio.
\newblock Uncertainty quantification in reacting-flow simulations through
  non-intrusive spectral projection.
\newblock {\em Combustion and Flame}, 132(3):545--555, 2003.

\bibitem{nationalSystems}
Graham-cedar: New national heterogeneous {HPC} clusters managed by {C}ompute
  {C}anada, 2017.

\bibitem{keyes1987comparison}
David~E Keyes and William~D Gropp.
\newblock A comparison of domain decomposition techniques for elliptic partial
  differential equations and their parallel implementation.
\newblock {\em {SIAM} Journal on Scientific and Statistical Computing},
  8(2):s166--s202, 1987.

\bibitem{ghanem1999ingredients}
Roger Ghanem.
\newblock Ingredients for a general purpose stochastic finite elements
  implementation.
\newblock {\em Computer Methods in Applied Mechanics and Engineering},
  168(1):19--34, 1999.

\bibitem{ghanemSFEM1991}
Roger Ghanem and Pol Spanos.
\newblock {\em Stochastic finite elements: a spectral approach}.
\newblock Springer-Verlag, New York, 1991.

\bibitem{desai2010analysis}
Ajit Desai and Sunetra Sarkar.
\newblock Analysis of a nonlinear aeroelastic system with parametric
  uncertainties using polynomial chaos expansion.
\newblock {\em Mathematical Problems in Engineering}, 2010, 2010.

\bibitem{mathew2008domain}
Tarek Mathew.
\newblock {\em Domain decomposition methods for the numerical solution of
  partial differential equations}, volume~61.
\newblock Springer Science \& Business Media, 2008.

\bibitem{billingsley2008probability}
Patrick Billingsley.
\newblock {\em Probability and measure}.
\newblock John Wiley \& Sons, 2008.

\bibitem{gropp1999MPI}
William Gropp, Ewing Lusk, and Anthony Skjellum.
\newblock {\em Using {MPI}: portable parallel programming with the
  message-passing interface}, volume~1.
\newblock MIT press, 1999.

\bibitem{petsc2016}
Satish Balay, Shrirang Abhyankar, Mark Adams, Jed Brown, Peter Brune, Kris
  Buschelman, Lisandro Dalcin, Victor Eijkhout, William Gropp, Dinesh Kaushik,
  Matthew Knepley, Dave May, Lois McInnes, Karl Rupp, Patrick Sanan, and Barry
  Smith.
\newblock {PETS}c users manual 3.7.
\newblock Technical report, Argonne National Laboratory, 2017.

\bibitem{geuzaine2009gmsh}
Christophe Geuzaine and Fran{\c{c}}ois Remacle.
\newblock {GMSH}: A {3-D} finite element mesh generator with built-in pre-and
  post-processing facilities.
\newblock {\em International Journal for Numerical Methods in Engineering},
  79(11):1309--1331, 2009.

\bibitem{gmshWeb2017}
{GMSH}- a three-dimensional finite element mesh generator, 2017.

\bibitem{karypis1995metis}
George Karypis and Vipin Kumar.
\newblock {METIS}--unstructured graph partitioning and sparse matrix ordering
  system, version 2.0, 1995.

\bibitem{metisWeb2017}
{METIS}- serial graph partitioning, 2017.

\bibitem{debusschere2013uqtk}
Bert Debusschere, Khachik Sargsyan, and Cosmin Safta.
\newblock {UQTk} version 2.1 user manual.
\newblock Technical report, Sandia National Laboratory (SNL), 2013.

\bibitem{ahrens2005ParaView}
James Ahrens, Berk Geveci, and Charles Law.
\newblock {ParaView}: An end-user tool for large-data visualization, 2005.

\bibitem{paraviewWeb2017}
{ParaView}- multi-platform data analysis and visualization application, 2017.

\bibitem{matlabGuide}
The {M}athworks {I}nc: {M}atlab user's guide, 2015.

\bibitem{nationalSystemsNia}
Niagara supercomputer: New national homogeneous {HPC} clusters managed by
  {C}ompute {C}anada, 2018.

\bibitem{doostan2007stochastic}
Alireza Doostan, Roger~G Ghanem, and John Red-Horse.
\newblock Stochastic model reduction for chaos representations.
\newblock {\em Computer Methods in Applied Mechanics and Engineering},
  196(37-40):3951--3966, 2007.

\bibitem{ghanem2008probabilistic}
Roger~G Ghanem, Alireza Doostan, and John Red-Horse.
\newblock A probabilistic construction of model validation.
\newblock {\em Computer Methods in Applied Mechanics and Engineering},
  197(29-32):2585--2595, 2008.

\bibitem{klawonn2006dual}
Axel Klawonn and Olof~B Widlund.
\newblock Dual-primal {FETI} methods for linear elasticity.
\newblock {\em Communications on Pure and Applied Mathematics: A Journal Issued
  by the Courant Institute of Mathematical Sciences}, 59(11):1523--1572, 2006.

\bibitem{klawonn2006parallel}
Axel Klawonn and Oliver Rheinbach.
\newblock A parallel implementation of dual-primal {FETI} methods for
  three-dimensional linear elasticity using a transformation of basis.
\newblock {\em SIAM Journal on Scientific Computing}, 28(5):1886--1906, 2006.

\end{thebibliography}

\end{document}